\documentclass[twocolumn]{aastex61}
\usepackage{comment}

\newcommand{\myemail}{konyvestoth.reka@csfk.mta.hu}

\shorttitle{Constraints on ejecta parameters of SNe Ia}
\shortauthors{K\"onyves-T\'oth et al.}

\begin{document}

\title{Constraints on the physical properties of Type Ia supernovae from photometry}

\author{R. K\"onyves-T\'oth}
\affiliation{Konkoly Observatory, CSFK, Konkoly-Thege M. ut 15-17, Budapest, Hungary}
\email{\myemail}

\author{J. Vink\'o}
\affiliation{Konkoly Observatory, CSFK, Konkoly-Thege M. ut 15-17, Budapest, Hungary}
\affiliation{Department of Optics and Quantum Electronics, University of Szeged, Hungary}

\author{A. Ordasi}
\affiliation{Konkoly Observatory, CSFK, Konkoly-Thege M. ut 15-17, Budapest, Hungary}

\author{K. S\'arneczky}
\affiliation{Konkoly Observatory, CSFK, Konkoly-Thege M. ut 15-17, Budapest, Hungary}

\author{A. B\'odi}
\affiliation{Konkoly Observatory, CSFK, Konkoly-Thege M. ut 15-17, Budapest, Hungary}
\affiliation{CSFK Lend\"ulet Near-Field Cosmology Research Group}

\author{B. Cseh}
\affiliation{Konkoly Observatory, CSFK, Konkoly-Thege M. ut 15-17, Budapest, Hungary}

\author{G. Cs\"ornyei}
\affiliation{Konkoly Observatory, CSFK, Konkoly-Thege M. ut 15-17, Budapest, Hungary}

\author{Z. Dencs}
\affiliation{Konkoly Observatory, CSFK, Konkoly-Thege M. ut 15-17, Budapest, Hungary}

\author{O. Hanyecz}
\affiliation{Konkoly Observatory, CSFK, Konkoly-Thege M. ut 15-17, Budapest, Hungary}

\author{B. Ign\'acz}
\affiliation{Konkoly Observatory, CSFK, Konkoly-Thege M. ut 15-17, Budapest, Hungary}

\author{Cs. Kalup}
\affiliation{Konkoly Observatory, CSFK, Konkoly-Thege M. ut 15-17, Budapest, Hungary}

\author{L. Kriskovics}
\affiliation{Konkoly Observatory, CSFK, Konkoly-Thege M. ut 15-17, Budapest, Hungary}

\author{A. P\'al}
\affiliation{Konkoly Observatory, CSFK, Konkoly-Thege M. ut 15-17, Budapest, Hungary}

\author{B. Seli}
\affiliation{Konkoly Observatory, CSFK, Konkoly-Thege M. ut 15-17, Budapest, Hungary}

\author{\'A. S\'odor}
\affiliation{Konkoly Observatory, CSFK, Konkoly-Thege M. ut 15-17, Budapest, Hungary}

\author{R. Szak\'ats}
\affiliation{Konkoly Observatory, CSFK, Konkoly-Thege M. ut 15-17, Budapest, Hungary}

\author{P. Sz\'ekely}
\affiliation{Department of Experimental Physics,University of Szeged, D\'om t\'er 9., Szeged, Hungary}

\author{E. Varga-Vereb\'elyi}
\affiliation{Konkoly Observatory, CSFK, Konkoly-Thege M. ut 15-17, Budapest, Hungary}

\author{K. Vida}
\affiliation{Konkoly Observatory, CSFK, Konkoly-Thege M. ut 15-17, Budapest, Hungary}

\author{G. Zsidi}
\affiliation{Konkoly Observatory, CSFK, Konkoly-Thege M. ut 15-17, Budapest, Hungary}

\begin{abstract}
We present a photometric study of 17 Type Ia supernovae (SNe) based on multi-color (Johnson-Cousins-Bessell $BVRI$) data taken at Piszk\'estet\H{o} mountain station of Konkoly Observatory, Hungary between 2016 and 2018. We analyze the light curves (LCs) using the publicly available LC-fitter {\tt SNooPy2} to derive distance and reddening information. The bolometric LCs are fit with a radiation-diffusion Arnett model to get constraints on the physical parameters of the ejecta: the optical opacity, the ejected mass and  the initial nickel mass in particular. 
We also study the pre-maximum, de-reddened $(B-V)_0$ color evolution by comparing our data with standard delayed detonation and pulsational delayed detonation models, and show that the $^{56}$Ni masses of the models that fit the $(B-V)_0$ colors are consistent with those derived from the bolometric LC fitting.
We find similar correlations between the ejecta parameters (e.g. ejecta mass, or $^{56}$Ni mass vs decline rate) as published recently by \citet{scalzo18}.
\end{abstract}


\section{Introduction}\label{intro}

 Type Ia supernovae (SNe) are especially important objects for measuring extragalactic distances as their peak absolute magnitudes can be inferred via fitting their observed, multi-color LCs.  Normal SN Ia events obey the empirical Phillips-relation \citep{psk,phillips93},
 which states that the LCs of intrinsically fainter objects decline faster than those of brighter ones. The decline rate is often parametrized by $\Delta m_{15}$, i.e. the magnitude difference between the peak and the one measured at 15 days after maximum in a given (often the $B$) band. 
  For example, the  earlier version of the {\tt SNooPy} code \citep{burns11}
 applied $\Delta m_{15}$ as a fitting parameter for the decline rate. 
  In the new version, {\tt SNooPy2} \citep{burns14, burns18}, a new parameter ($s_{BV}$) that measures the time difference between the maxima of the $B$-band light curve and the $B-V$ color curve, was introduced. 
  Other parametrizations also exist: for example the {\tt SALT2} code \citep{guy07,guy10,beto14}
 applies the $x_1$ (stretch) parameter, while {\tt MLCS2k2} \citep{riess98,jha99,jha07}
 uses $\Delta$  that corresponds to the magnitude difference between the actual $V$-band maximum brightness and that of a fiducial Type Ia SN. All of them are based on the same Phillips-relation, thus, $\Delta m_{15}$, $s_{BV}$, $x_1$ or $\Delta$ are related to each other. 
 
Studying SNe Ia opens a door for constraining the Hubble-parameter $H_0$ \citep{riess12,riess16,dhaw18}
by getting accurate distances to their host galaxies. Such absolute distances are the
quintessential cornerstones of the cosmic distance ladder.
Via constraining $H_0$, SNe Ia play a major role in investigating the expansion of the Universe
\citep{riess98,pearl99,astier06,riess07,wv07,kessler09,guy10,conley11,beto14,rest14,scolnic14, bengaly15,jones15,li16,zhang17}
and testing the most recent cosmological models
\citep[e.g.][]{bh13,beto14}. 
 
 Even though they are extensively used to estimate distances, the improvement of the precision as well as the  accuracy of the method is still a subject of recent studies \citep[see e.g.][]{vinko18}. In order to achieve the desired 1\% accuracy, it is important to  understand the physical properties of the progenitor system and the explosion mechanism better.
 
 The actual progenitor that explodes as a SN Ia, as well as the explosion mechanism, is still an issue.
 There are two main proposed progenitor scenarios: single-degenerate (SD) 
 \citep{whelan73}
 and double-degenerate (DD)
 \citep{iben84}.
 The SD scenario presumes that a carbon-oxygen white dwarf (C/O WD) has a non-degenerate companion star, e.g. a red giant, which, after overflowing its Roche-lobe, transfers mass to the WD. When the WD approaches the Chandrasekhar mass ($M_{Ch}$), spontaneous fusion of C/O to $^{56}$Ni develops that quickly engulfs the whole WD, leading to a thermonuclear explosion.
 
 The  details on the onset and the progress of the C/O fusion is still debated, and many possible mechanisms have been proposed in the literature. 
 The most successful one is the delayed detonation explosion (DDE) model, in which the burning starts as a deflagration, but later it turns into a detonation wave \citep{nomoto84, khok91, dessart,maoz14}. 
 A variant of that is the pulsational delayed detonation explosion (PDDE): during the initial deflagration phase the expansion of the WD expels some material from its outmost layers, which pulsates, expands and avoids burning. After that, the bound material falls back to the WD that leads to a subsequent detonation \citep{dessart}.
  
There is a theoretical possibility for a sub-Chandrasekhar double-detonation scenario, where the WD accretes a thin layer of helium onto its surface, which is compressed by its own mass that leads to He-detonation This triggers the thermonuclear explosion of the underlying C/O WD \citep{woosley94,fink10,kromer10,sim10,sim12}.

 In the DD scenario two WDs merge or collide that results in a subsequent explosion \citep{maoz14, vanrossum16}.

 It may be possible to distinguish between these scenarios e.g. by constraining the mass of the progenitor.  While the traditional SD and DD scenarios require nearly $M_{Ch}$ WD masses, explosion of sub-$M_{Ch}$ WDs may be possible via the ``violent merger'' or the double detonation mechanisms \citep[see e.g.][for a detailed review]{maoz14}. 
 Thus, the ejecta mass is an extremely important physical quantity, which can be inferred by fitting LC models to the observations. 
 
  The idea that the bolometric LC of SNe Ia can be used to infer the ejecta mass via a semi-analytical model, was introduced by \citet{arnett82} and developed further by \citet{jeff99}. \citet{arnett82} showed that the ejecta mass correlates with the rise time to maximum light, provided the expansion velocity and the mean optical opacity of the ejecta are known. Later, \citet{jeff99} suggested the usage of the rate of the deviation of the observed LC from the rate of the Co-decay during the early nebular phase (i.e. the transparency timescale, $t_\gamma$), which measures the leakage of $\gamma$-photons from the diluting ejecta. The advantage of using $t_\gamma$ for constraining the ejecta mass is that $t_\gamma$ is proportional to the gamma-ray opacity, which is much better known than the mean optical opacity. This technique was applied to real data by \citet{strici06}, \citet{scalzo14a} and more recently by \citet{scalzo18}. The conclusion of all these studies was that most SNe Ia seem to have sub-$M_{Ch}$ ejecta, which was also confirmed recently by theoretical models \citep{dhawan18, gold18, wilk18, papa19}.  Other studies lead to similar conclusions: for example, \citet{dhawan16,dhawan17} studied the correlation of the bolometric peak luminosity with the phase of the second maximum in NIR, and found that sub-$M_{Ch}$ explosions can be separated into two categories in respect of this property, however super-$M_{Ch}$ explosions do not follow such a relation.
 
 The usage of the transparency timescale as a proxy for the ejecta mass has some caveats, though.
 $t_\gamma$ also depends on the characteristic velocity ($v_e$) of the expanding ejecta, which is not easy to constrain as it is related to the velocity of a layer deep inside the ejecta. \citet{strici06}, for example, assumed that $v_e$ is uniform for all SNe Ia (they adopted $v_e \sim 3000$ km~s$^{-1}$), which may not be true in reality, because it is known that a diversity in expansion velocities exists for most SNe Ia \citep{wang13}. Another issue is the distribution of the radioactive $^{56}$Ni, encoded by the $q$ parameter by \citet{jeff99}, which is usually taken from models: $q \sim 1/3$, as given by the W7 model, is often assumed. These assumptions, although may not be too far from reality, might introduce some sort of systematic uncertainties in the inferred ejecta masses, which may be worth for further studies.   
 
 Additionally, \citet{levanon19} investigated the influence of  asymmetric $^{56}$Ni distribution to $t_\gamma$, and found that the full range of $t_\gamma$-values cannot be explained by variations in this asymmetry. Their suggested that both $M_{Ch}$ and sub-$M_{Ch}$ explosions are needed to account for the diversity of the observed physical parameters of SN Ia explosions.

 The main motivation of the present paper is to give constraints on the ejecta mass and some other physical parameters for a sample of 17 recent SNe Ia  (Figure \ref{fig:100} and \ref{fig:100_2}) observed from Piszk\'estet\H{o} station of Konkoly Observatory, Hungary.  We generalize the prescription 
 of inferring the ejecta masses by combining the LC rise time and the transparency timescale within the framework of the constant-density Arnett model. 
 
 In the following we present the description of the photometric sample (Section \ref{obs}), then we show the results from  multi-color LC modeling (Section \ref{anal}). We construct  and fit  the bolometric LCs in order to derive the ejecta mass, and other parameters such as the diffusion- and gamma-leakage timescales, the optical opacity and the expansion velocity.
 
 In Section \ref{dis}, we first discuss the early (de-reddened) $(B-V)_0$ color evolution, which might also provide some constraints on the explosion mechanism and the progenitor system \citep[e.g.][]{hosse17, miller18, strici18}.
 There is a growing number of evidence for the appearance of blue excess light during the earliest phase of some SNe Ia \citep{marion16, hosse17, dim19, li18, shap19, strici18}.
 At present the cause of this excess emission is debated, and a number of possible explanations were proposed recently \citep{hosse17, miller18, strici18}. These include $i)$ SN shock cooling, $ii)$ interaction with a non-degenerate companion, $iii)$ presence of  high velocity $^{56}$Ni in the outer layers of the ejecta,
$iv)$ interaction with the circumstellar matter (CSM) and $v)$ differences in the composition or variable opacity.

 Furthermore, we compare our measured $(B-V)_0$ colors with the predictions of various explosion models \citep{dessart},
 and examine the possible correlations between the derived physical parameters following
 \citet{scalzo14a}, \citet{scalzo18} and \citet{kk18}.
 
Finally  Section \ref{sum}  summarizes the results of this paper.   

\section{Observations}\label{obs}

\begin{figure*}
\begin{center}
\includegraphics[width=0.7\textwidth]{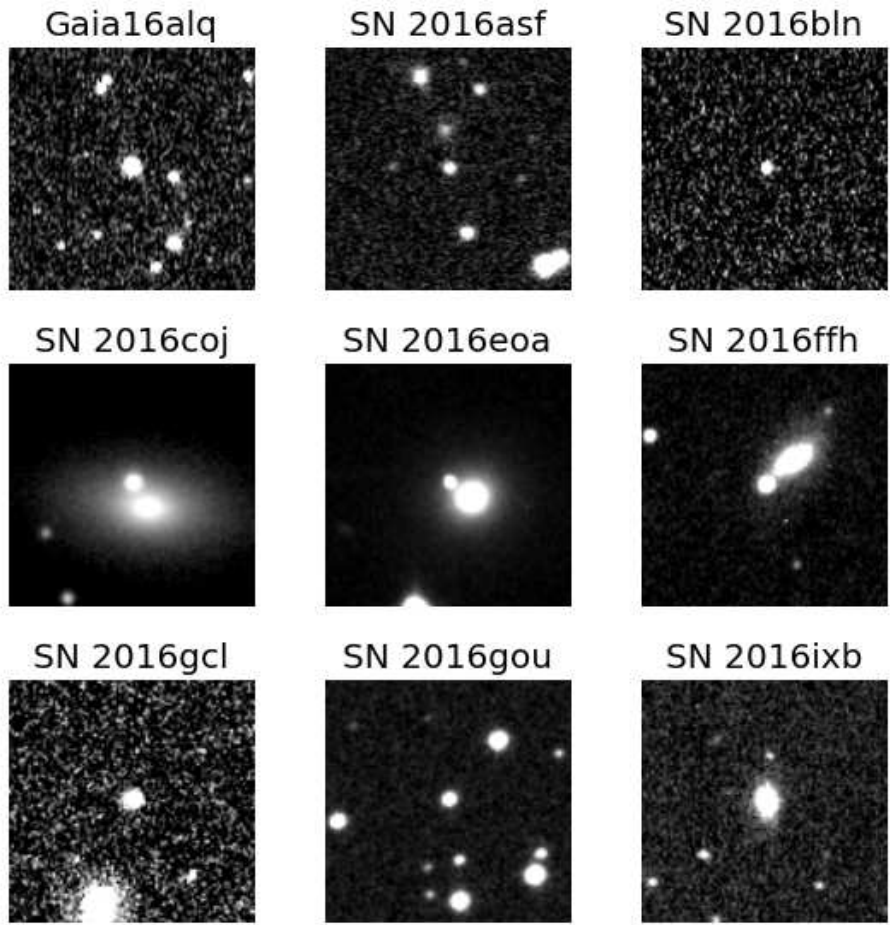}
\caption{Images of the program SNe observed in 2016. The size of each subframe is $1.7 \times 1.7$ arcmin$^2$. 
The supernova is the central object, while North is up and East is to the left. }
\end{center}
\label{fig:100}
\end{figure*}

\begin{figure*}
\begin{center}
\includegraphics[width=0.7\textwidth]{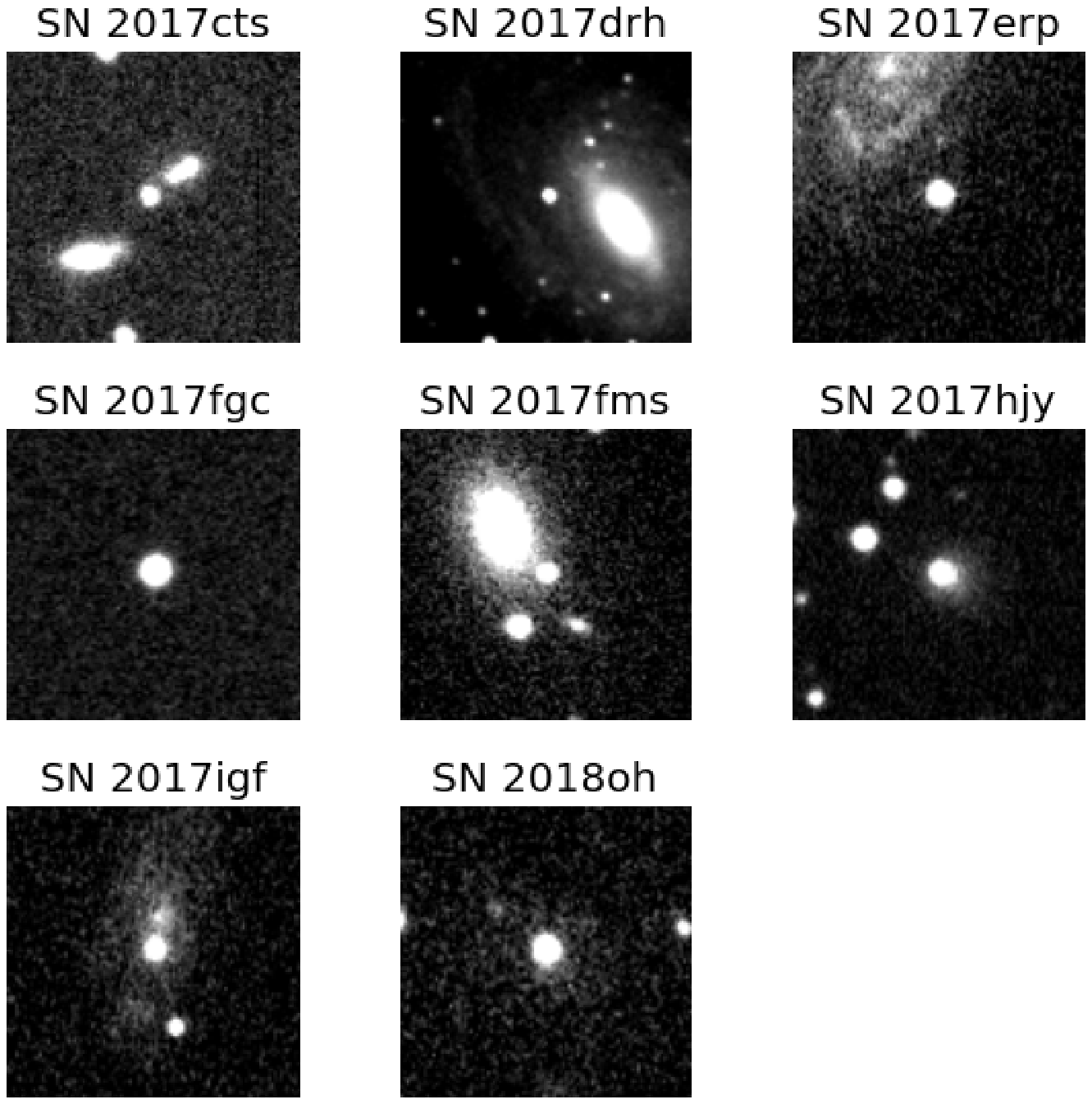}
\caption{The same as Figure~1 but for the SNe observed in 2017 and 2018.}
\end{center}
\label{fig:100_2}
\end{figure*}

We obtained multi-band photometry for 17 bright Type Ia SNe from the Piszk\'estet\H{o} station of Konkoly Observatory, Hungary between 2016 and 2018. The selection criteria for the sample were as follows: $i)$ accessibility from the site (i.e. declination above $-15$ degree), $ii)$ sufficiently early (pre-maximum) discovery, $iii)$ ability for follow-up beyond $t \sim +40$ days after maximum, and $iv)$ low redshift ($z \lesssim 0.05$).

All data were taken with the 0.6/0.9 m Schmidt-telescope equipped with a $4096 \times 4096$ FLI CCD and Bessell $BVRI$ filters, thus, providing a homogeneous, high signal-to-noise data sample of nearby SNe Ia. 

Data reduction and photometry was done the same way as described in \citet{vinko18}.
The raw data were reduced using IRAF\footnote{IRAF is distributed by the National Optical Astronomy
Observatories, which are operated by the Association of Universities for Research in Astronomy, Inc., under cooperative agreement with the National Science Foundation. http://iraf.noao.edu} (Image Reduction and Analysis Facility) by completing bias, dark and flatfield corrections.
Geometric registration of the sky frames was made in two steps. First, we used {\tt SExtractor} \citep{sex} for identifying point sources on each frame. Second, the {\tt imwcs} routine from the {\tt wcstools}\footnote{http://tdc-www.harvard.edu/wcstools/} package was applied to assign R.A. and Dec. coordinates to pixels on the CCD frames. 

Photometry on each SN and several other local comparison stars was made via PSF-fitting  with DAOPHOT in IRAF. Note that due to the strong, variable background from the host galaxy, image subtraction was unavoidable in the case of SN~2016coj, 2016gcl, 2016ixb, 2017drh and 2017hjy. For subtraction we used template frames taken with the same telescope and instrumental setup more than 1 year after the discovery of the SN. In these cases the photometry of the comparison stars was computed on the unsubtracted frames, while for the SN it was done on the host-subtracted frames. Particular attention was paid to keep the flux zero point of the subtracted frame the same as that of the unsubtracted one, thus, getting consistent photometry from both frames. Simple PSF fitting gave acceptable results in the case of the other SNe that suffered less severe contamination from their hosts.

The magnitudes of the local comparison stars were determined from their PS1-photometry\footnote{\tt https://archive.stsci.edu/panstarrs/} after transforming the PS1 $g_P,r_P, i_P$ magnitudes to the Johnson-Cousins $BVRI$ system. The zero points of the standard transformation were tied to these magnitudes. 

 Uncertainties of the magnitudes in each filter are estimated by combining the photometric errors reported by DAOPHOT with the uncertainties of the standard transformation represented by the rms of the residuals between the measured and catalogued magnitudes of the local comparison stars. The latter is found as the leading  source of error in most cases, resulting in 0.071, 0.045, 0.048 and 0.053 mag for the median uncertainty in $B$-, $V$-, $R$- and $I$-bands, respectively. 

The basic data of the observed SNe are collected in Table~\ref{tab:basic}. 
Plots of the $V$-band sub-frames centered on the SNe are shown in Fig \ref{fig:100} and \ref{fig:100_2}.  After acceptance, all photometric data will be made available via the {\it Open Supernova Catalog}\footnote{\tt https://sne.space}.

\begin{table*}
\caption{Basic data of the observed SNe}
\centering
\begin{tabular}{lccccccc}
\hline
\hline
SN name & Type & R.A.& Dec. & Host galaxy & $z$ & Discovery & Ref. \\ 
\hline
Gaia16alq & Ia-norm & 18:12:29.36 & +31:16:47.32 & PSO J181229.441+311647.834 & 0.023 & 2016-04-21 & a\\ 
SN~2016asf & Ia-norm & 06:50:36.73 & +31:06:45.36 & KUG 0647+311 & 0.021 & 2016-03-06 & b\\
SN~2016bln & Ia-91T & 13:34:45.49 & +13:51:14.30 & NGC 5221 & 0.0235 & 2016-04-04 & c \\
SN~2016coj & Ia-norm & 12:08:06.80 & +65:10:38.24 & NGC 4125 & 0.005 & 2016-05-28 & d \\
SN~2016eoa &Ia-91bg &  00:21:23.10 & +22:26:08.30 & NGC 0083 & 0.021 & 2016-08-02 & e \\
SN~2016ffh & Ia-norm & 15:11:49.48 & +46:15:03.22 & MCG +08-28-006 & 0.018 & 2016-08-17 & f\\
SN~2016gcl & Ia-91T & 23:37:56.62 & +27:16:37.73 & AGC 331536 & 0.028 & 2016-09-08 & g\\
SN~2016gou & Ia-norm & 18:08:06.50 & +25:24:31.32 & PSO J180806.461+252431.916 & 0.016 & 2016-09-22 & h\\
SN~2016ixb & Ia-91bg & 04:54:00.04 & +01:57:46.62 & NPM1G +01.0158 & 0.028343 & 2016-12-17 & i\\
SN~2017cts & Ia-norm & 17:03:11.76 & +61:27:26.06 & CGCG 299-048 & 0.02 & 2017-04-02 & j\\
SN~2017drh & Ia-norm & 17:32:26.05 & +07:03:47.52 & NGC 6384 & 0.005554 & 2017-05-03 & k \\
SN~2017erp & Ia-norm & 15:09:14.81 & -11:20:03.20 & NGC 5861 & 0.0062 & 2017-06-13 & l\\
SN~2017fgc & Ia-norm & 01:20:14.44 & +03:24:09.96 & NGC 0474 & 0.008 & 2017-07-11  & m\\
SN~2017fms & Ia-91bg & 21:20:14.60 & -04:52:51.30 & IC 1371 & 0.031 & 2017-07-17  & n\\
SN~2017hjy & Ia-norm & 02:36:02.56 & +43:28:19.51 & PSO J023602.146+432817.771 & 0.007 & 2017-10-14 & o\\
SN~2017igf & Ia-91bg & 11:42:49.85 & +77:22:12.94 & NGC 3901 & 0.006 & 2017-11-18  & p\\
SN~2018oh & Ia-norm & 09:06:39.54 & +19:20:17.77 & UGC 04780 & 0.012 & 2018-02-04  & q\\
\hline
\end{tabular}
\tablecomments{a: \citet{16alq}; b: \citet{16asf}; c: \citet{16bln}; d: \citet{16coj}; e: \citet{16eoa}; f: \citet{16ffh}; g: \citet{16gcl}; h: \citet{16gou}; i: \citet{16ixb}; j: \citet{17cts}; k: \citet{17drh}; l: \citet{17erp}; m: \citet{17fgc}; n: \citet{17fms}; o: \citet{17hjy}; p: \citet{17igf}; q: \citet{18oh} }
\label{tab:basic}
\end{table*}


\section{Analysis}\label{anal}

\subsection{Distance and reddening estimates via multi-color light curve modeling}\label{snpy}

\begin{table*}[h!]
\caption{Best-fit parameter from SNooPy2 assuming $R_V = 3.1$}
\label{tab:snpy}
\centering
\begin{tabular}{lccccccc}
\hline
\hline
Name & $E(B-V)_{MW}$ & $E(B-V)_{host}$ & $T_{max}$ & $\mu_0$ & $s_{BV}$ & $\Delta m_{15}$ & $\chi^2$  \\
  & mag & mag & MJD & mag & & mag & \\
\hline
SN~2011fe & 0.0075 & 0.048 (0.060) & 55815.31 (0.06) & 29.08 (0.082) & 0.940 (0.031) & 1.175 (0.008) & 1.058 \\
Gaia16alq & 0.0576 & 0.089 (0.061) & 57508.11 (0.14) & 35.05 (0.084) & 1.132 (0.033) & 0.939 (0.024) & 1.358 \\
SN~2016asf & 0.1149 & 0.076 (0.097) & 57464.66 (0.11) & 34.69 (0.083) & 1.001 (0.031) & 1.102 (0.023) & 2.722 \\
SN~2016bln & 0.0249 & 0.213 (0.061) & 57499.40 (0.34) & 34.78 (0.116) & 1.058 (0.032) & 1.005 (0.024) & 3.839 \\
SN~2016coj & 0.0163 & 0.000 (0.061) & 57548.30 (0.34) & 32.08 (0.083) & 0.788 (0.030) & 1.438 (0.020) & 1.954 \\
SN~2016eoa & 0.0633 & 0.242 (0.062) & 57615.66 (0.35) & 34.24 (0.101) & 0.775 (0.032) & 1.486 (0.036) & 1.867 \\
SN~2016ffh & 0.0239 & 0.198 (0.061) & 57630.48 (0.36) & 34.61 (0.094) & 0.926 (0.032) & 1.082 (0.018) & 2.392 \\
SN~2016gcl & 0.0630 & 0.056 (0.061) & 57649.48 (0.36) & 35.45 (0.090) & 1.155 (0.043) & 0.901 (0.031) & 2.792 \\
SN~2016gou & 0.1095 & 0.258 (0.060) & 57666.40 (0.34) & 34.25 (0.104) & 0.984 (0.031) & 0.925 (0.032) & 6.560 \\
SN~2016ixb & 0.0520 & 0.077 (0.061) & 57745.11 (0.37) & 35.620 (0.087) & 0.758 (0.032) & 1.657 (0.039) & 2.072 \\
SN~2017cts & 0.0265 & 0.158 (0.061) & 57856.83 (0.35) & 34.56 (0.088) & 0.931 (0.030) & 1.208 (0.020) & 1.394 \\
SN~2017drh & 0.1090 & 1.396 (0.062) & 57890.94 (0.36) & 32.29 (0.326) & 0.838 (0.031) & 1.352 (0.065) & 2.075 \\
SN~2017erp & 0.0928 & 0.210 (0.061) & 57934.40 (0.35) & 32.34 (0.097) & 1.174 (0.030) & 1.129 (0.011) & 1.627 \\
SN~2017fgc & 0.0294 & 0.162 (0.061) & 57959.77 (0.37) & 32.61 (0.092) & 1.137 (0.037) & 1.086 (0.027) & 1.761 \\
SN~2017fms & 0.0568 & 0.022 (0.061) & 57960.09 (0.35) & 35.52 (0.084) & 0.746 (0.033) & 1.425 (0.026) & 1.875 \\
SN~2017hjy & 0.0768 & 0.211 (0.061) & 58056.02 (0.35) & 34.05 (0.096) & 0.949 (0.031) & 1.137 (0.011) & 1.519 \\
SN~2017igf & 0.0456 & 0.158 (0.061) & 58084.61 (0.36) & 32.80 (0.096) & 0.608 (0.032) & 1.757 (0.006) & 3.461 \\
SN~2018oh & 0.0382 & 0.071 (0.061) & 58162.96 (0.37) & 33.40 (0.083) & 1.089 (0.034) & 0.989 (0.013) & 1.231 \\
\hline
\end{tabular}
\end{table*}

To fit and analyze the observed LCs, we used the {\tt SNooPy2}\footnote{http://csp.obs.carnegiescience.edu/data/snpy/snpy} \citep{burns11,burns14} LC fitter, which is based on the Phillips-relation \citep{phillips93}. It can be categorized as a "distance calculator" \citep{conley08}, since it provides the absolute distance as a fitting parameter.  This was an important criterion for choosing {\tt SNooPy2}, because precise distance estimates were necessary for the goals of this paper.

We applied both the EBV-model that fits the template LCs by \citet{prieto06} to 
the data in $BVRI$ filters, and the EBV2-model that is based on the $uBVgriYJH$ light curves by \citet{burns11}. 
Both of these models relate the distance modulus of a normal SN Ia to its decline rate and color as
\begin{eqnarray}
m_X (\varphi_o) ~=~ T_Y(\varphi, p) + M_Y(p) + \mu_0 + K_{XY} + \nonumber \\
 + R_X \cdot E(B-V)_{MW} + R_Y \cdot E(B-V)_{host},
\label{eq:snpy}
\end{eqnarray}
where X and Y represent the filter of observed data and the template LC,  $\varphi$ is the rest-frame phase of the SN (defined as rest-frame days since the moment of $B$-band maximum light), $\varphi_o = \varphi \cdot (1+z)$ is the phase in the observer's frame, $m_X (\varphi_o)$ is the observed magnitude in filter X, $T_Y(\varphi, p)$ is the template LC as a function of  phase, $p$ is a generalized decline rate parameter ($p = \Delta m_{15}$ or $p = s_{BV}$ in the EBV and EBV2 models, respectively), $M_Y(p)$ is the peak absolute magnitude of the SN having $p$ decline-rate, $\mu_0$ is the reddening-free distance modulus in magnitudes, $E(B-V)_{MW}$ is the color excess due to interstellar extinction in the Milky Way (“MW”), ($E(B-V)_{host}$ in the host galaxy), $R_{X, Y}$ are the
reddening slopes in filter X or Y and $K_{XY} = K_{XY} (\varphi, z, E(B-V)_{MW}, E(B-V)_{host})$ is the cross-band K-correction that matches the observed broad-band magnitudes of a redshifted SN taken with filter X to a template SN LC taken in filter Y. Since our SNe have very low redshifts (see Table~\ref{tab:basic}), K-corrections are often negligible compared to the observational uncertainties.  

 The motivation behind using the EBV2 model, despite the difference between the photometric system of its LC templates and our data, is that the EBV2 model allows the fitting of the $s_{BV}$ color-stretch parameter, which is more suitable for describing fast-decliner (91bg-like) SNe Ia than the canonical $\Delta m_{15}$ \citep{burns14}\footnote{We thank the referee for this suggestion.}.  Also, the EBV2 model is based on a more recent, better photometric calibration than the EBV model.

 First, we fit the observed $BVRI$ LCs, adopting $R_V = 3.1$  (utilizing the {\tt calibration=1} mode) for the reddening slope both in the MW and in the host , with $\chi^2$-minimization using the built-in MCMC routine in {\tt SNooPy2}.  {\tt SNooPy2} has a capability of K-correcting the data while fitting them with templates from a different photometric system, which was also applied in this case.
The inferred parameters were the following ones:
\begin{itemize}
\item{ $E(B-V)_{host}$ : interstellar extinction of the host galaxy (in magnitude); }
\item{ $T_{max}$ : moment of the maximum light in the B-band (in MJD); }
\item{ $\mu_0$ :  extinction-free distance modulus (in magnitude);}
\item{$s_{BV}$: color-stretch parameter in the EBV2 model;}
\item{ $\Delta m_{15}$ :  decline rate parameter in the EBV model (in magnitude).}
\end{itemize}
 We also attempted to include $R_V$ as a fitting parameter, but the results were close the original $R_V =3.1$ value, indicating that our photometry is not suitable for constraining this parameter. This is not surprising, given the lack of UV- or NIR-data in our sample. 

Note that since the EBV2 model is based on a different photometric system than our Johnson-Cousins $BVRI$ data, the inferred best-fit values for $E(B-V)_{host}$ and $\mu_0$ are expected to be slightly different from those obtained from the EBV model that is based on $BVRI$ data. After comparing the best-fit parameters taken from both models, it is found that the $E(B-V)_{host}$ values are consistent with each other within $1 \sigma$, while there is a systematic shift between the distance moduli of $\mu_0(EBV) - \mu_0(EBV2) \sim 0.1$ mag.  For consistency, we decided to adopt the $E(B-V)_{host}$ and $\mu_0$ parameters from the EBV2 model as final, but added $\sim 0.1$ mag systematic uncertainty to the distance modulus (corresponding to $\sim 5$ percent relative error in the distances). 

The final best-fit parameters are shown in Table~ \ref{tab:snpy}.  The reported errors include the systematic uncertainties of the template vectors as given by {\tt SNooPy2}.
Plots of the observed LCs and their best-fit {\tt SNooPy2} templates can be found in the Appendix.  The overall fitting quality is good; most of the reduced $\chi^2$ values (column 8 in Table~\ref{tab:snpy}) are in between 1 and 2, and the highest $\chi^2 \sim 6.6$ is that of SN~2016gou. As Figure \ref{18sne_snpy} of the Appendix demonstrates, the best model fits the near-maximum data of SN~2016gou adequately, and the relatively high $\chi^2$ value is likely caused by the scattering in the last few data points after +40 days. 
 
 As seen in Table~\ref{tab:snpy}, the host reddening of  SN~2017drh turned out to be extremely high ($E(B-V)_{host} \sim 1.4$ mag) compared to the rest of the sample. Since this may add an extra uncertainty in the derived distance and other parameters, SN~2017drh was excluded from the sample and was not analyzed further.  From the remaining 17 SNe, 8 of them have low $E(B-V)_{host}$ values that do not exceed their $2 \sigma$ uncertainties ($\sim 0.12$ mag for most of them), but their $3 \sigma$ upper limits still indicate low reddening for them. This suggests that $E(B-V)_{host}$ cannot be constrained for these SNe from our data due to the photometric uncertainties. On the other hand, none of the other SNe in our sample show $ E(B-V)_{host} \gtrsim 0.3$ mag. 
 Thus, even if we adopt the upper limits for the 8 low-reddened objects, our sample is only moderately reddened, which significantly reduces the extinction-related issues while converting the observed data to physical quantities.
 
\begin{figure}
\begin{center}
\includegraphics[width=8.5cm]{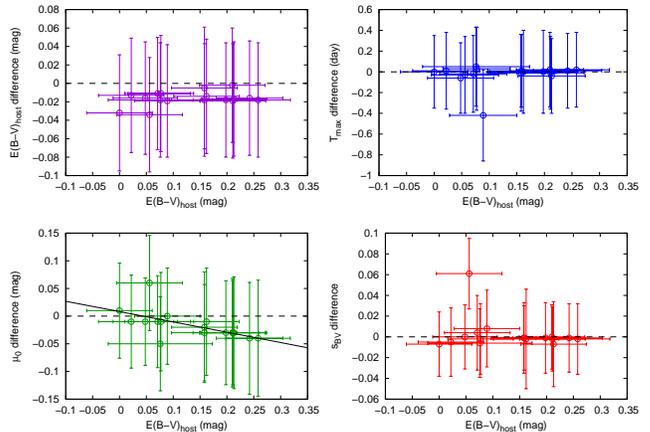}
\caption{Differences between the best-fit {\tt SNooPy2} parameters from the $R_V = 1.5$ model and those from the $R_V = 3.1$ model. Only the distance modulus (bottom left panel) shows a slight trend against reddening, but is statistically insignificant (see text).}
\end{center}
\label{fig:rv15}
\end{figure}
 
  Nevertheless, we also tested the effect of the adopted reddening law on the inferred LC parameters. The motivation for this kind of test was the fact that
 the validity of the $R_V = 3.1$ reddening law have been questioned for larger samples of SNe Ia in several studies \citep[e.g.][]{reindl05, fola11, ama15}, but see \citet{scolnic14} for a different conclusion. In order to test the bias caused by the assumption of the $R_V = 3.1$ reddening law on the inferred SNooPy2 parameters, especially on the distance, all fits were re-computed with the built-in {\tt calibration = 3} mode (corresponding to $R_V = 1.5$) in SNooPy2. Since our sample contains only moderately reddened SNe, no major differences in the fit parameters are expected. Indeed, it was found that all the fit parameters from the $R_V = 1.5$ model are consistent with their counterparts from the $R_V = 3.1$ model within $1 \sigma$. This is illustrated in Figure~\ref{fig:rv15} where the differences between the best-fit parameters for the $R_V = 1.5$ model and those of the $R_V = 3.1$ model are plotted against $E(B-V)_{host}$. 
 Disregarding a single deviant point, the inferred $E(B-V)_{host}$, $T_{max}$ and $s_{BV}$ parameters do not show any trend with the $E(B-V)_{host}$ parameter, suggesting negligible dependence of these fit parameters on the assumed reddening law. However, the $\Delta \mu_0$ distance modulus difference ($\mu_0(1.5) - \mu_0(3.1)$) exhibits a slight trend with reddening as $\Delta \mu_0 ~=~ -0.186 (\pm 0.062) \cdot E(B-V)_{host} ~+~ 0.008 (\pm 0.010)$, but none of the $\Delta \mu_0$ residuals exceed their uncertainties more than $1 \sigma$. Thus, we adopt the parameters from the $R_V = 3.1$ model for further studies, but we also investigate the effect of the lower reddening slope on the bolometric LC and the final inferred physical parameters (see Sections \ref{bol1}, \ref{bol2} and \ref{dis}).


\subsection{Construction of the bolometric light curve}\label{bol1}

\begin{figure*}
\begin{center}
\includegraphics[width=8.5cm] {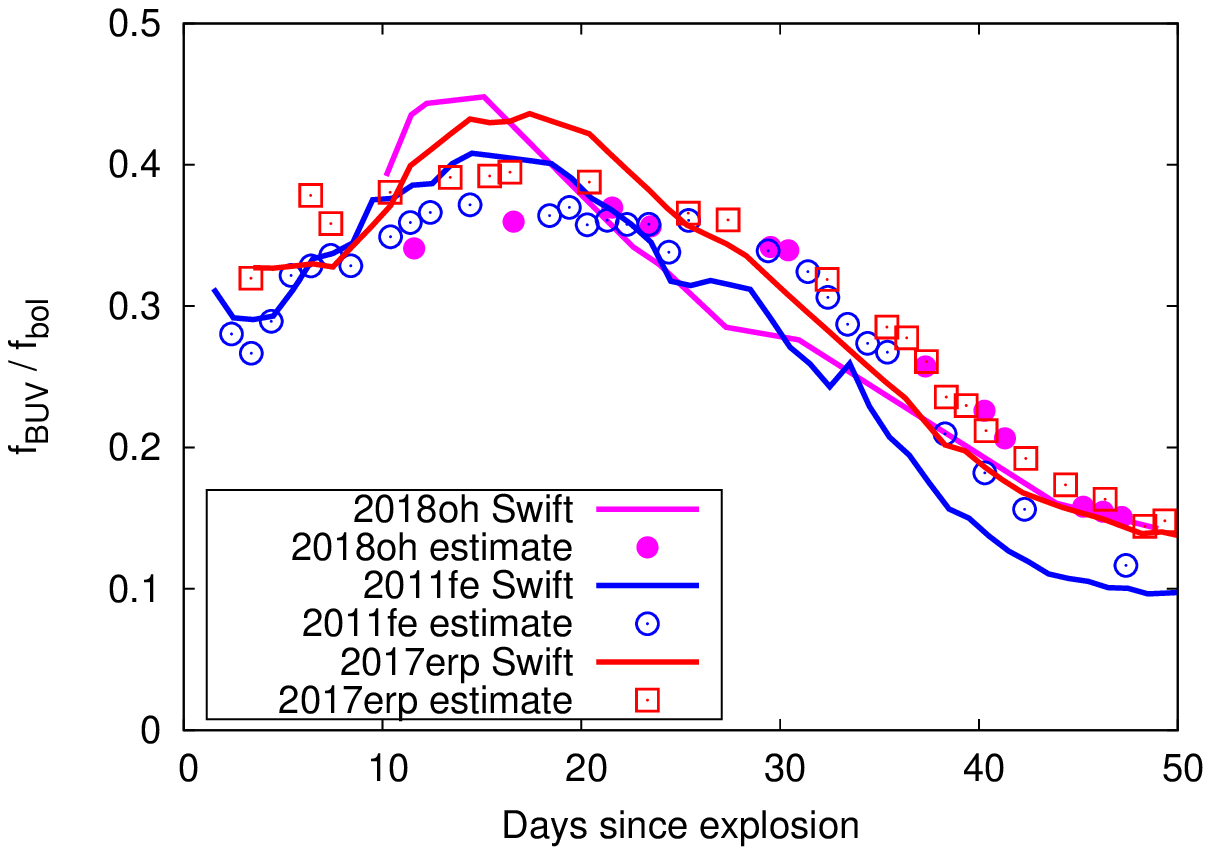}
\includegraphics[width=8.5cm]{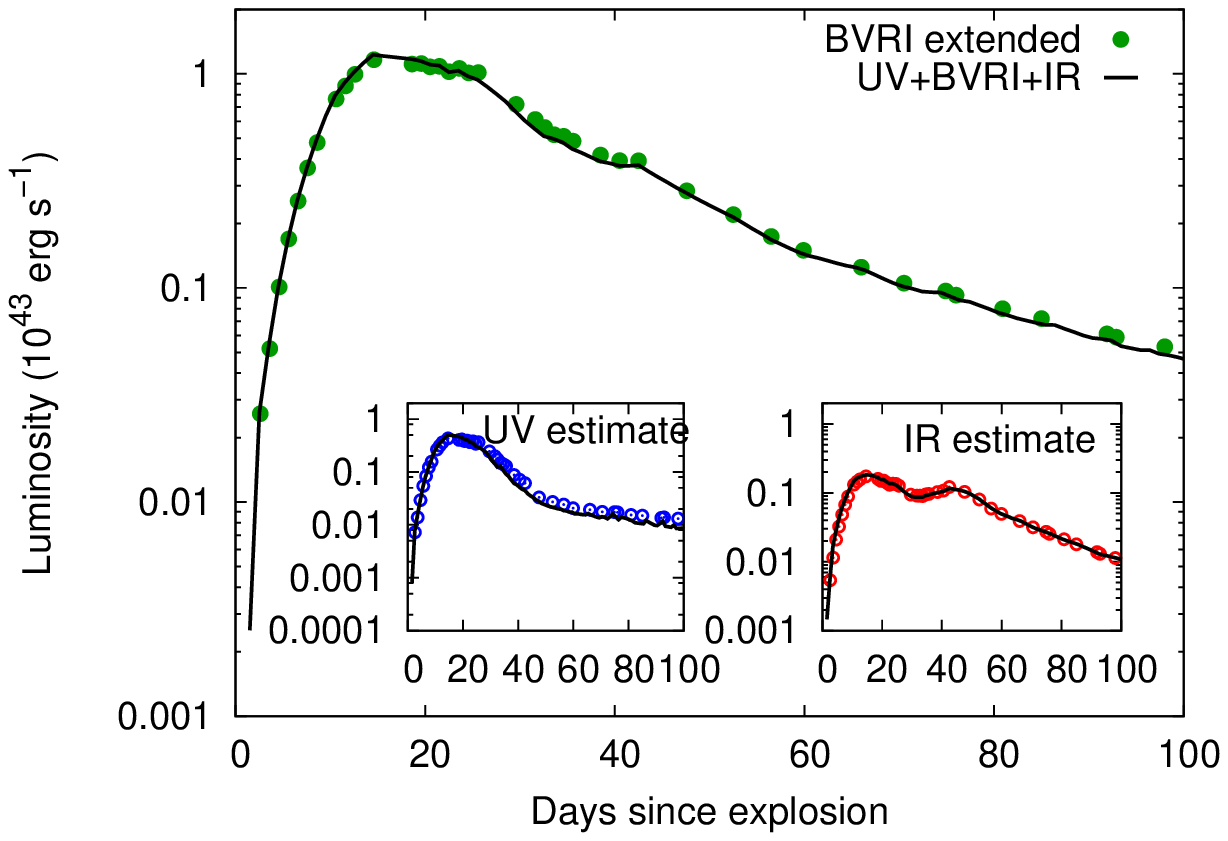}
\includegraphics[width=8.5cm] {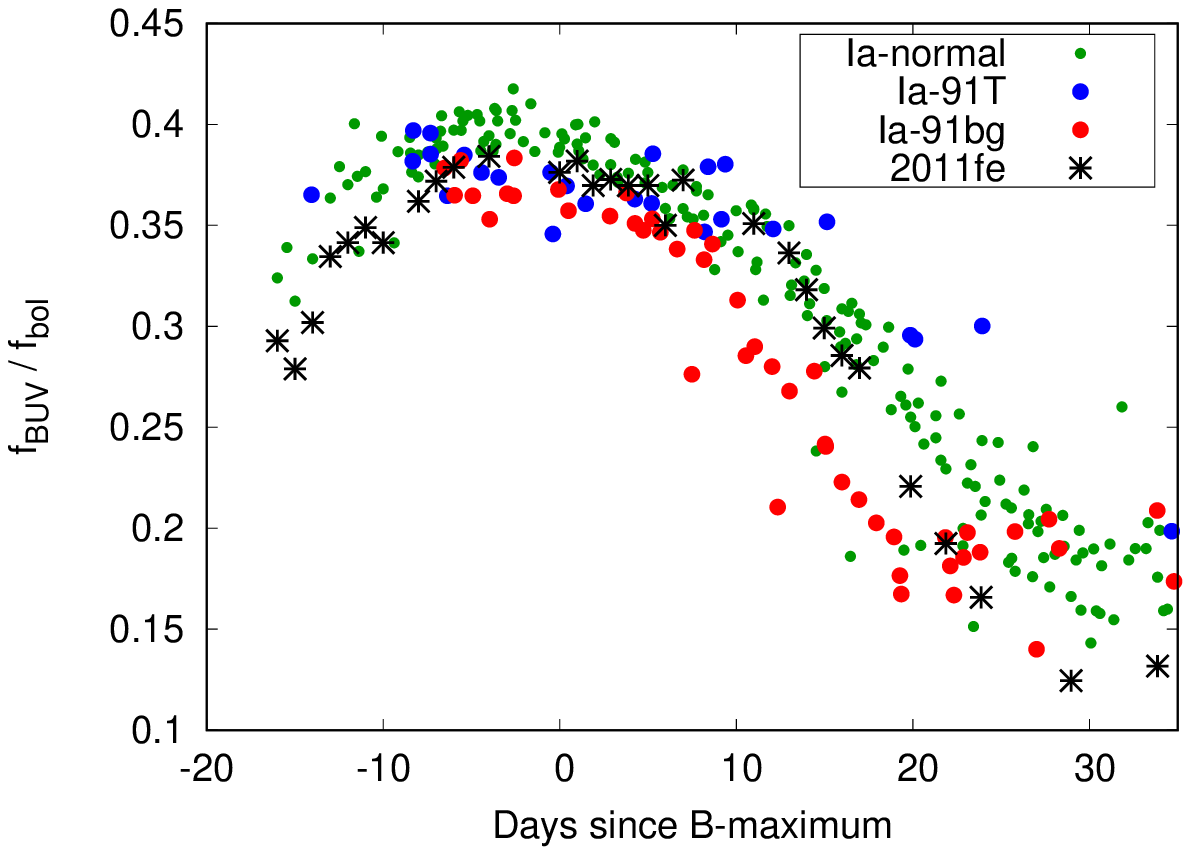}
\includegraphics[width=8.5cm] {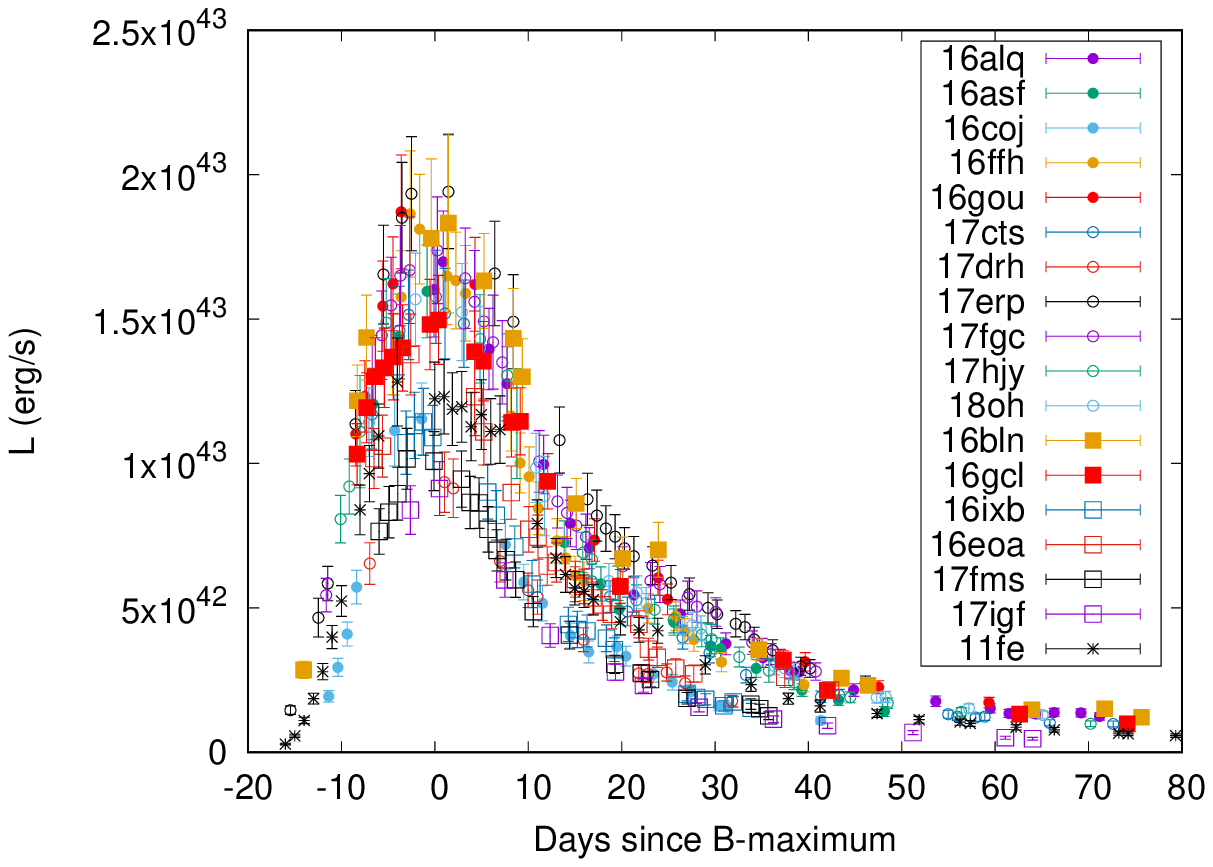}

\caption{Top left panel: the ratio of BUV flux to the total bolometric flux as a function of time since explosion for SN~2018oh (magenta), SN~2011fe (blue) and SN~2017erp (red).   Data obtained by direct integration of {\it Swift} fluxes are plotted with lines, while the symbols correspond to data estimated from {\it BVRI} observations (see text). 
Top right panel: 
Comparison of the pseudo-bolometric LCs of SN~2011fe derived from the extrapolated $BVRI$ SED (symbols) and by direct integration of the observed UV + optical + NIR data (black line). The two insets show the UV and the IR contributions separately.
Bottom left panel: the same as the top left panel but for the whole observed sample. Colors code the different SN subtypes as indicated in the legend. 
Bottom right panel: the derived pseudo-bolometric light curves for the sample SNe.
}
\label{fig:bolo-valid}
\end{center}
\end{figure*}

While constructing the bolometric LCs we followed the same procedure as applied recently by \citet{li18} for SN~2018oh. Briefly, after correcting for extinction within the Milky Way and the host galaxy (Table~\ref{tab:basic} and \ref{tab:snpy}),  the observed $BVRI$ magnitudes were converted to physical fluxes via the calibration of \citet{bessell98}.  Uncertainties of the fluxes were estimated in the same way using the magnitude uncertainties obtained via PSF fitting (Sect.~\ref{obs}). Note that because of the relatively low reddening of the sample, the conversion of broadband magnitudes to quasi-monochromatic fluxes is not expected to be significantly affected by the alteration of the spectral energy distribution (SED) due to interstellar extinction.

After that, the fluxes were integrated against wavelength via the trapezoidal rule  to get a pseudo-bolometric ($L_{BVRI}$) LC.

 Since the UV- and IR regimes are known to have significant contribution to the true bolometric flux of SNe Ia, they cannot be neglected. Because neither the UV, nor the IR bands were covered by our observations, they were estimated by extrapolations  in the following way.

In the UV regime the flux was assumed to decrease linearly between 2000 \AA\ and $\lambda_B$, and the UV-contribution was estimated from the extinction-corrected $B$-band flux $f_B$ as $f_{bol}^{UV} ~=~ 0.5 f_B (\lambda_B - 2000)$ \citep{li18}.  This estimate is tied directly to the $B$-band flux, and its validity is tested below.

In the IR, to get an estimate for the contribution of the unobserved flux, a Rayleigh-Jeans tail was fit to the corrected $I$-band fluxes ($f_I$) and integrated between $\lambda_I$ and infinity.  This resulted in the formula of $f_{bol}^{IR} ~=~ 1.3 f_I \lambda_I / 3$ that we adopted as the estimate of the contribution of the missing IR-bands.
The factor 1.3  was found by matching the IR contribution estimated this way to the one that was obtained via direct integration of real observed near-IR $JHK$ photometry  taken from literature for a few well-observed objects (see below). 

Finally, the bolometric fluxes are corrected for distances  inferred from the distance moduli from the {\tt SNooPy2} fits (Section~\ref{snpy}) as $D (\mathrm{Mpc}) = 10^{ 0.2 \cdot (\mu_0 - 25)}$ .

This procedure was validated by comparing the estimated UV and IR-contributions to those calculated from existing data for three well-observed normal Type Ia SNe: 2011fe \citep{brown14, vinko12, mathe12}, 2017erp \citep{erp} and 2018oh \citep{li18}. All these data were downloaded directly from the {\it Open Supernova Catalog}\footnote{https://sne.space} \citep{james17}.

Fig.~\ref{fig:bolo-valid} shows the comparison of the $BVRI$-extrapolated pseudo-bolometric fluxes (colored symbols) with the ones obtained by direct integrations from the UV to the NIR. In the latter case, the integral of a Rayleigh-Jeans tail was also added to the final bolometric flux, but the tail was fit to the $K$-band flux ($f_K$) instead of $f_I$ and the factor 1.3 was not applied. 

 The top-left panel exhibits the comparison of the flux ratio $f_{BUV}$ to $f_{bol}$ as a function of phase, where $f_{BUV}$ is the integrated flux between the $B$-band and 2000 \AA\, and $f_{bol}$ is the total bolometric flux. The filled circles are based on  extrapolated fluxes (as described above), while the lines are from direct integration using the UV data from the {\it Neil Gehrels Swift Observatory}. It is seen that there is reasonable agreement between the  extrapolated and the directly integrated blue-UV fluxes:  the differences between the $f_{BUV} / f_{bol}$ flux ratios from the extrapolated and the directly integrated data (marked by circles and lines in Fig.\ref{fig:bolo-valid}, respectively) do not exceed $\pm 5$ percent for most of the data, which is less than the estimated uncertainty of the total bolometric flux (see below).
 
 The bottom-left panel shows the same flux ratio computed from the interpolated data but for all SNe in our sample. This illustrate that there is no major difference in the flux ratios between the slow- and fast-decliner SNe Ia around maximum light, despite a $\sim 5$ percent relative flux uncertainty (estimated from the scattering of the data, see above) that we include in the final flux uncertainty estimate. 

In the top-right panel the full bolometric LC from direct integration (solid line) and the one based on extrapolation (circles) is shown for the extremely well-observed SN~2011fe. The insets illustrate the same but only for the UV (left) and NIR (right) regimes. Again, there seems to be good agreement between the directly integrated and the extrapolated bolometric fluxes. In the bottom-right panel the calculated luminosity evolution is plotted for all SNe in our sample.  

It is concluded that the pseudo-bolometric LCs obtained from extrapolations described above are reliable representations of the true bolometric data, and the systematic errors due to the missing bands  do not exceed $\sim 5$ percent. 
 Together with the errors due to uncertainties in the distances (see above), the final relative uncertainty of the bolometric fluxes are estimated as $\sim 10$ percent.
 
In order to cross-check the sensitivity of the bolometric LCs to the assumed 
reddening law, all LCs were re-computed using $R_V = 1.5$ for estimating the extinction within the host galaxy (cf. Section~\ref{anal}). As expected, this resulted in bolometric LCs having reduced amplitudes with respect to the ones calculated by assuming $R_V=3.1$. The amplitude ratio, $L(1.5)/L(3.1)$, as a function of $E(B-V)_{host}$ turned out to be 
$ L(1.5)/L(3.1) = 1 - 1.349 (\pm 0.046) \cdot E(B-V)_{host}$,  
which gives $L(1.5) \approx 0.65 \cdot L(3.1)$ for SN~2016gou that had the highest $E(B-V)_{host} \sim 0.26$ mag in our sample (see Table~\ref{tab:snpy}). This directly influences the derived $^{56}$Ni masses, predicting lower values for those SNe that suffered higher reddening within their host (cf. Section~\ref{bol2}).


\subsection{Fitting the bolometric light curve}\label{bol2} 

\begin{table*}[h!]
\caption{Best-fit and inferred parameters from the bolometric LC fitting using $R_V~=~3.1$}.
\scriptsize
\label{tab:lc-bol}
\centering
\begin{tabular}{lcccccccccc}
\hline
\hline
Name & $t_0$ & $t_{lc}$ & $t_\gamma$ & $M_{Ni}$ & $\kappa^-$ & $\kappa^+$ & $M_{ej}$ & $v_{exp}$ & $E_{kin}$ & $\chi^2$ \\
  & (day) & (day) & (day) & (M$_\odot$) & \multicolumn{2}{c}{(cm$^2$g$^{-1}$)} & (M$_\odot$) & (km s$^{-1}$) & ($10^{51}$ erg) \\
\hline
2011fe &16.59 (0.061) & 14.87 (0.321) & 37.60 (0.670) & 0.567 (0.042) & 0.166 & 0.232 & 1.002 (0.070) & 11660 (542) & 0.817 (0.407) & 0.517\\
Gaia16alq &19.92 (0.418) & 13.78 (0.899) & 46.669 (0.717) & 0.744 (0.055) & 0.115 & 0.129 & 1.274 (0.330) & 10594 (1421) & 0.858 (0.269) & 1.755 \\
2016asf &15.08 (2.579) & 11.35 (1.140) & 39.192 (1.329) & 0.597 (0.149) & 0.093 & 0.124 & 1.051 (0.430) & 11459 (2429) & 0.828 (0.491) & 0.502 \\
2016bln &17.42 (0.342) & 14.00 (1.219) & 44.508 (1.125) & 0.789 (0.097) & 0.124 & 0.146 & 1.211 (0.430) & 10833 (1964) & 0.852 (0.363) & 0.662 \\
2016coj &14.17 (0.261) & 10.57 (0.367) & 32.967 (0.863) & 0.401 (0.053) & 0.095 & 0.152 & 0.856 (0.130) & 12298 (1070) & 0.777 (0.557) & 0.487 \\
2016eoa &14.07 (3.053) & 10.55 (1.092) & 39.038 (0.935) & 0.482 (0.103) & 0.080 & 0.108 & 1.047 (0.440) & 11483 (2439) & 0.828 (0.498) & 0.617 \\
2016ffh &14.03 (1.16) & 9.736 (0.521) & 40.521 (0.926) & 0.573 (0.078) & 0.067 & 0.088 & 1.035 (0.230) & 11298 (1316) & 0.836 (0.363) & 0.653 \\
2016gcl &17.79 (2.184) & 15.85 (1.177) & 43.623 (1.182) & 0.689 (0.164) & 0.162 & 0.195 & 1.185 (0.360) & 10931 (1728) & 0.849 (0.343) & 0.280 \\
2016gou &15.03 (0.525) & 11.05 (0.798) & 45.793 (1.412) & 0.678 (0.063) & 0.075 & 0.086 & 1.250 (0.30) & 10696 (1681) & 0.856 (0.304) & 0.263 \\
2016ixb &15.64 (2.088) & 13.12 (2.364) & 30.520 (1.345) & 0.483 (0.064) & 0.159 & 0.274 & 0.777 (0.30) & 12653 (3050) & 0.747 (0.539) & 0.177 \\
2017cts &14.23 (1.101) & 9.97 (1.259) & 42.485 (0.959) & 0.539 (0.063) & 0.066 & 0.081 & 1.151 (0.58) & 11062 (2838) & 0.845 (0.505) & 0.506 \\
2017erp &17.96 (0.092) & 17.40 (0.518) & 37.603 (1.084) & 0.975 (0.083) & 0.227 & 0.317 & 1.002 (0.13) & 11661 (966) & 0.817 (0.414) & 0.413 \\
2017fgc &16.21 (0.239) & 12.50 (0.386) & 45.398 (0.941) & 0.692 (0.047) & 0.097 & 0.112 & 1.237 (0.16) & 10734 (799) & 0.855 (0.210) & 0.253 \\
2017fms &14.04 (0.504) & 10.58 (0.575) & 34.612 (0.731) & 0.360 (0.029) & 0.091 & 0.138 & 0.909 (0.20) & 12066 (1407) & 0.794 (0.523) & 0.264 \\
2017hjy &16.29 (0.492) & 12.83 (0.823) & 39.484 (0.840) & 0.688 (0.057) & 0.117 & 0.156 & 1.060 (0.270) & 11425 (1545) & 0.830 (0.406) & 0.213 \\
2017igf &19.58 (0.933) & 15.30 (1.353) & 34.554 (1.193) & 0.420 (0.051) & 0.191 & 0.291 & 0.906 (0.320) & 12070 (2291) & 0.792 (0.572) & 0.779 \\
2018oh &14.86 (0.864) & 11.17 (1.098) & 44.654 (0.928) & 0.598 (0.059) & 0.078 & 0.092 & 1.217 (0.480) & 10824 (2175) & 0.856 (0.394) & 0.187 \\
\hline
\end{tabular}
\end{table*}

We estimated the physical parameters of the SN ejecta via applying the radiation-diffusion model of \citet{arnett82} by fitting the bolometric LCs with the {\tt Minim} code \citep{minim}.  This is a  Monte-Carlo code utilizing the Price-algorithm that intends to find the position of the absolute minimum on the $\chi^2$-hypersurface within the allowed volume of the parameter space. Parameter uncertainties are estimated from the final distribution of $N = 200$ test points that probe the parameter space around the $\chi^2$ minimum where $\Delta \chi^2 \leq 1$. See \citet{minim} for more details.

The fitted parameters were the following: the epoch of first light  with respect to the moment of the $B$-band maximum ($t_0$), the light curve time scale ($t_{lc}$), the gamma-ray leaking time scale ($t_\gamma$), and the initial nickel mass ($M_{Ni}$). These parameters can be found in Table \ref{tab:lc-bol}. 
 The final uncertainty of the nickel mass also contains the error of the distance (as given by {\tt SNooPy2}) which is added to the fitting uncertainty reported by {\tt Minim} in quadrature.

Plots of the the best-fit bolometric LCs corresponding to the smallest $\chi^2$ are shown in the Appendix. 

The crucial parameter in the semi-analytic LC codes is the effective optical opacity ($\kappa$) that is assumed to be constant both in space and time. 
To estimate the effective optical opacity for our sample, we applied the same technique as done by  \citet{li18} for SN~2018oh recently. This technique is based on the combination of the light curve time scale, $t_{lc}$ and that of the gamma-ray leakage, $t_\gamma$. These parameters can be expressed with the physical parameters of the ejecta as 
\begin{equation}
t_{lc} ^2~={~2 \kappa M_{ej} \over \beta c v_{exp}}
\quad\text{and}\quad 
t_\gamma ^2~={~3 \kappa_\gamma M_{ej} \over 4 \pi v_{exp}^2}
\label{eq:t}
\end{equation}
\citep{arnett82, cw97, valenti08, manos12, li18},
where $M_{ej}$ is the ejecta mass, $\beta = 13.8$ is a fixed LC parameter related to the density distribution \citep{arnett82},  $v_{exp}$ is the expansion velocity and $\kappa_\gamma = 0.03$ cm$^2$g$^{-1}$ is the opacity for $\gamma$-rays \citep{wheeler15}.

Since $t_{lc}$ and $t_\gamma$ are measured quantities, the two formulae in Equation~\ref{eq:t} contains 
three unknowns: $M_{ej}$, $v_{exp}$, and $\kappa$. 
Following \citet{li18}, we apply two additional constraints for $M_{ej}$ and $v_{exp}$ to get upper and lower limits for $\kappa$.
Assuming that the ejecta mass cannot exceed the Chandrasekhar limit ($M_{ej}\leq M_{Ch}$) we get a lower limit for the optical opacity, $\kappa^{-}$, while assuming a lower limit for $v_{exp}$ as $v_{exp}\geq 10,000$ km~s$^{-1}$ we get an upper limit, $\kappa^{+}$ (see Equation \ref{eq:t}.).

In the first case the lower limit for $\kappa$  can be calculated  as 
\begin{equation}
\kappa^{-} ~=\sqrt{~{3 \kappa_\gamma t_{lc}^4 \beta^2 c^2 }\over{ 16 \pi  t_\gamma^2 M_{Ch} }} .
\label{eq:kappam}
\end{equation}

Second, the lower limit for the expansion velocity, $v_{exp} = 10,000$ km~s$^{-1}$, implies
\begin{equation}
\kappa^{+} ~=~ { {3 \kappa_\gamma t_{lc}^2 \beta c} \over {8 \pi v_{exp} t_\gamma^2}} .
\label{eq:kappap}
\end{equation}

Finally we estimate $\kappa$ as the average of $\kappa^{-}$ and $\kappa^{+}$, and derive $M_{ej}$ and $v_{exp}$ via the following expressions
\begin{equation}
 M_{ej}~=~{3 \kappa_\gamma t_{lc} ^4 \beta^2 c^2 \over 16 \pi  t_\gamma^2 \kappa ^2}
 \quad\text{and}\quad 
 v_{exp}~=~{3 \kappa_\gamma t_{lc} ^2 \beta c \over 8 \pi \kappa  t_\gamma^2}.
\label{eq:mejvexp}
\end{equation}

Having $M_{ej}$ and $v_{exp}$ evaluated, we express the kinetic energy of ejecta as 
$E_{kin} = 0.3 \cdot M_{ej} \cdot v_{exp}^2$ \citep{arnett82,manos12}.

The results of these calculations are collected in Table~\ref{tab:lc-bol}, where the
 best-fit LC parameters for the $R_V = 3.1$ reddening model are shown.
Uncertainties (given in parentheses, as previously) are calculated via error propagation taking into account the uncertainties of the fitted timescales and the mean optical opacity, the latter approximated as $\sigma_\kappa \approx 0.5 ( \kappa^{+} - \kappa^{-})$. 

 The same fitting process was repeated on the set of bolometric LCs computed with the $R_V=1.5$ reddening law (Section~\ref{bol1}). Since only the amplitude of the LCs are affected by the choice of the reddening law, and the timescales are not, it is expected that only the $^{56}$Ni values will be different from those listed in Table~\ref{tab:lc-bol}. Indeed, the resulting best-fit parameters remained almost the same, i.e. within their $1 \sigma$ uncertainties, except for the $^{56}$Ni masses that were found to be systematically lower. The nickel mass ratio between the two different reddening models had the same dependence on $E(B-V)_{host}$ as the luminosity ratio shown in Section~\ref{bol1}, which is expected since the amplitude of the LC is directly related to the calculated nickel mass.

 Note that the lack of spectroscopic (i.e. velocity) data makes the kinetic energies poorly constrained. This can be seen from the large relative errors of the $E_{kin}$ values given in Table~\ref{tab:lc-bol} that may exceed $\sim 50$ percent in some cases. Therefore, we do not use the kinetic energies derived from the LC fitting for constraining the physics in these Type Ia SNe.

\begin{figure*}
\begin{center}
\includegraphics[width=8.5cm]{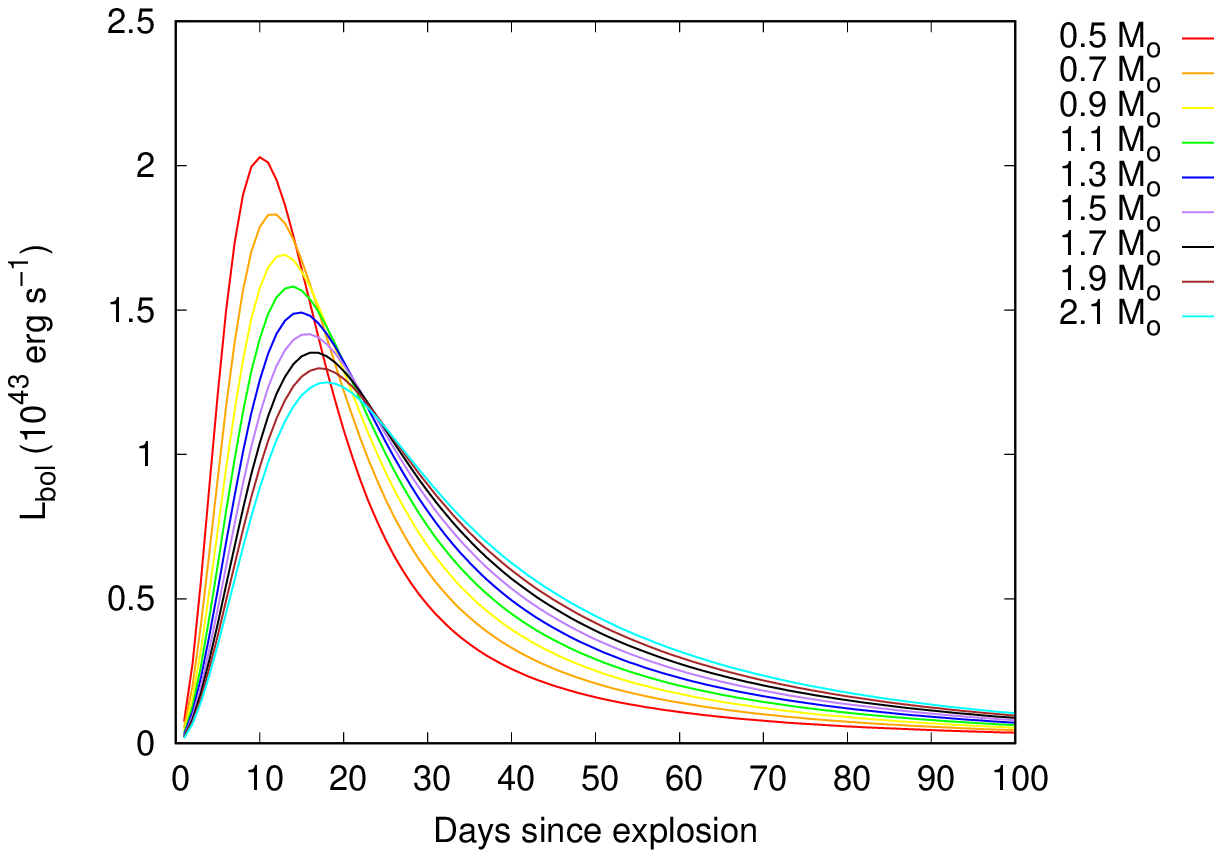}
\includegraphics[width=8.5cm]{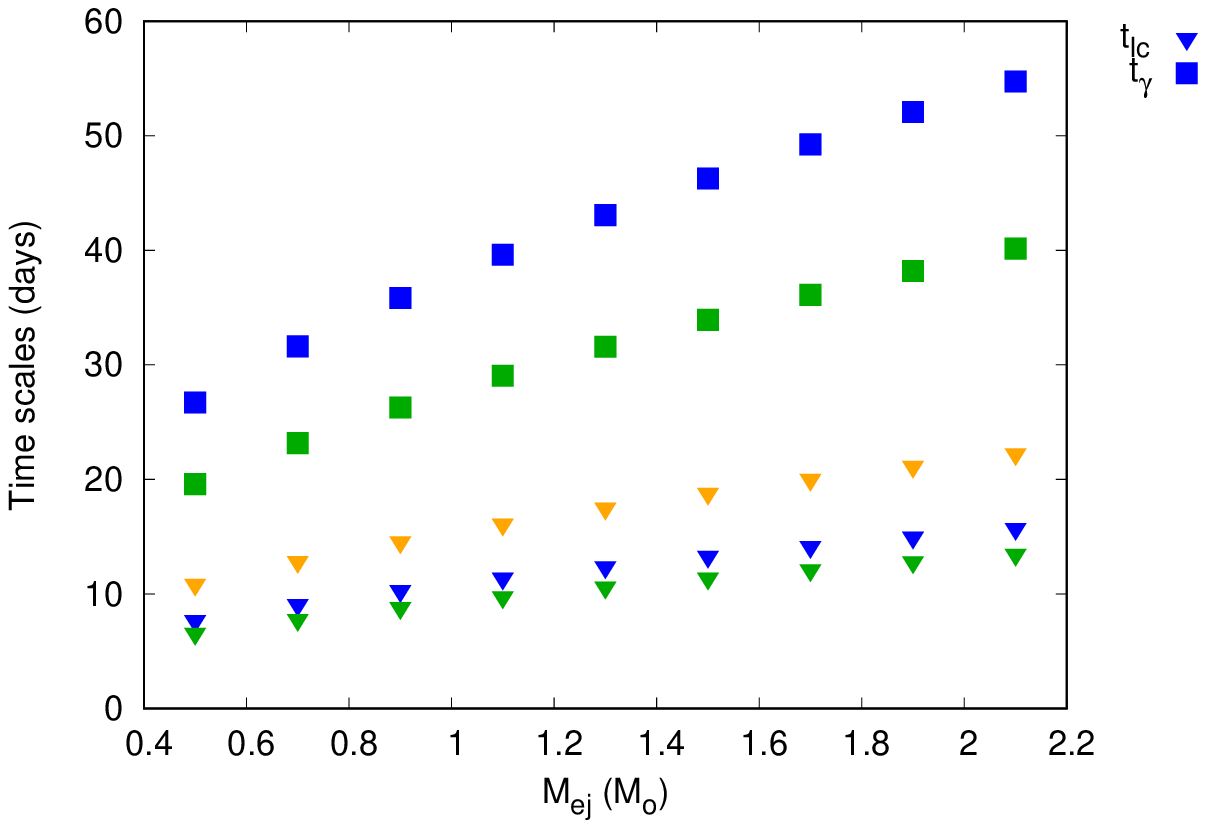}
\end{center}
\caption{Left panel: bolometric LCs for models having $M_{ej}$ in between 0.5 and 2.1 M$_\odot$. All other input parameters for the models were fixed as $R_0 = 0.01 R_\odot$, $v_{exp} = 11~000$ km~s$^{-1}$, $\kappa$ = 0.1 cm$^2$g$^{-1}$, and $M_{Ni}~=~0.6$ M$_\odot$.
Right panel: $t_{lc}$ (triangles) and $t_\gamma$ (squares) as a function of $M_{ej}$.   The values corresponding to $v_{exp} = 11000$ km~s$^{-1}$ and $\kappa$ = 0.1 cm$^2$g$^{-1}$ are plotted with blue,  $v_{exp} =15000$ km~s$^{-1}$ and $\kappa = 0.1$ cm$^2$g$^{-1}$ with green, and $v_{exp}=11000$ km/s, $\kappa=0.2$ cm$^2$g$^{-1}$ with orange symbols.
}
\label{fig:mejtlctgamma}
\end{figure*}

 Instead, in this paper we focus on constraining the ejecta masses.
In order to illustrate that the ejecta mass can be inferred from the combination of $t_{lc}$ and $t_\gamma$ via Eq. \ref{eq:mejvexp}, we constructed model LCs using the same Arnett model as above, assuming various ejecta parameters: progenitor radius ($R_0$), ejecta mass ($M_{ej}$), initial nickel mass ($M_{Ni}$), expansion velocity ($v_{exp}$) and optical opacity ($\kappa$). The left panel of Fig.~\ref{fig:mejtlctgamma} shows some of them corresponding to 
$\kappa = 0.1$ cm$^2$g$^{-1}$, 
$v_{exp} = 11,000$ km~s$^{-1}$, and $M_{ej}$ in between 0.5 and 2.1 M$_\odot$. 
It can be seen that larger $M_{ej}$ implies both longer rise time and slower decline rate consistently with Eq. \ref{eq:t}. 
The right panel indicates the correlation between the ejecta mass and the two timescale parameters, $t_{lc}$ and $t_\gamma$.
Here $t_\gamma$ is plotted with squares while $t_{lc}$ with  triangles, and different colors code different physical parameters as follows: blue means $v_{exp}=11,000$ km~s$^{-1}$, $\kappa = 0.1$ cm$^2$g$^{-1}$, green denotes $v_{exp}=15,000$ km~s$^{-1}$, $\kappa = 0.1$ cm$^2$g$^{-1}$, and orange is $v_{exp}=11,000$ km~s$^{-1}$, $\kappa = 0.2$ cm$^2$g$^{-1}$.
As it is expected, the shorter the time scales, the lower the model ejecta mass, suggesting that the combination of $t_{lc}$ and $t_\gamma$ outlined above may indeed provide realistic estimates for $M_{ej}$  within the framework of the Arnett model.

We use two SNe as reference objects in order to test the consistency of our LC modeling described above with those presented in other studies. SN~2011fe is chosen as the first test object due to the availability of precise, high-cadence observations spanning from the near-UV to the near-IR regimes (see Section~\ref{bol1}). \citet{scalzo14a} modeled the bolometric LC of SN~2011fe and obtained $M_{ej} = 1.19 \pm 0.12$ M$_\odot$ and $M_{Ni} = 0.42 \pm 0.08$ M$_\odot$ for the ejected mass and the nickel mass, respectively.
Using $R_V~=~3.1$, our best-fit results are $M_{ej}=1.00 \pm 0.070$ M$_\odot$ and $M_{Ni} = 0.567 \pm 0.042$ M$_\odot$ (see Table \ref{tab:lc-bol}). 
 It is seen that these two estimates  
are only marginally consistent: the difference between the two ejecta masses slightly exceeds $1 \sigma$, while the $^{56}Ni$-masses differ by $\sim 2 \sigma$. 
In addition to the sensitivity of the $^{56}Ni$-mass to the uncertainties in the distance, this highlights the possible systematic differences between the two modeling schemes applied by \citet{scalzo14a} and in this paper: \citet{scalzo14a} used only the late-time bolometric LC to constrain the ejecta and nickel masses via $t_{\gamma}$ based on the method of \citet{jeff99}, while we fit the full Arnett model to the entire LC.  
Note that the $^{56}Ni$ mass of SN~2011fe has also been determined in several other papers, including \citet{pereira13} ($0.53 \pm 0.11$ M$_\odot$), \citet{mazzali15} ($0.47 \pm 0.07$ M$_\odot$) and \citet{zhang16} ($0.57$ M$_\odot$). The scattering of these various estimates suggest a value of $M_{Ni} \sim 0.5 \pm 0.1$ M$_\odot$ for SN~2011fe, which makes both our result and that of \citet{scalzo14a} consistent. 

 On the other hand, good agreement is found between the parameters of the other test object, SN~2018oh.
Recently \citet{li18} derived $M_{ej} = 1.27 \pm 0.15$ M$_\odot$ and $M_{Ni} = 0.55 \pm 0.04$ M$_\odot$, which are very similar to our best-fit values (Table~\ref{tab:lc-bol}), $M_{ej} = 1.22 \pm 0.48$ M$_\odot$ and $M_{Ni} = 0.60 \pm 0.06$ M$_\odot$. Even though \citet{li18} applied the same method for the LC fitting as we use in this paper, their bolometric LC was assembled from a much denser, more extended dataset, including observed near-UV and near-IR photometry. The good agreement between our best-fit parameters and theirs suggests that our parameters are not unrealistic, and probably do not suffer from severe systematic errors.  Note, however, that our estimate for the ejecta mass of SN~2018oh has higher uncertainty ($\sim 40 \%$) than that of \citet{li18}, thus, our data are less constraining regarding the ejecta mass. This is true for most SNe listed in Table~\ref{tab:lc-bol}.


\section{Discussion}\label{dis}
\subsection{Early color evolution}\label{ece}

\begin{figure*}
\begin{center}
\includegraphics[width=8.5cm]{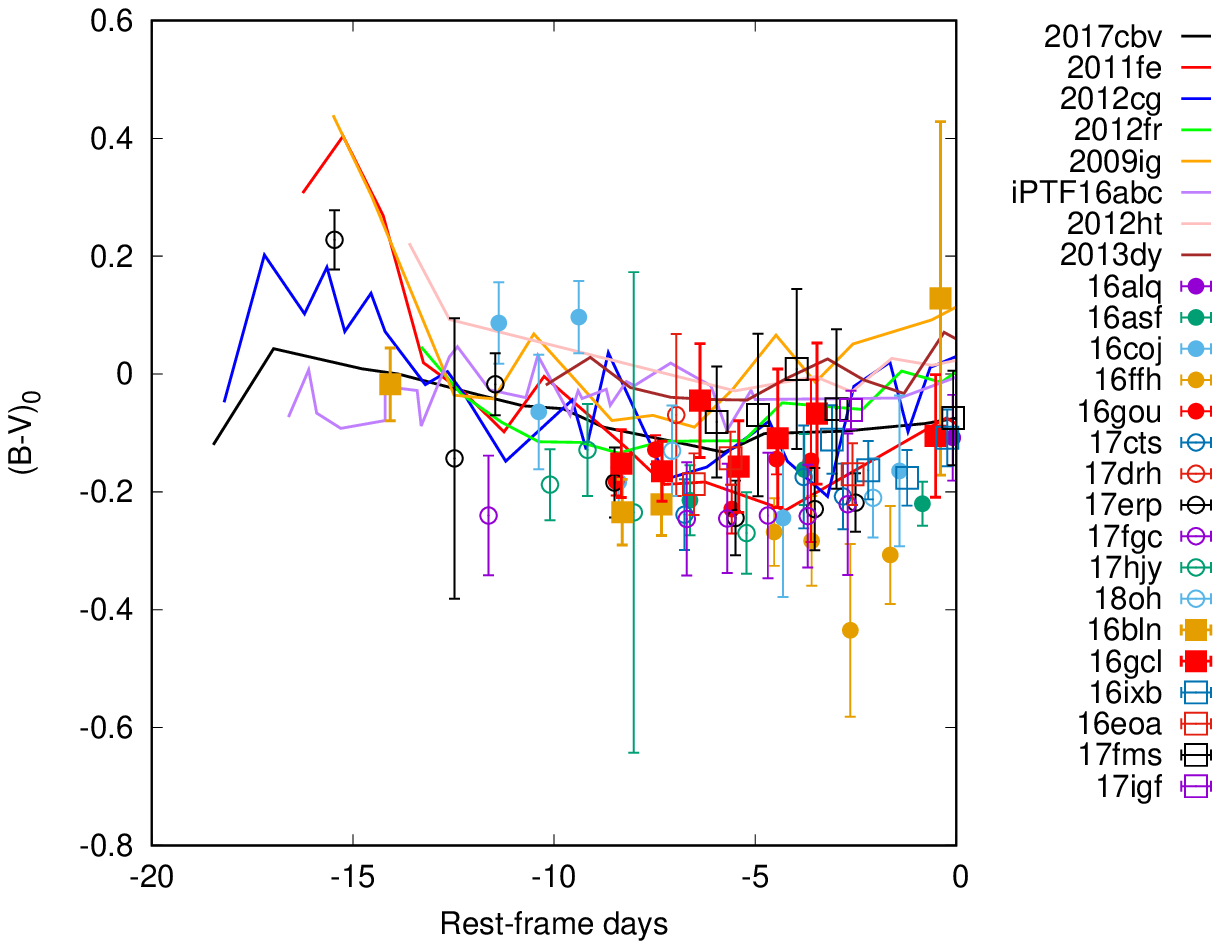}
\includegraphics[width=8.5cm]{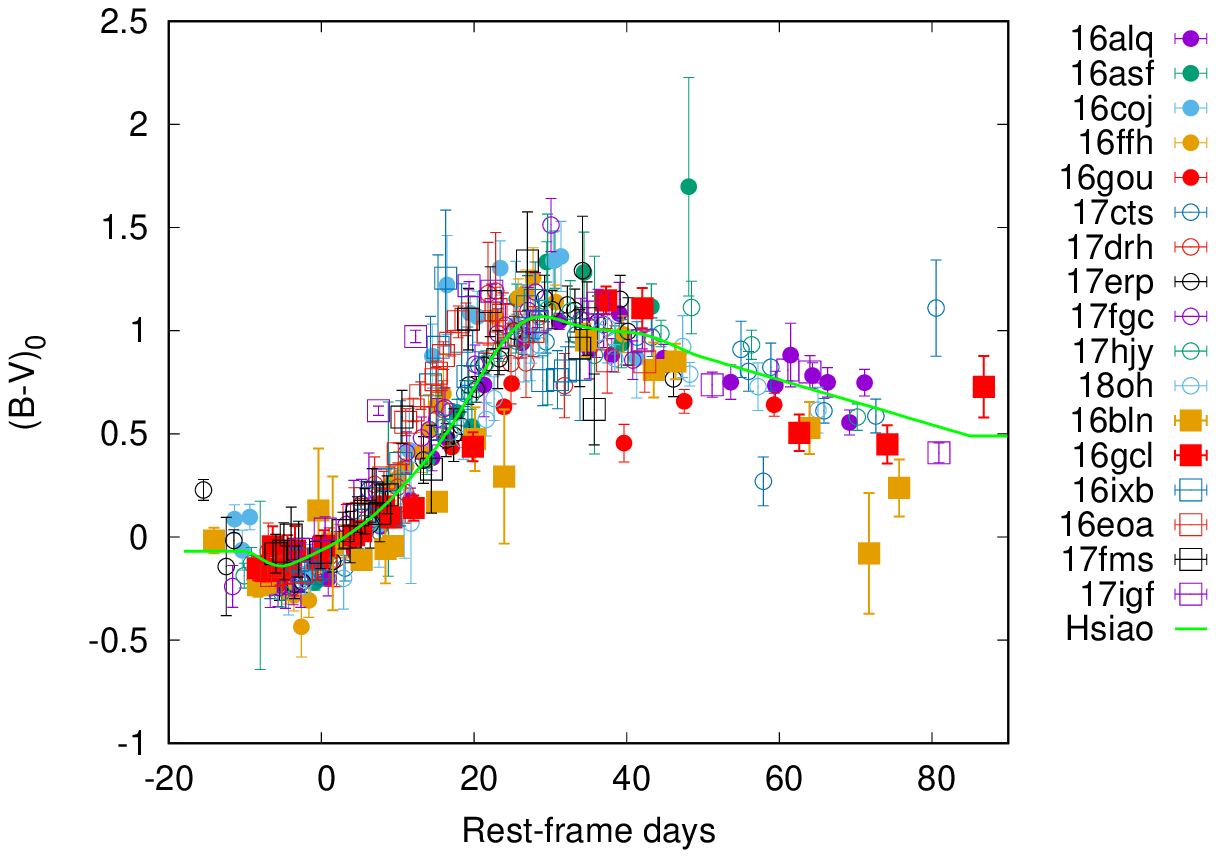}
\end{center}
\caption{The evolution of the reddening-corrected $(B-V)_0$ colors for the sample SNe (colored symbols). Left panel: the pre-maximum color evolution compared to other well-observed SNe collected from literature (colored lines).  Right panel:  $B-V_0$ colors of the sample plotted together with Hsiao template (green curve) up to 90 days after maximum.  }
\label{ebv_neg}
\end{figure*}

In Fig.~\ref{ebv_neg} we plot 
the early $(B-V)_0$ colors, corrected for both Milky Way and host galaxy reddening (Table \ref{tab:snpy}), of our sample together with the data for several other well-observed SNe Ia: SN~2017cbv \citep{hosse17}, SN~2011fe \citep{vinko12}, SN~20112cg \citep{vinko18}, SN~2017fr \citep{contreras18}, SN~2009ig \citep{marion16, foley12}, iPTF16abc \citep{miller18}, SN~2012ht \citep{vinko18} and SN~2013dy \citep{vinko18}).
In the following we investigate the pre-maximum color evolution of our observed sample.

Early-phase $(B-V)_0$ observations of SNe Ia suggest that they can be divided into two categories: early-red and early-blue type \citep[e.g.][]{strici18}. The cause of this dichotomy is still debated (see Section~\ref{intro}). For example, \citet{miller18} proposed physical models for the progenitor system and the explosion of iPTF16abc. This SN Ia showed blue, nearly constant $(B-V)_0$ color starting from t $\sim$ -10 days, which was thought to be caused by strong $^{56}$Ni-mixing in the ejecta. 

SNe Ia experience a dark phase after shock breakout (SB), before the heating from radioactive decay diffuses through the photosphere. The duration of this dark phase depends on how much $^{56}$Ni is mixed into the outer layers of the ejecta. If $^{56}$Ni is confined to the innermost layers, the dark phase lasts for a few days, so that weak mixing leads to redder colors and moderate luminosity rise. On the contrary, strong mixing results in higher luminosities and bluer optical colors. In the latter case the dark phase does not exist or very short, because the $\gamma$-photons originating from the Ni-decay rapidly diffuse out from the ejecta. \citet{shap19} found that strong mixing accounts for the early excess light in the LC of SN~2018oh observed by {\it Kepler} during the K2-C16 campaign, although its color could not be constrained by the unfiltered K2 observations. 

Another possibility for the early blue flux is the collision of the ejecta into a nearby companion star or some kind of a circumstellar envelope. In this case a strong, quickly declining ultraviolet pulse is thought to be the root cause of the excess blue emission during the earliest phases. However, the observability of this emission requires a favorable geometric configuration, i.e. the companion being in front of the SN toward the observer, so it is expected to occur in less than 10\% of the actually observed SNe Ia \citep[e.g.][]{kasen10}.

With the use of our new photometric
data we attempt to investigate the evolution of the early $(B-V)_0$ color for our sample SNe Ia.  
Fig.\ref{ebv_neg} illustrates that our data are consistent with the colors of other SNe Ia collected from recent literature (plotted as continuous lines in the left panel of Fig. \ref{ebv_neg}). Also shown is the prediction based on the Hsiao-template (plotted with a green line in the right panel) that represents an empirical description of the time-evolving spectral energy distribution of a fiducial SN Ia \citep{hsiao07}. In this case the colors are derived by synthetic photometry using Bessell $B$ and $V$ filter functions \citep{bessell90} on the spectrum templates given by \citet{hsiao07}. 

In order to classify SNe Ia into the
early-red and early-blue groups, photometric data taken between $-20$ and $-10$ days before maximum is necessary. Between $t = -10$ days and $t_{max}$ the $(B-V)_0$ colors are so similar for most SNe Ia that it is almost 
impossible to make such a distinction. Unfortunately, most of the SNe Ia in our sample do not have photometry taken early enough for this purpose.
During our campaign this very early phase ($-20 < t < -10$ days) have been observed only in two cases: 
SN~2016bln and 2017erp (see the left panel in Figure \ref{ebv_neg}).

SN~2016bln is a 1991T-like or slow-decliner Ia belonging to the Type SS (Shallow Silicon) subclass on the Branch-diagram \citep{bln}. Such SNe Ia are recently found to be associated with the early-blue group by \citet{strici18}: they pointed out that such SNe tend to be located in between the Core Normal and the Shallow Silicon types on the Branch diagram, similar to 1991T-like events. Their finding suggests that SN~2016bln may also belong to the early-blue group. As seen on the left panel of Figure~\ref{ebv_neg}, the earliest data on SN~2016bln are indeed close to those of iPTF16abc and SN~2017cbv, two well-known members of the early-blue group, thus, they are consistent with the expected early-color evolution. However, our data are too sparse to draw a more definite conclusion.

The other object in our sample that was observed sufficiently early, 
SN~2017erp, shows an early $(B-V)_0$ color that is similar to that of SN~2011fe, thus, it belongs to the early-red group. It is interesting that \citet{erp} found that SN~2017erp was a near-UV (NUV)-red object, while SN~2011fe was a NUV-blue event. This difference between the NUV-colors seems to be independent from the early-phase optical colors, because
2011fe and 2017erp both had similarly red early $(B-V)_0$ color. 

This result may suggest that the observed spread in the early NUV- and optical $(B-V)_0$ colors of SNe Ia has different physical reasons. \citet{erp} concluded that the diversity in the NUV colors is likely due to the metallicity of the progenitor that affects the NUV-continuum and the strength of the Ca H{\&}K features. The fact that SN~2017erp and SN~2011fe have similar early $(B-V)_0$ but different NUV colors may suggest that the early $(B-V)_0$  diversity might not be directly related to the progenitor metallicity. 


\subsection{Comparison with explosion models}\label{pdd}

\begin{table}[]
\centering
\caption{Parameters of the DDE and PDDE models by \citet{dessart}}
\begin{tabular}{ccc|ccc}
\hline
\hline
Model & $E_{kin}$ & $M_{Ni}$ &  Model & $E_{kin}$ & $M_Ni$ \\
      & ($10^{51}$ erg) & (M$_\odot$) &   & ($10^{51}$ erg) & (M$_\odot$) \\
\hline
DDC0 & 1.573 & 0.869 & PDDEL1 & 1.398 & 0.758 \\
DDC6 & 1.530 & 0.722 & PDDEL3 & 1.353 & 0.685 \\
DDC10 & 1.520 & 0.623 & PDDEL7 & 1.336 & 0.604 \\
DDC15 & 1.465 & 0.511 & PDDEL4 & 1.344 & 0.529 \\
DDC17 & 1.459 & 0.412 & PDDEL9 & 1.342 & 0.408 \\
DDC20 & 1.442 & 0.300 & PDDEL11 & 1.236 & 0.312 \\
DDC22 & 1.345 & 0.211 & PDDEL12 & 1.262 & 0.268 \\   
DDC25 & 1.185 & 0.119 & & & \\
\hline
\end{tabular}
\label{tab:demodels}
\end{table}

\begin{table}[]
    \centering
     \caption{$^{56}$Ni masses of the best-fit explosion models. The second column contains the $^{56}$Ni masses inferred from the bolometric LCs assuming $R_V=3.1$ reddening law. Uncertainties are given in parentheses. }
    \label{tab:demodels-31}
    \begin{tabular}{lccc}
    \hline
    \hline
    Name & $M_{Ni}^\mathrm{bol}$ & $M_{Ni}^\mathrm{DDE}$ & $M_{Ni}^\mathrm{PDDE}$\\
     & ($M_\odot$) & ($M_\odot$) & ($M_\odot$) \\
     \hline
    Gaia16alq & 0.744 (0.055) & 0.623 & 0.604 \\
    2016asf & 0.597 (0.149) & 0.623 & 0.604 \\
    2016bln & 0.789 (0.097) & 0.869 & 0.758 \\ 
    2016coj & 0.401 (0.053) & 0.412 & 0.604 \\
    2016eoa & 0.482 (0.103) & 0.412 & 0.604 \\
    2016ffh & 0.573 (0.078) & 0.511 & 0.529 \\
    2016gcl & 0.689 (0.164) & 0.623 & 0.758 \\
    2016gou & 0.678 (0.063) & 0.623 & 0.758 \\
    2016ixb & 0.483 (0.064) & 0.511 & 0.685 \\
    2017cts & 0.539 (0.063) & 0.511 & 0.685 \\
    2017erp & 0.975 (0.083) & 0.722 & 0.685 \\
    2017fgc & 0.692 (0.047) & 0.623 & 0.604 \\
    2017fms & 0.360 (0.029) & 0.511 & 0.685 \\
    2017hjy & 0.688 (0.057) & 0.623 & 0.685 \\
    2017igf & 0.420 (0.051) & 0.300 & 0.408 \\
    2018oh & 0.598 (0.059) & 0.623 & 0.685 \\
    \hline
    \end{tabular}
\end{table}

\begin{table}[]
    \centering
     \caption{The same as Table~\ref{tab:demodels-31} but assuming $R_V=1.5$ reddening law.}
    \label{tab:demodels-15}
    \begin{tabular}{lccc}
    \hline
    \hline
    Name & $M_{Ni}^\mathrm{bol}$ & $M_{Ni}^\mathrm{DDE}$ & $M_{Ni}^\mathrm{PDDE}$\\
     & ($M_\odot$) & ($M_\odot$) & ($M_\odot$) \\
     \hline
    Gaia16alq & 0.651 (0.055) & 0.623 & 0.604 \\
    2016asf & 0.492 (0.149) & 0.511 & 0.604 \\
    2016bln & 0.560 (0.097) & 0.869 & 0.758 \\ 
    2016coj & 0.397 (0.053) & 0.412 & 0.604 \\
    2016eoa & 0.333 (0.103) & 0.412 & 0.604 \\
    2016ffh & 0.410 (0.078) & 0.511 & 0.604 \\
    2016gcl & 0.641 (0.164) & 0.623 & 0.758 \\
    2016gou & 0.450 (0.063) & 0.623 & 0.685 \\
    2016ixb & 0.417 (0.064) & 0.511 & 0.685 \\
    2017cts & 0.411 (0.063) & 0.511 & 0.685 \\
    2017erp & 0.686 (0.083) & 0.722 & 0.685 \\
    2017fgc & 0.538 (0.047) & 0.623 & 0.604 \\
    2017fms & 0.338 (0.029) & 0.511 & 0.685 \\
    2017hjy & 0.489 (0.057) & 0.623 & 0.685 \\
    2017igf & 0.327 (0.051) & 0.300 & 0.408 \\
    2018oh & 0.489 (0.059) & 0.623 & 0.685 \\
    \hline
    \end{tabular}
\end{table}

\begin{figure*}
\begin{center}
\includegraphics[width=8.5cm] {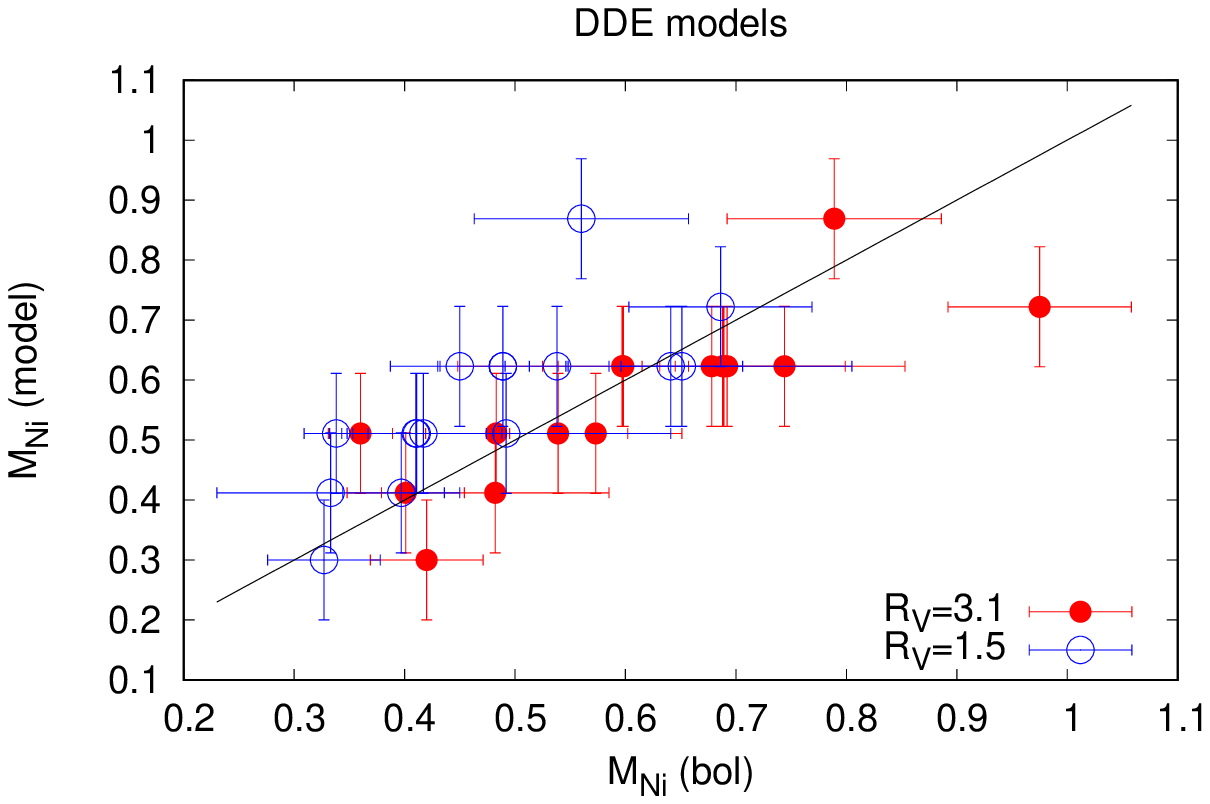}
\includegraphics[width=8.5cm] {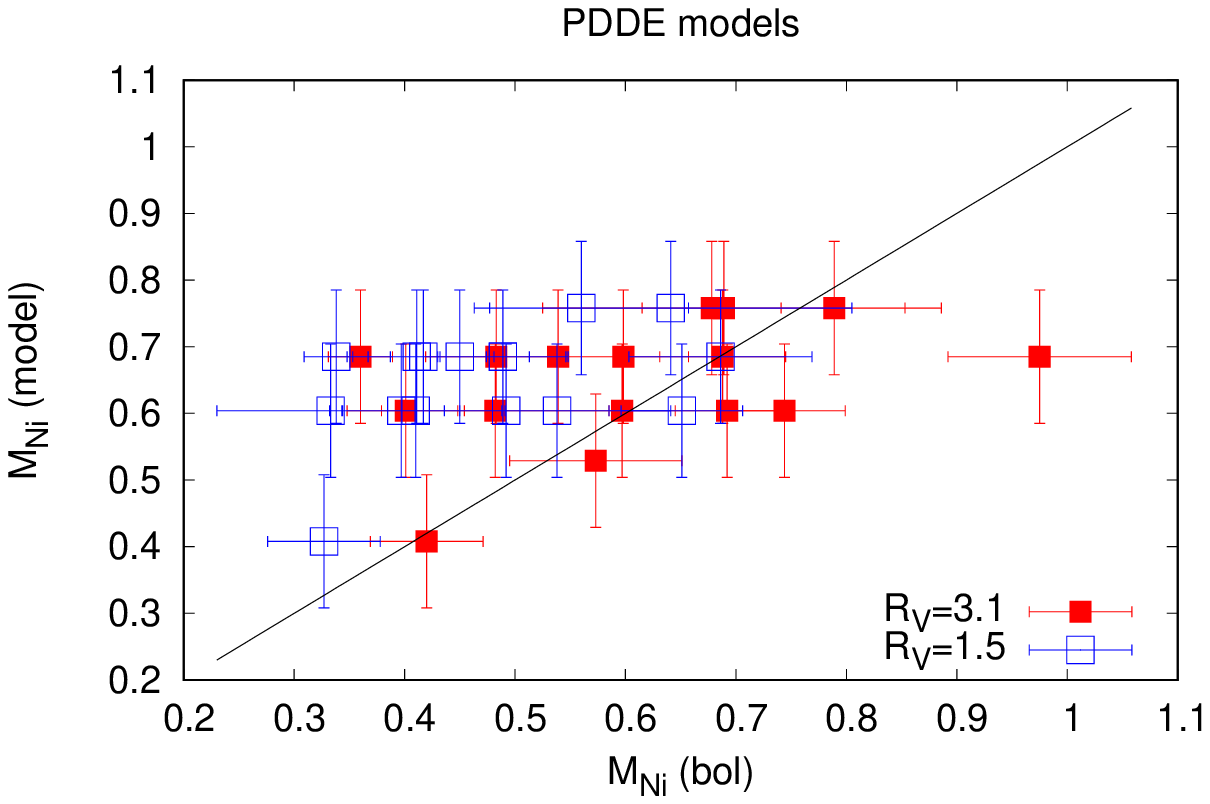}
\end{center}
\caption{Comparison of the Ni-masses derived from the bolometric LCs (plotted on the horizontal axis) with those from DDE (left panel) and PDDE (right panel)  models. Data inferred by assuming $R_V=3.1$ reddening law are plotted with filled red symbols, while those from the $R_V=1.5$ reddening law are shown by open blue ones. Solid lines represents the expected 1:1 relation on both panels. }\label{ni}
\end{figure*}

In this subsection we compare parameters derived from the bolometric LC fitting, in particular the nickel mass, to those taken from several explosion models. First, we consider the DDE and PDDE models computed by \citet{dessart}, as listed in Table~\ref{tab:demodels} (we keep the original naming scheme of these models, such as DDC$nn$ models use the DDE scenario, while the PDDEL$nn$ models assume the PDDE mechanism). It is seen that the Ni-masses of these models span the same range than the ones inferred from the bolometric LC fitting (see Table~\ref{tab:lc-bol}), but the kinetic energies of the models are higher by about a factor of $\sim 2$.

PDDE models exhibit strong C~II lines shortly after explosion, which are formed in the outer, unburned material. Furthermore,  the collision with the previously ejected unbound material surrounding the WD results in the heat-up of the outer layers of the ejecta. Thus, in the PDDE scenario the early color of the SN is bluer, and the luminosity rises faster than in the conventional DDE models.

On the contrary, standard DDE models typically leave no unburned material. Instead, at 1-2 days after explosion they  show red optical colors ($(B-V)_0 \sim 1$ mag ), which gets bluer continuously as the SN evolves toward maximum light. After maximum both the DDE and the PDDE scenarios show nearly the same $(B-V)_0$ color. 

We compare the observed, reddening-corrected $(B-V)_0$ colors of our sample with the predictions from these explosion models. The colors from the models were derived via synthetic photometry applying the standard Bessell $B$ and $V$ filters \citep{bessell90}, as above.
Note that the in the redshift range  of the observed SNe ($z \leq 0.031$) the K-corrections for the $(B-V)_0$ color indices does not exceed 0.06 mag, which is comparable to the uncertainty of our color measurements in these bands. Thus, the K-corrections were neglected when comparing the observed $(B-V)_0$ colors to the synthetic colors inferred from the the explosion models. 
A plot comparing the observed and synthetic color curves, for both DDE and PDDE models, can be found in the Appendix. 

After computing the synthetic $B-V$ curves as a function of phase, the one that  has the lowest $\chi^2$ with respect to the 
observed $(B-V)_0$ colors was chosen as the most probable explosion model that describes the observed SN. 
 These best-fit models and their nickel masses are shown in Table \ref{tab:demodels-31} and \ref{tab:demodels-15} for the $R_V=3.1$ and $R_V=1.5$ reddening laws, respectively.
Figure \ref{ni} compares the Ni-masses of the best-fit models to those derived directly from bolometric LC fitting (Table~\ref{tab:lc-bol}).

Although the grid resolution of the models is inferior, it is seen that the nickel masses  from the DDE models nicely correlate with the ones from the bolometric LC fitting; the only outlier, SN~2017erp, may have an overestimated $M_{Ni} \sim 1$ M$_\odot$ due to a reddening issue \citep{erp}, even though its early $(B-V)_0$ color was similar to that of SN~2011fe, cf. Section~\ref{ece}).
The agreement is  worse for the PDDE models: most of the best-fit models have practically the same Ni-mass, $M_{Ni} \sim$ 0.6 - 0.7 M$_\odot$.

\begin{figure}
\begin{center}
\includegraphics[width=8.5cm] {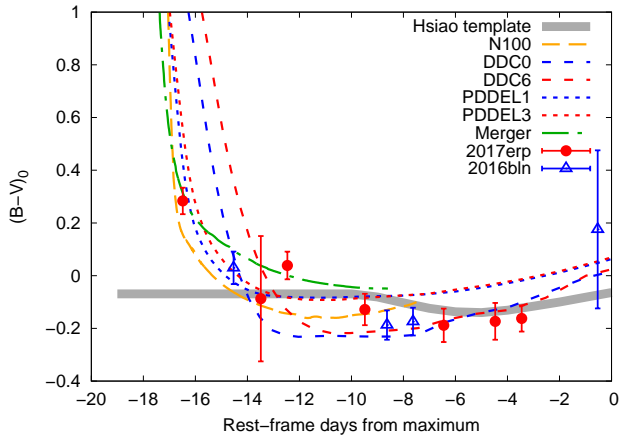}
\end{center}
\caption{Comparison of the early (dereddened) $(B-V)_0$ colors of SN~2017erp (red circles) and 2016bln (blue triangles) with the synthetic colors predicted by several Ia explosion models (DDE, PDDE and Violent Merger mechanisms, see text). The color evolution from the empirical Hsiao template is also shown with a thick grey line.}
\label{ddmodels}
\end{figure}

In Figure~\ref{ddmodels} a similar comparison of the synthetic $(B-V)$ colors from different explosion models with observations is shown, but only for the pre-maximum phases. Beside the best-fit DDE and PDDE models by \citet{dessart}, synthetic colors from two other theoretical models are also plotted: the N100 explosion model (DDE in a Chandrasekhar-mass WD) by \citet{seite13} and the Violent Merger (VM) model by \citet{pakmor12}. The color evolution from the empirical Hsiao template \citep[thick grey curve;][]{hsiao07} is also shown for comparison.  

It is seen that three of these model families (N100, PDDEL and VM)  show similar pre-maximum $(B-V)$ colors to the observed data at the earliest ($t < -14$ days) phases, although the PDDEL models are somewhat redder than the observations after $-10$ days. The DDC0 model fits the data of SN~2016bln very well. The DDC6 model by \citet{dessart} predicts too red $(B-V)$ color at the earliest phases, while those from the Hsiao template look being too blue. Although the models shown here do not seem to constrain the observed early-blue and early-red events, at least the two SNe in our sample (2016bln and 2017erp) that have been sampled at $t < -14$d have  similar $(B-V)_0$ colors to those of the models considered here.

It is concluded that the observed, de-reddened $(B-V)_0$ color evolution of SNe Ia seems to be more-or-less reproduced by current models of DDE and/or VM mechanisms.  The Ni-masses of the DDE models by \citet{dessart} that match the observed $(B-V)_0$ colors are consistent with the Ni-masses inferred from the bolometric LC fitting  assuming $R_V=3.1$ for the reddening law, but they are systematically higher than those inferred by applying the $R_V=1.5$ reddening law. This is not true for the PDDE models, since they predict too high nickel masses for SNe that have $M_{Ni} < 0.6$ M$_\odot$ from their bolometric LC fitting,  regardless of the adopted reddening law.

\subsection{Ejecta parameters}

In this section we examine the relations between the inferred ejecta parameters (Table~\ref{tab:lc-bol}) following \citet{scalzo14a} and \citet{scalzo18}.  First, we use the parameters derived from the $R_V=3.1$ reddening law, then we discuss how the assumption of a different reddening law ($R_V = 1.5$) affects the conclusions.

\subsubsection{Comparison with \citet{scalzo14a, scalzo18}}

\begin{figure*}[]
\begin{center}
\includegraphics[width=8cm] {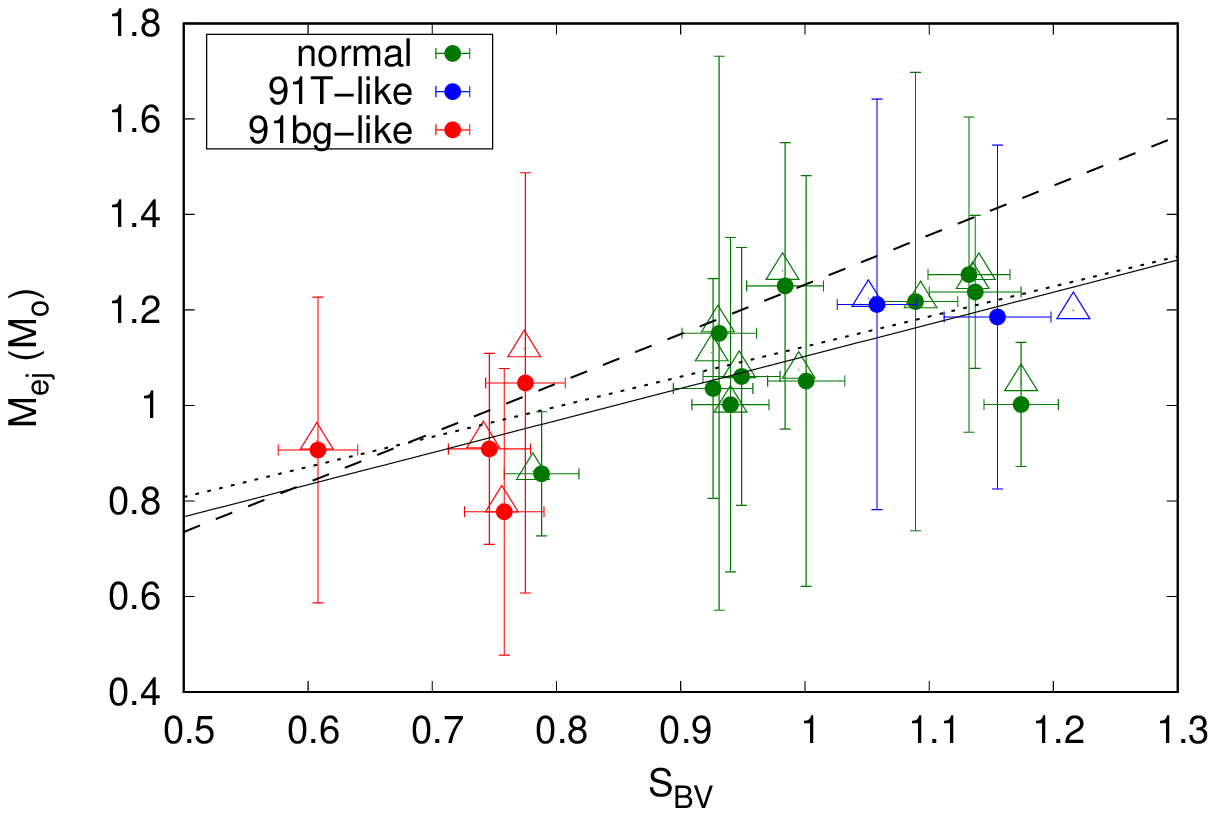}
\includegraphics[width=8cm] {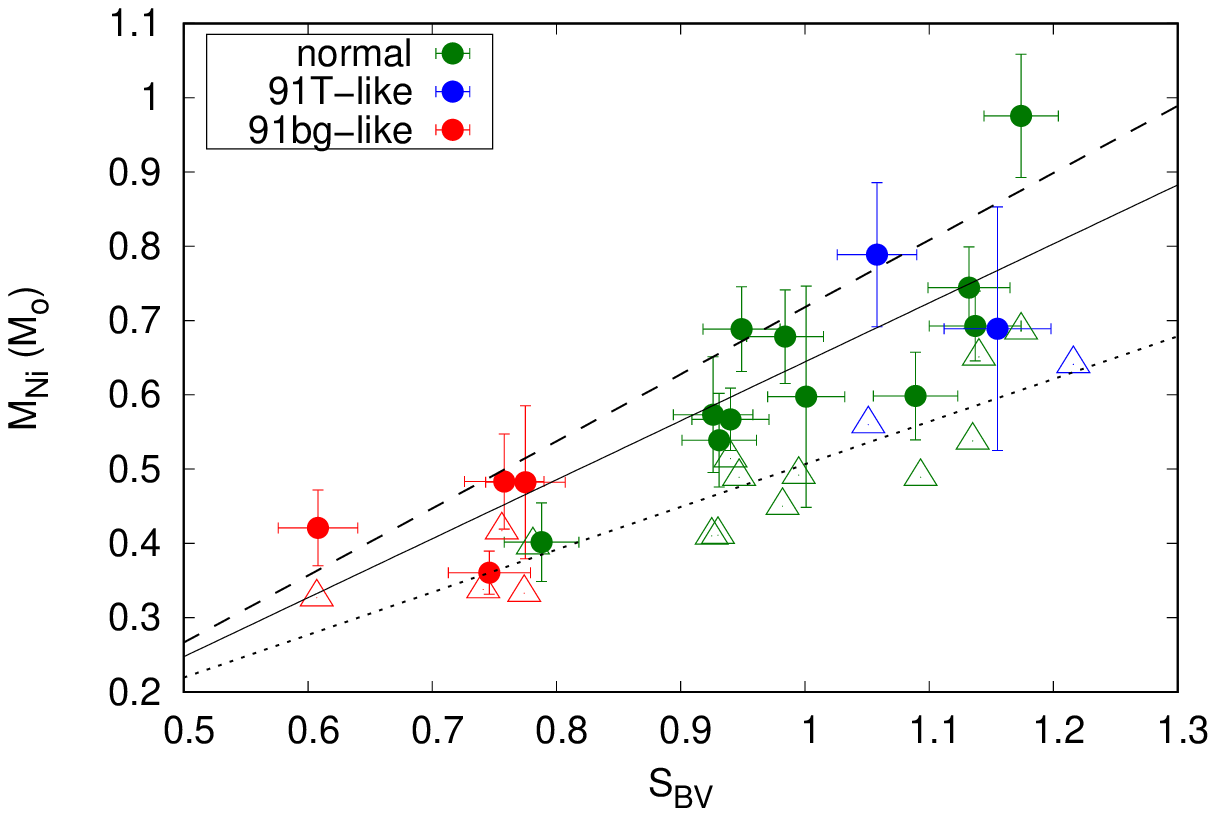}
\includegraphics[width=8cm] {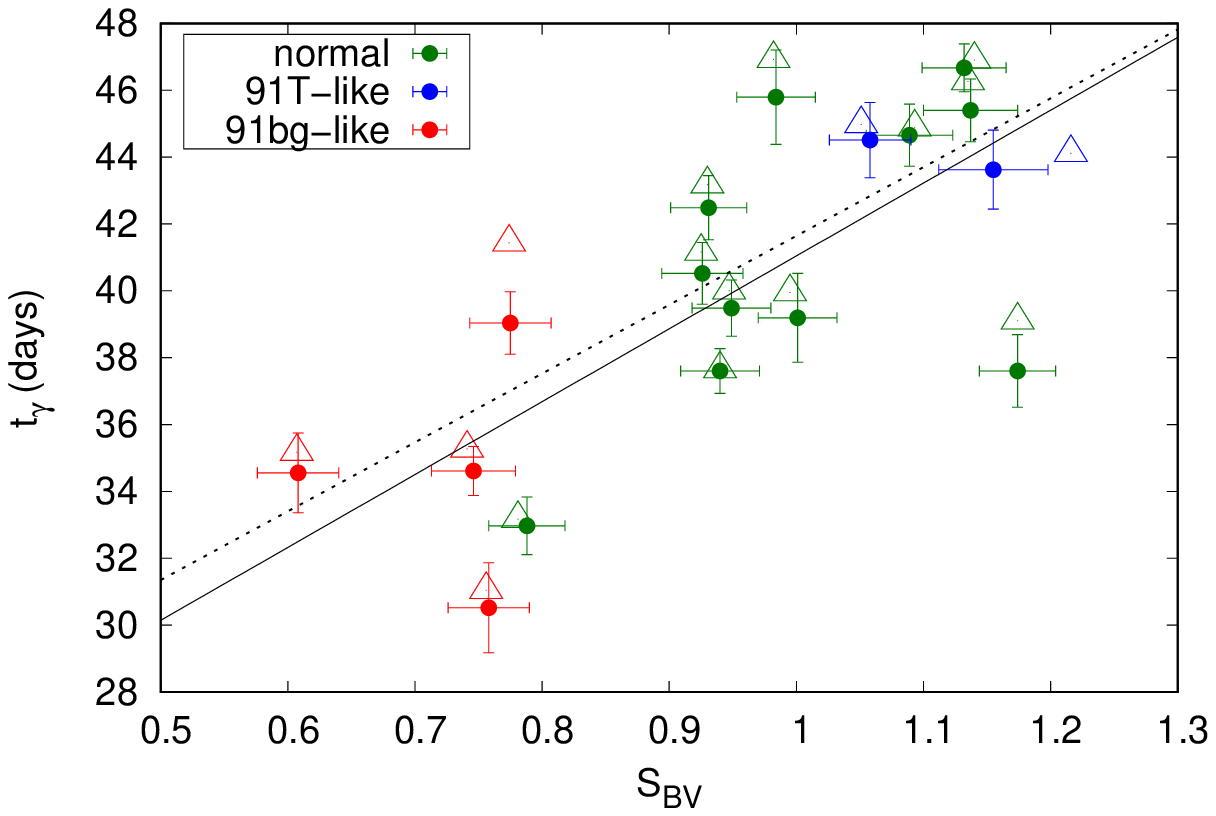}
\includegraphics[width=8cm] {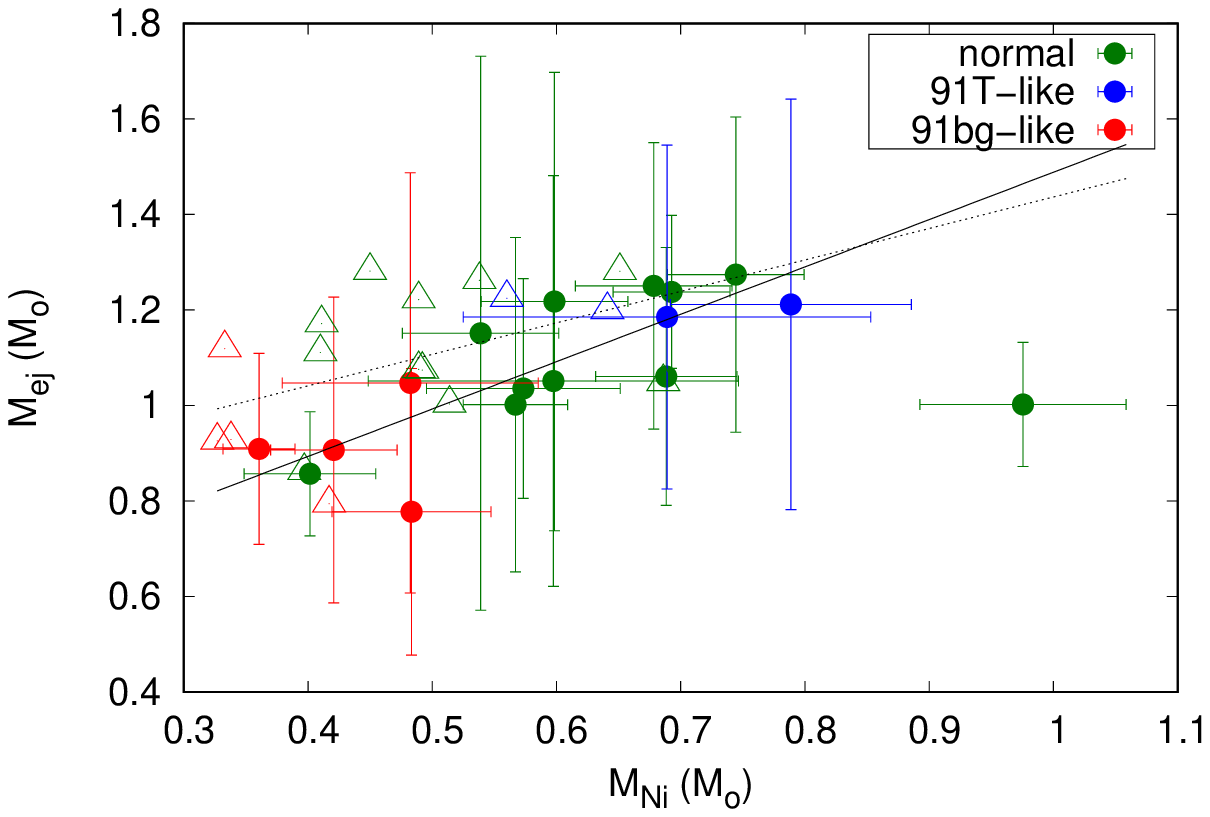}
\caption{Correlation between $s_{BV}$ and $M_{ej}$ (top left panel), $s_{BV}$ and $M_{Ni}$ (top right panel), $s_{BV}$ and $t_\gamma$ (bottom left panel), and $M_{Ni}$ and $M_{ej}$ (bottom right panel).  Filled dots correspond to parameters derived from the $R_V~= ~3.1$ reddening law, while open triangles denote the results from the $R_V=1.5$ reddening law.
Normal SNe are plotted with green symbols, while  blue and red symbols denote 91T-like and 91bg-like SNe, respectively. 
Dashed lines in the top panels represent the relations found by   \citet{scalzo18}, while continuous ($R_V=3.1$) and dotted ($R_V=1.5$) lines indicate fits to the present data. }
\end{center}
\label{scalzo1}
\end{figure*}

 The top left panel of Figure \ref{scalzo1} shows the dependence between the SNooPy2 color-stretch parameter, $s_{BV}$ and the ejecta mass. Different colors represent different SN Ia subtypes: normal SNe are plotted with green symbols, while 91T-like events are blue and 91bg-like objects are red. 
 
 The dashed line indicates the correlation found by \citet{scalzo18} between $M_{ej}$ and $s_{BV}$:
 $M_{ej} = (1.253 \pm 0.021) + (1.036 \pm 0.095) \cdot (s_{BV} - 1)$.
 It is seen that their finding is consistent with our data, at least for those that have $s_{BV} \lesssim 1$. 
 
 Fitting a straight line to the observed data resulted in the following empirical relation 
 \begin{equation}
 M_{ej} ~=~ (1.102 \pm 0.087) ~+~ (0.661 \pm 0.455) \cdot (s_{BV} - 1),
  \label{ktr_mej}
 \end{equation}
which is also  plotted in  Figure \ref{scalzo1} as a continuous line. 
It is seen that even though the new fitting parameters predict a less steep correlation between $s_{BV}$ and $M_{ej}$, they are consistent in $\pm 1 \sigma$ with the ones given by \citet{scalzo18}. As shown by Figure~\ref{scalzo1}, both the continuous and the dashed lines run within the errorbars of the data in the observed range of the $s_{BV}$ parameter.

The statistical significance of the correlation is investigated via deriving the Pearson correlation coefficient from all data shown in the top left panel of Figure~\ref{scalzo1}. 
Its value is $r = 0.751$, suggesting at first glance that $s_{BV}$ and $M_{ej}$ might be correlated. However, this estimate does not take into account the relatively high errorbars of our data, which likely make the real correlation statistically less significant. In order to test the effect of errors on the correlation coefficient, we computed 5,000 random realizations of the observed data by adding Gaussian random noise (having FWHM values equal to the errors) to each data point, then the Pearson correlation coefficient was computed for each random sample. The final value of the correlation coefficient and its uncertainty was estimated as the average and the standard deviation of the r-values from all of these randomized samples. This resulted in $r' = 0.301 \pm 0.200$, which is significantly less than the value above that does not take into account the uncertainties. Since the uncertainty of the correlation coefficient is similar to the value of the $r'$ parameter itself, it is concluded that the correlation between $s_{BV}$ and $M_{ej}$ is not significant in the present sample due to the relatively high uncertainties of the inferred ejecta masses. 
 
The top right panel of Figure \ref{scalzo1} plots $M_{Ni}$ versus $s_{BV}$. The dashed and continuous lines show the same correlations as found by \citet{scalzo18} and this paper, respectively.  The former was given as $M_{Ni} = (0.718 \pm 0.027) + (0.903 \pm 0.108) \cdot (s_{BV}-1)$, while the latter can be expressed as
\begin{equation}
 M_{Ni} = (0.643 \pm 0.023) + (0.768 \pm 0.122)  \cdot (s_{BV} - 1). 
 \label{ktr_ni}
 \end{equation}
The Pearson correlation coefficient, after taking into account the uncertainties as above, is $r' = 0.745 \pm 0.083$. Its conventional value (from the data alone without the errorbars) would be $r = 0.852$. The correlation coefficient suggests that $M_{Ni}$ indeed correlates with $s_{BV}$, even though the parameters given in Equation~\ref{ktr_ni} are only marginally consistent with those found by \citet{scalzo18}.

Equation~\ref{ktr_ni} suggests that  91T-like objects with slower decline rate (i.e. higher $s_{BV}$) tend to have larger Ni-masses, while 91bg-like SNe having lower $s_{BV}$ show smaller $M_{Ni}$. 


The bottom left panel in Figure~\ref{scalzo1} illustrates the dependence between $s_{BV}$ and $t_\gamma$ that is similar to the one found by \citet{scalzo14a} between their ``transparency time scale'' and the color-stretch parameter.
The line represents the fit to the data as 
\begin{equation}
    t_\gamma ~=~ (41.04 \pm 0.85) ~+~ (21.79 \pm 4.98) \cdot (s_{BV} - 1), 
\end{equation}
while the Pearson correlation coefficient is $r' = 0.722 \pm 0,048$ ($r = 0.749$ without the uncertainties). 

The results above suggests that the nickel masses also depend on the ejecta masses. The correlation between the derived $M_{Ni}$ and $M_{ej}$ masses is shown in the bottom right panel in Figure~\ref{scalzo1}. Except one outlier (SN~2017erp that likely has an overestimated $M_{Ni}$ due to its overestimated reddening), these two quantities seem to be connected, as the SNe with higher $M_{Ni}$ tend to have larger $M_{ej}$ as well.
However, the Pearson correlation coefficient, after removing the outlying SN~2017erp, but taking into account the uncertainties as above, is only 
$r'= 0.280 \pm 0.224$, which suggests that the correlation between the data shown is not statistically significant, even though the correlation coefficient without the errorbars would be $r = 0.822$.

Nevertheless, after fitting a straight line to the remaining 16 data points, the result can be expressed as
\begin{equation}
    M_{ej} ~=~ (0.728 \pm 0.586) \cdot M_{Ni} ~+~ (0.651 \pm 0.343).
\end{equation}
Since the uncertainty of the slope parameter is comparable to its best-fit value itself ($\pm 0.586$ vs. $0.728$), this result is also consistent with the low value of the $r'$ correlation coefficient determined above. We conclude that although the correlation between $M_{Ni}$ and $M_{ej}$ cannot be ruled out, it cannot be constrained from the present dataset due to mainly the relatively high errorbars of these two inferred parameters.

 From Figure~\ref{scalzo1} it is seen that the parameters derived from the $R_V=1.5$ reddening law (plotted with open triangles) do not differ significantly from the ones inferred from the $R_V=3.1$ model. The only exception is the $^{56}$Ni-mass (top right panel in Figure~\ref{scalzo1}) that is directly related to the systematic shift of the distance due to the different reddening law. In this case the relation between $s_{BV}$ and $M_{Ni}$ (again, taking the errorbars into account) is found as 
\begin{equation}
M_{Ni} ~=~ (0.506 \pm 0.023) ~+~ (0.554 \pm 0.125) \cdot (s_{BV}-1 ),
\end{equation}
which predicts systematically lower nickel masses for the same $s_{BV}$ compared to either Equation~\ref{ktr_ni} or the relation given by \citet{scalzo18}. 

Similarly, the relation between $M_{ej}$ and $M_{Ni}$ changes to 
\begin{equation}
M_{ej} ~=~ (0.472 \pm 0.868) \cdot M_{Ni} ~+~ (0.866 \pm 0.445).
\end{equation}
As above, the high uncertainty of the slope parameter indicates that the correlation between $M_{Ni}$ and $M_{ej}$, if exists, is not statistically significant from the present sample, because of the too large uncertainties of the inferred mass parameters. Nevertheless, the trend that higher ejecta masses seem to be coupled with higher nickel masses remains visible in the bottom right panel of Figure~\ref{scalzo1}. A plot published by \citet{scalzo14a} (see their Fig.~10) also shows a similar increase in $M_{ej}$ for increasing $M_{Ni}$, even though \citet{scalzo14a} did not present any analytical fit to their data. 

\begin{figure}
\begin{center}
\includegraphics[width=8.5cm] {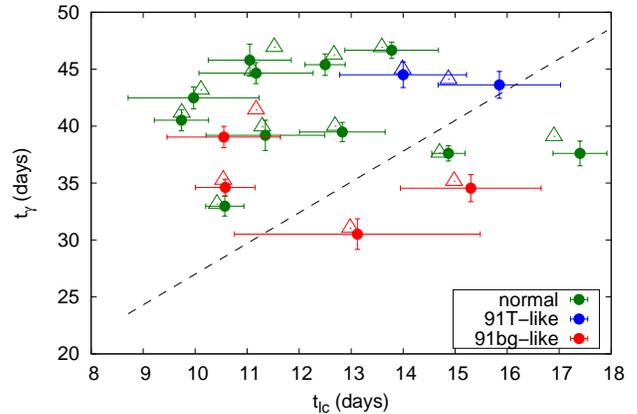}
\end{center}
\caption{The best-fit $t_{lc}$ and  $t_\gamma$ values plotted together with the linear relation suggested by \citet{scalzo18}. Colors  and symbol shapes} have the same meaning as in Fig.\ref{scalzo1}.
\label{tlctgamma}
\end{figure}

\citet{scalzo18} argued that within the framework of the Arnett model the ratio of $t_{lc}$ and $t_\gamma$ ($\tau_m / t_0$ in their nomenclature) should be nearly constant, at least for SNe Ia having $M_{ej} < M_{Ch}$, i.e. $t_\gamma \sim t_{lc}$. Figure~\ref{tlctgamma} shows the dependence between our best-fit $t_\gamma$ and $t_{lc}$ values (Table~\ref{tab:lc-bol}) together with the simple linear relation suggested by \citet{scalzo18} (plotted as a dashed line). It is seen that the parameters inferred directly from the bolometric LC fitting do not follow the linear trend proposed by \citet{scalzo18}. Instead, $t_\gamma$ seems to be nearly independent of $t_{lc}$. In fact, this finding agrees with the  assumption by \citet{scalzo18} that the ejecta mass can be reliably estimated from $t_\gamma$  only.  It suggests that $t_\gamma$ is a better parameter for constraining the ejecta parameters. However, since $t_\gamma$ is more difficult to measure than $t_{lc}$, as it requires precise photometry extending up to several months after maximum light, the combination of the two timescales, as shown in this paper, might be useful to reduce the systematic errors that might occur during the inference of the ejecta parameters solely from a single timescale.

\subsubsection{Comparison with \citet{kk18}}

\begin{figure*}
\begin{center}
\includegraphics[width=8.5cm] {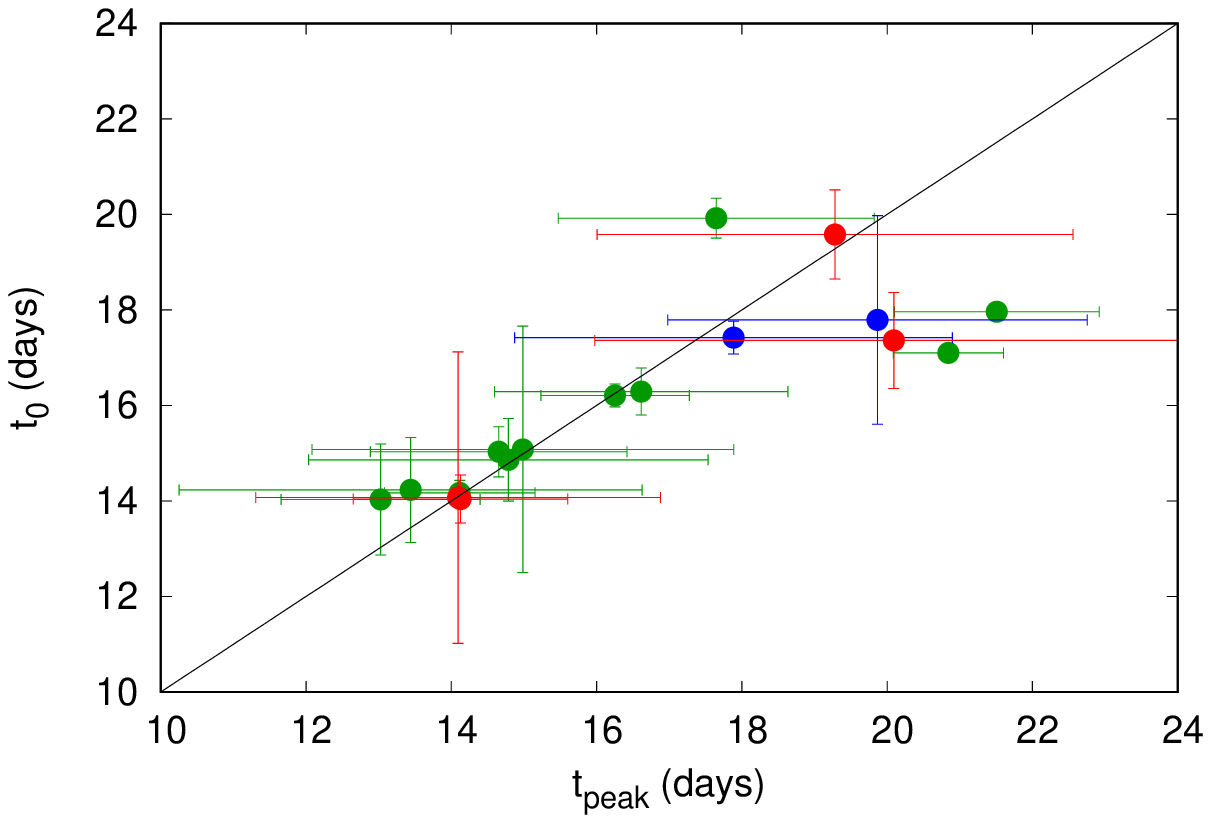}
\includegraphics[width=8.5cm] {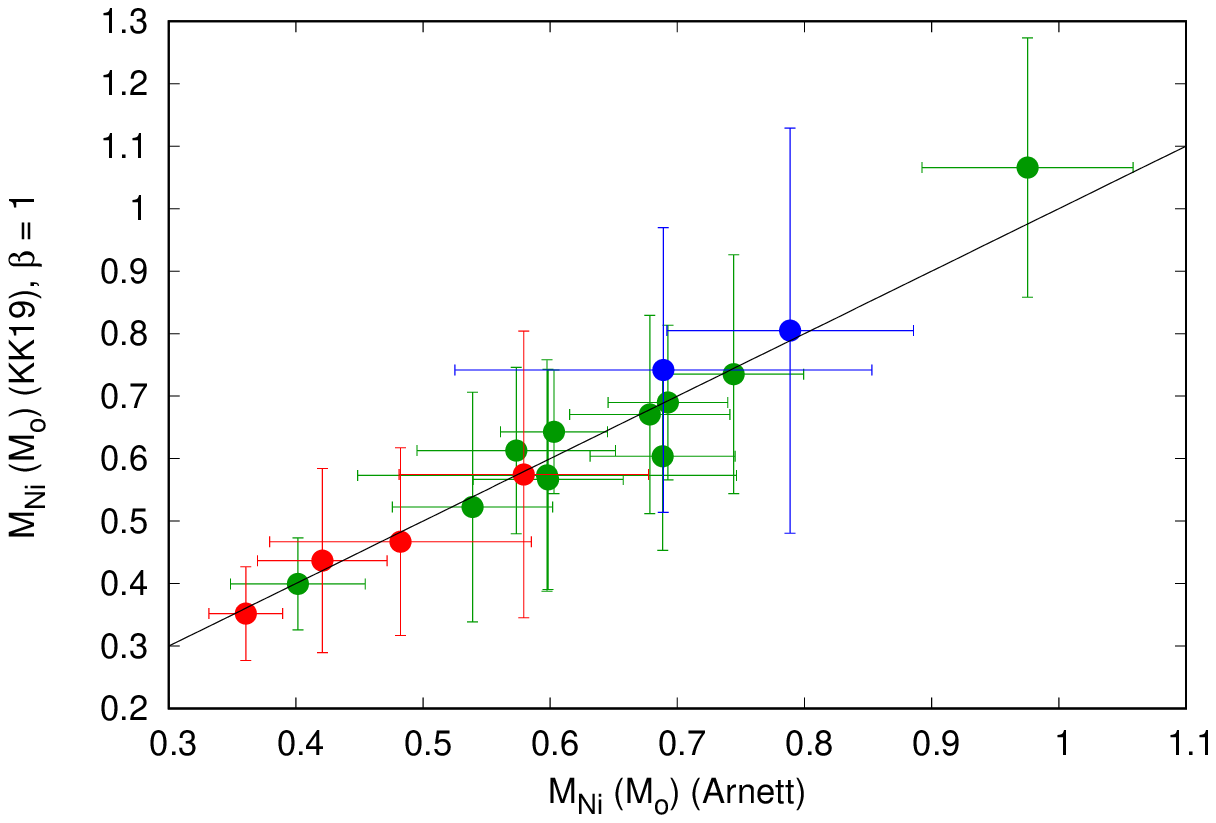}
\caption{Left panel: The LC rise time from the Arnett model ($t_0$ in Table~\ref{tab:lc-bol}) versus $t_{peak}$ inferred from Equation~\ref{eq:tpeak}. Right panel: $M_{Ni}$ inferred from Eq. \ref{eq:lpeak}. vs. $M_{Ni}$ from the Arnett model. The black line represents the 1:1 relation in both panels. Colors have the same meaning as in Figure~\ref{scalzo1}.}
\end{center}
\label{fig:kk19}
\end{figure*}

\citet{kk18} introduced a new analytic relation between peak time and luminosity. They also inferred a new equation connecting the peak time ($t_{peak}$) of the bolometric LC to the diffusion timescale ($t_d$) of the Arnett model defined similarly to $t_{lc}$ (see Equation~\ref{eq:t}). For a centrally located $^{56}$Ni distribution, 
\begin{equation}
 {t_{peak} \over t_d}~ = ~0.11 \cdot \ln (1+ {9t_s \over t_d}) + 0.36,
\label{eq:tpeak}
\end{equation}
where $t_d = (\kappa \cdot M_{ej} / (v \cdot c))^{1/2}$ and $t_s = t_{Ni} = 8.8$ day is the nickel decay time scale.  

We can test whether Equation~\ref{eq:tpeak} were applicable in the case of our SNe by comparing the best-fit $t_{0}$ parameter (i.e. the time between the moment of first light and the epoch of B-band maximum) from the Arnett model (Table~\ref{tab:lc-bol}) to the $t_{peak}$ values inferred from Equation~\ref{eq:tpeak}.
The left panel of Figure~\ref{fig:kk19} shows $t_{0}$ as a function of the corresponding $t_{peak}$ values. The solid line indicates the 1:1 relation, which suggests that the $t_{peak}$ values given by Equation~\ref{eq:tpeak} are consistent with the best-fit $t_{0}$ parameters.

\citet{kk18} also derived the peak luminosity ($L_{peak}$) as
\begin{equation}
 L_{peak} ~ =~ {2 ~\epsilon_{Ni} \cdot M_{Ni} t_s^2 \over \beta_K^2 t_{peak}^2} [ 1 - (1 + \beta_K t_{peak}/ t_s)e^{-\beta_K t_{peak} /t_s} ],
\label{eq:lpeak}
\end{equation}
where $\epsilon_{Ni} = 3.9 \cdot 10^{10} erg ~g^{-1} ~s^{-1}$ is the heating rate of Ni-decay, and $\beta_K$ is a dimensionless parameter, whose value depends on the heating mechanism powering the LC.  introduced by \citet{kk18} (not related to the $\beta \sim 13.8$ density distribution parameter in the Arnett model). \citet{kk18} showed that $\beta_K \sim 1$ for a centrally located heating source (i.e. $^{56}$Ni), while mixing the radioactive Ni toward the outer parts of the ejecta tends to increase the value of $\beta_K$. 

In the right panel of Figure~\ref{fig:kk19} we plot the nickel masses calculated via Equation~\ref{eq:lpeak} using the {\it observed} $L_{peak}$ values from the assembled bolometric light curves and choosing $\beta_K = 1$, versus the best-fit $M_{Ni}$ parameters from Table~\ref{tab:lc-bol}  assuming $R_V=3.1$. It is seen that these two parameters are nicely consistent. Overall, Figure~\ref{fig:kk19} illustrates that the LC timescales and Ni-masses inferred by the bolometric LC fits in this paper are in very good agreement with those resulting from the application of the new theoretical relations given by \citet{kk18}. This supports the validity of using best-fit parameters from our Arnett models for representing the real ejecta parameters in the studied SNe Ia. 

\section{Summary}\label{sum}

We presented a photometric investigation of 17 Type Ia supernovae observed with the 0.6/0.9 m Schmidt-telescope at
Piszk\'estet\H{o} station of Konkoly Observatory, Hungary. The reduced $BVRI$ LCs were analyzed using the {\tt SNooPy2} public LC-fitter code. The reddening of the host galaxy ($E(B-V)_{host}$), the moment of the B-band maximum light ($T_{max}$), the extinction-free distance modulus ($\mu_0$), and color-stretch parameter and the LC decline rate  ($s_{BV}$, $\Delta m_{15}$) were were obtained via LC-fitting.

After correcting for the extinction in the Milky Way and the host galaxy,  the fluxes of the missing UV- and IR-bands were estimated by extrapolations. The bolometric LCs were constructed applying the trapezodial integration rule, and our methods were tested and validated with NIR and UV data
of three well-observed normal Ia SNe: SN~2011fe, SN~2017erp and SN~2018oh. The integrated optical fluxes supplemented by extrapolations into the unobserved UV and IR-bands are found to be reliable representations of the true bolometric data, thus the systematic errors caused by the missing bands should be negligible. Finally, bolometric luminosities were calculated from the integrated bolometric fluxes and the distances derived via {\tt SNooPy2}. 

We applied the {\tt Minim} code \citep{manos12, li18}  to fit the bolometric LCs with the radiation diffusion model of \citet{arnett82}. The optimized parameters of this model were the moment of first light ($t_0$), the LC time scale ($t_{lc}$), the $\gamma$-ray leakage time scale  ($t_\gamma$), and the initial nickel-mass ($M_{Ni}$).  It was found that the fitting results are not particularly sensitive to the assumed reddening law within the host galaxy, except for the inferred Ni-masses. Adopting $R_V=1.5$ instead of $R_V=3.1$ in the hosts resulted in systematically lower Ni-masses depending on the $E(B-V)_{host}$ parameter: $M_{Ni}(1.5) / M_{Ni}(3.1) = 1 - (1.349 \pm 0.046) \cdot E(B-V)_{host}$.

One of the critical parameters of the Arnett model is the value of the optical opacity ($\kappa$), which is approximated as a constant in both space and time.
Upper and lower limits for the optical opacity were estimated by using the same method as in \citet{li18}. This method combines the  $t_{lc}$ and $t_\gamma$ parameters, and may give a reliable evaluation of the ejecta mass ($M_{ej}$) and the expansion velocity ($v_{exp}$).  As above, comparing the inferred parameters of SN~2011fe and SN~2018oh to those published by \citet{scalzo14a} and \citet{li18}, reasonable agreement has been found.

The evolution of the  reddening-corrected pre-maximum $(B-V)_0$ color was also studied for the sample SNe in order to decide whether they belong to the early red or early blue group defined recently by \citet{strici18}. Even though  our data are  consistent with the expected color evolution of SNe Ia, only two SNe in our sample, SN~2016bln and SN~2017erp, were observed at sufficiently early epochs ($-10$ - $-20$ days before maximum) for this purpose.  The early $(B-V)_0$ color of SN~2016bln looks like similar to those of the early blue group, which is consistent with its 91T/SS spectral type \citep{strici18}. SN~2017erp, a NUV-red object \citep{erp}, however, seems to belong to the early red group, although our data are sparse and have higher uncertainty that prevents drawing a more definite conclusion.

The early-phase $(B-V)_0$ colors were also compared to the synthetic colors from DDE, PDDE 
\citep{dessart} and other explosion models (e.g. N100 from \citet{noe17}) in order to test whether the nickel masses in these models were consistent with those derived from the bolometric LC modeling. We found good agreement between  the Ni-masses of DDE models whose color curves match the observed $(B-V)_0$ colors and the Ni-masses inferred from the bolometric LCs  assuming $R_V=3.1$ reddening law. The agreement is worse for the PDDE models, since those models that have synthetic colors most similar to the observed ones have nearly the same Ni-mass, $M_{Ni} \sim 0.65 \pm 0.1$ M$_\odot$.  Also, using the $R_V=1.5$ reddening law made the agreement between the observed and model Ni-masses worse, since in this case the observed Ni-masses were found to be systematically lower.

Finally, we examined the possible correlations between various physical parameters of the ejecta, such as $s_{BV}$ vs $M_{ej}$, $M_{Ni}$, and $t_\gamma$, as well as $M_{ej}$ vs $M_{Ni}$. Similar correlations were found as published recently by \citet{scalzo18}. 
Our results also turned out to be consistent with the predictions from the new formalism proposed by \citet{kk18},  again, while assuming $R_V=3.1$ in the host galaxies.

\begin{acknowledgements}
We express our thanks to an anonymous referee whose criticism and helpful suggestions led to a significant improvement of this paper. 
This work is part of the project ``Transient Astrophysical Objects" GINOP 2.3.2-15-2016-00033 of the National Research, Development and Innovation Office (NKFIH), Hungary, funded by the European Union. This study has also been partly supported by the Lend\"ulet LP2018-7/2019 grant of the Hungarian Academy of Sciences, and the K-130405 and K-131508 grants of the Hungarian National Research, Development and Innovation Office. KV is supported by the Bolyai J\'anos Research Scholarship of the Hungarian Academy of Sciences.
RKT is also supported by the \'UNKP-19-02 New National Excellence Program of the Ministry for Innovation and Technology.
\end{acknowledgements}

\software{{\tt IRAF} \citep{iraf1, iraf2}, {\tt SExtractor} \citep{sex}, {\tt wcstools} \citep{wcst}, {\tt SNooPy2} \citep{burns11, burns14}, {\tt Minim} \citep{minim}}

\section{Appendix}

In the following we present plots of the best-fit {\tt SNooPy2} templates 
to the multi-color data of our SN sample, the best-fit Arnett models to the 
assembled bolometric LCs ( Appendix) and the comparison between the 
dereddened $(B-V)_0$ color curves and the synthetic colors computed 
from the DDE and PDDE models by \citet{dessart}.

\begin{figure*}
\begin{center}
\includegraphics[width=5cm]{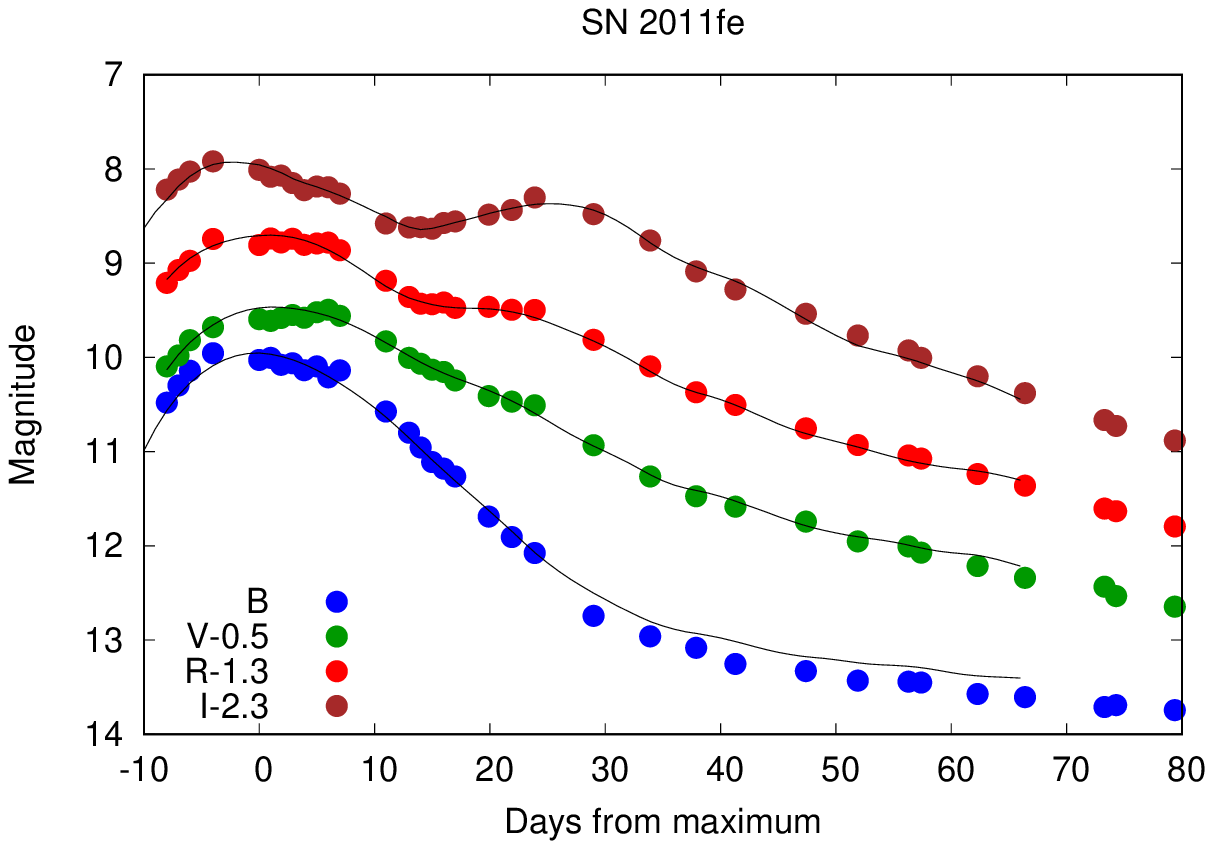}
\includegraphics[width=5cm]{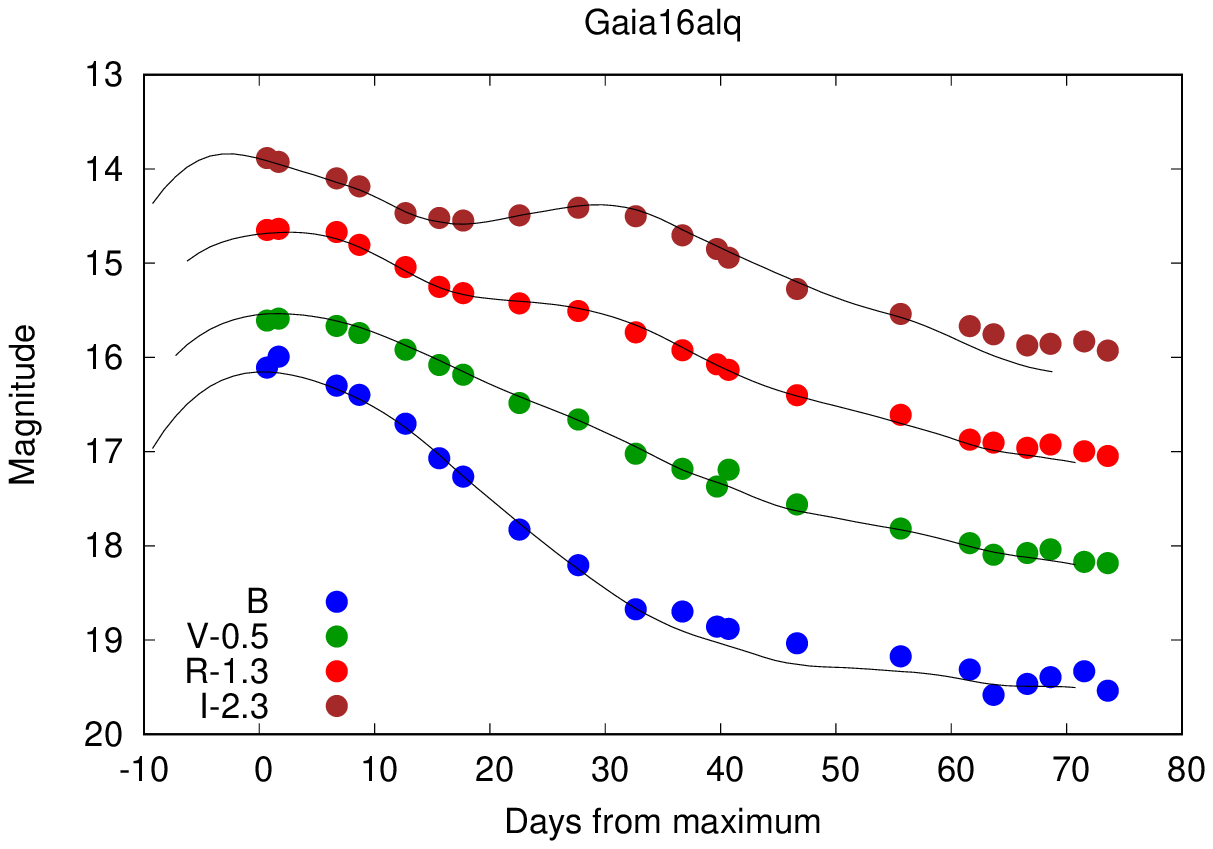}
\includegraphics[width=5cm]{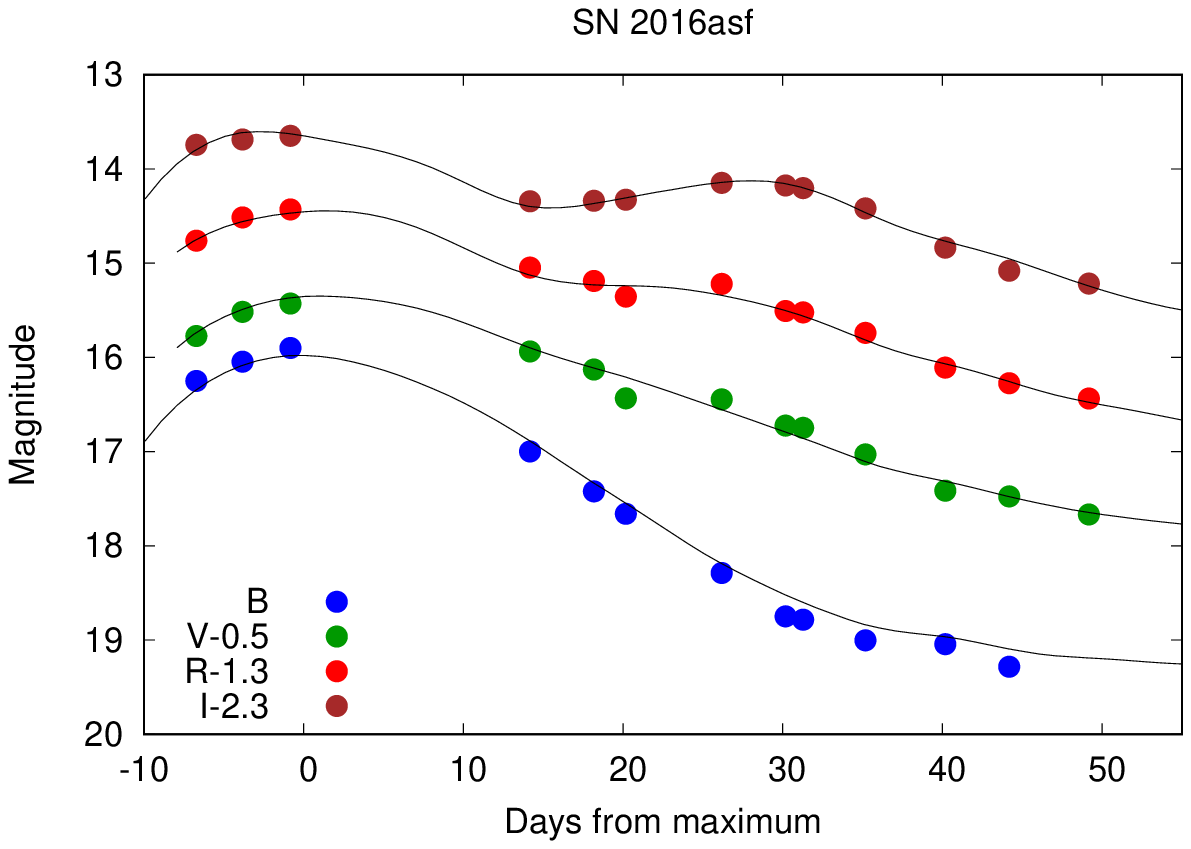}
\includegraphics[width=5cm]{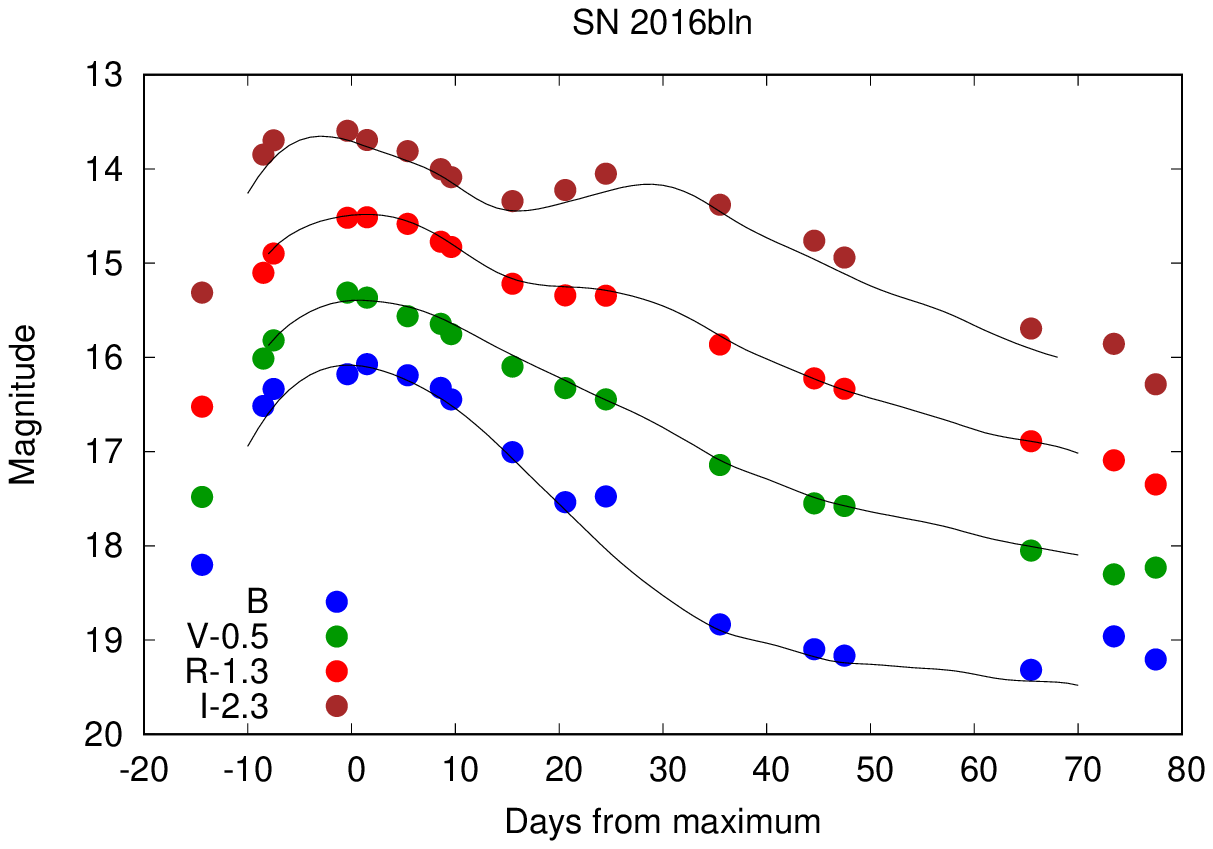}
\includegraphics[width=5cm]{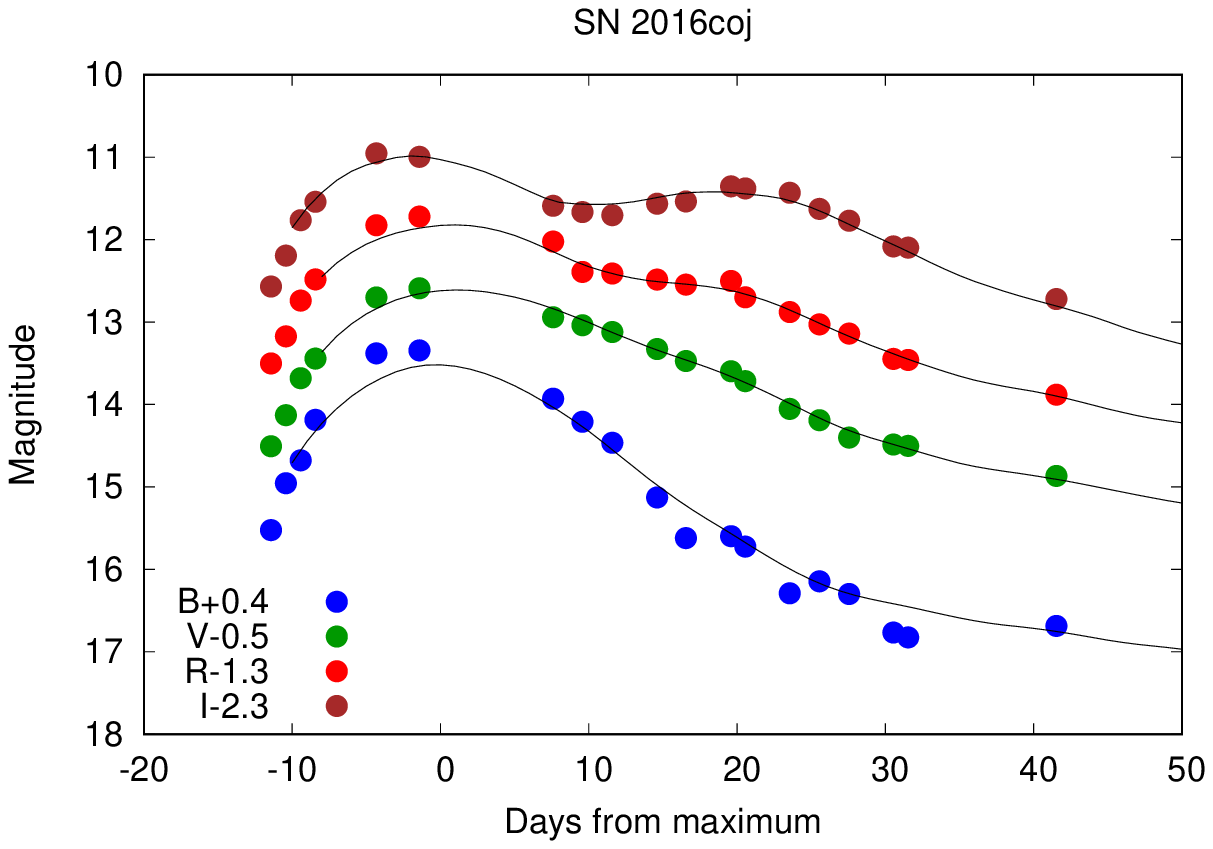}
\includegraphics[width=5cm]{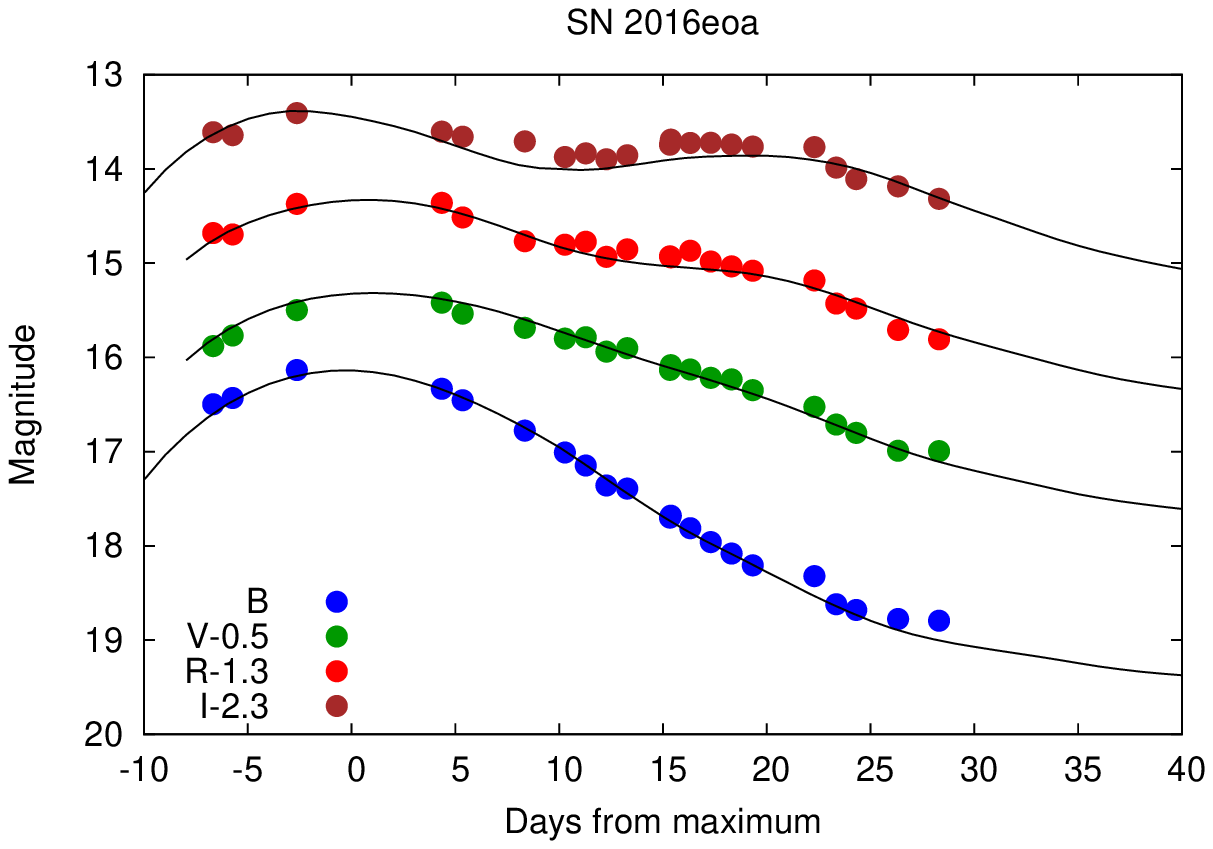}
\includegraphics[width=5cm]{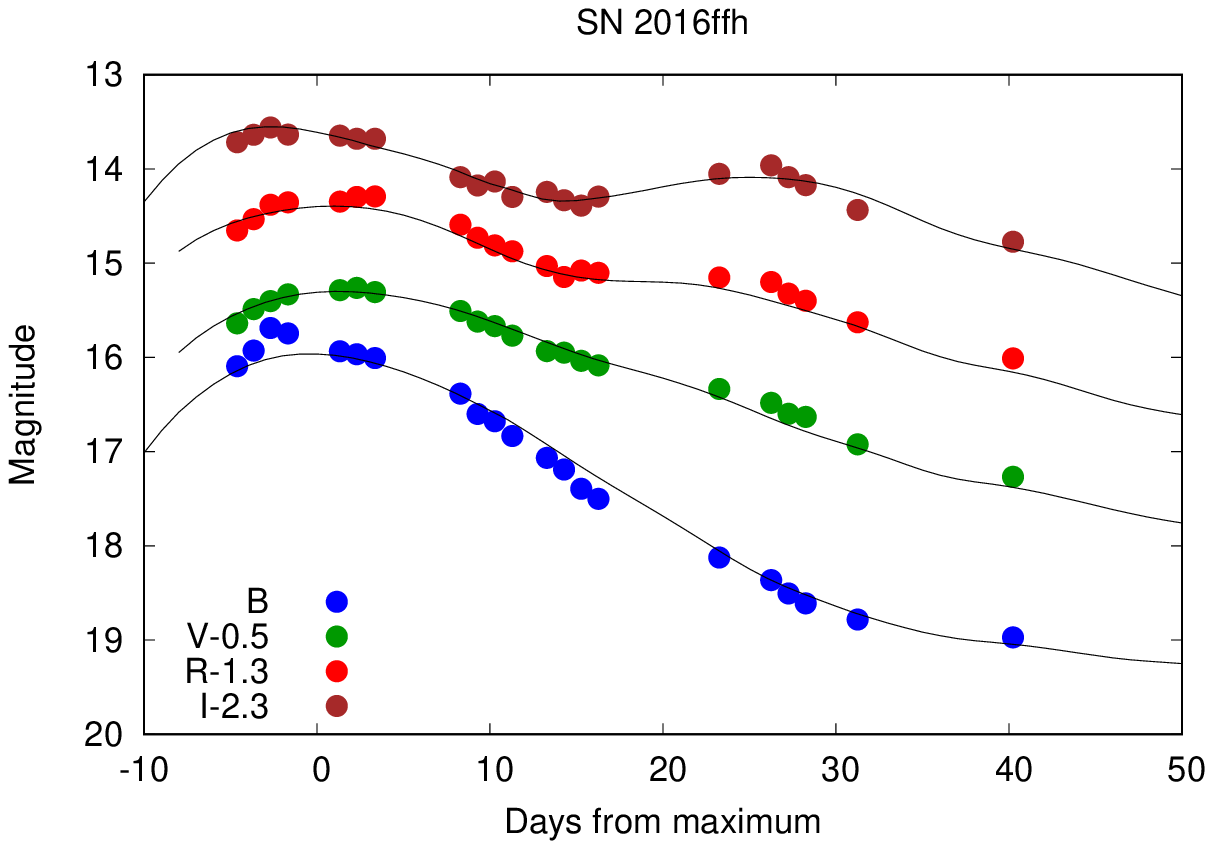}
\includegraphics[width=5cm]{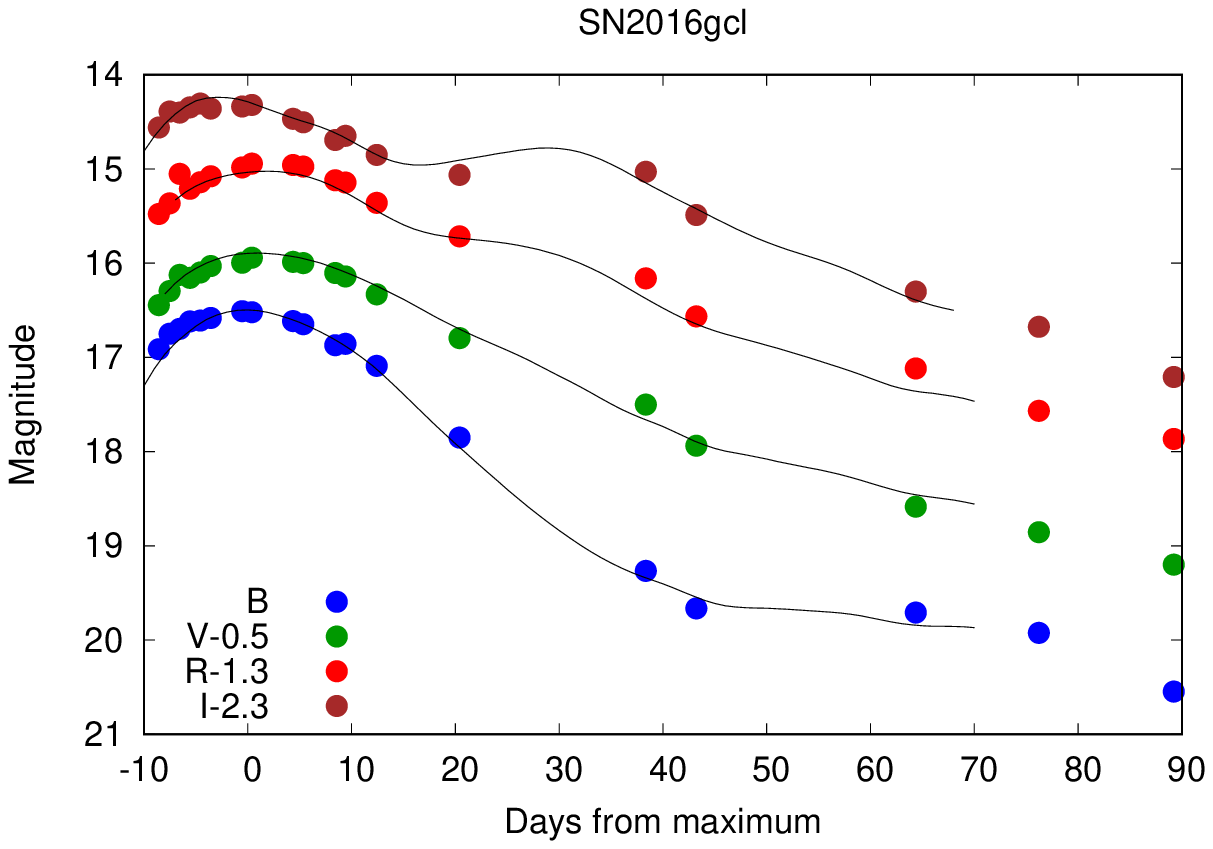}
\includegraphics[width=5cm]{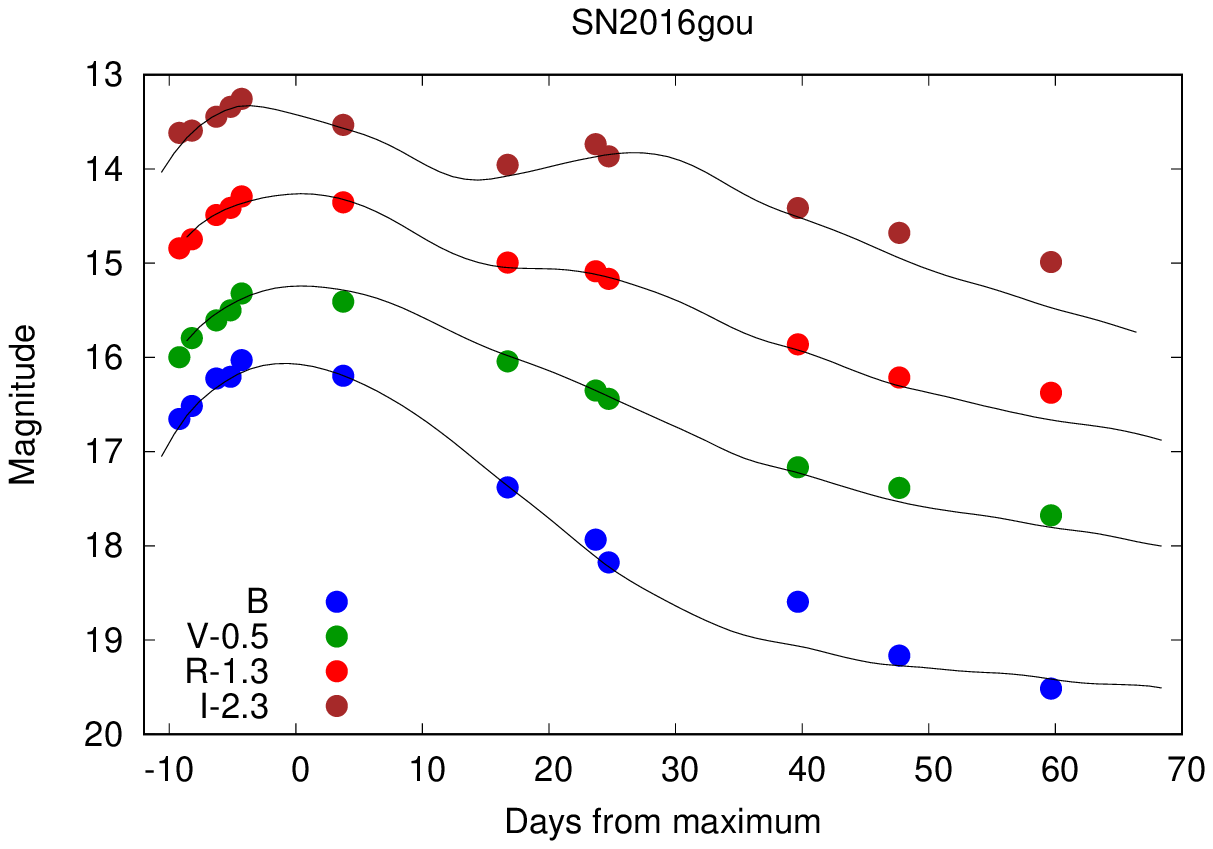}
\includegraphics[width=5cm]{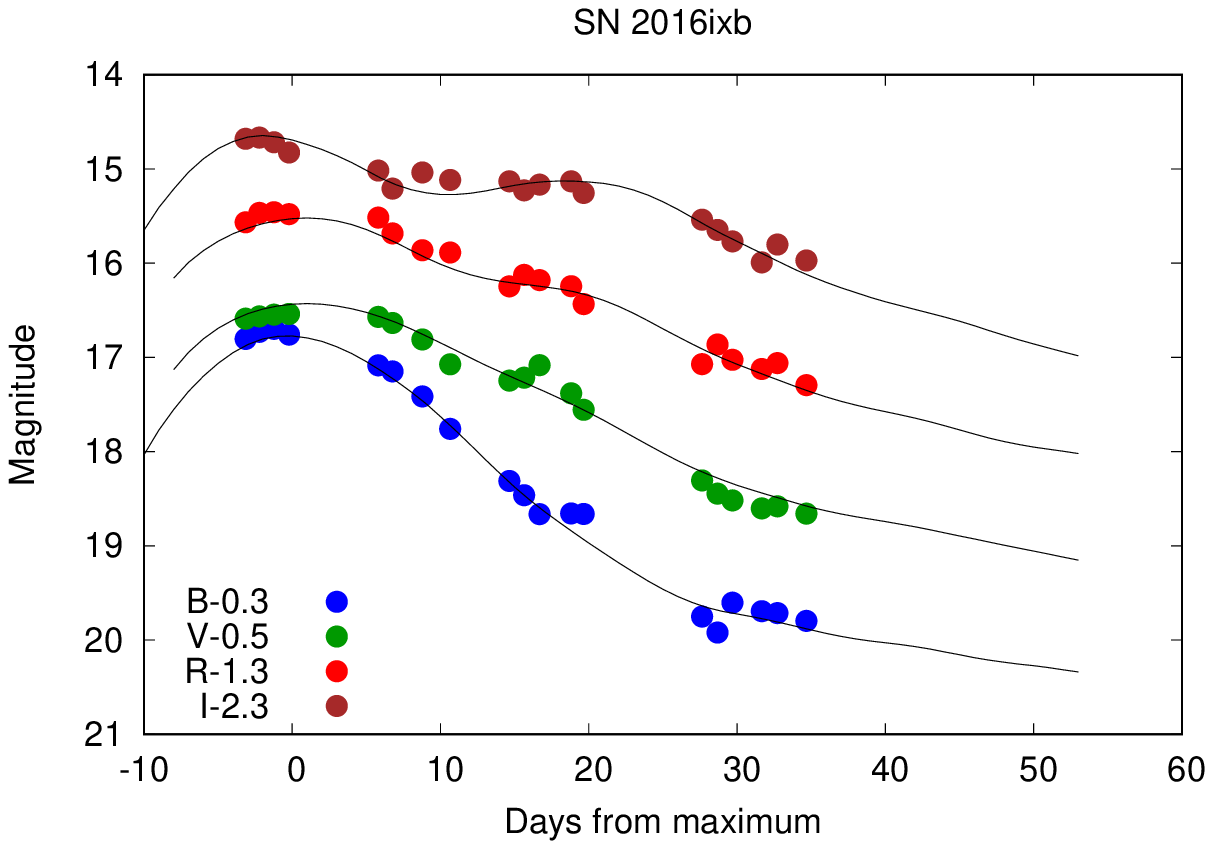}
\includegraphics[width=5cm]{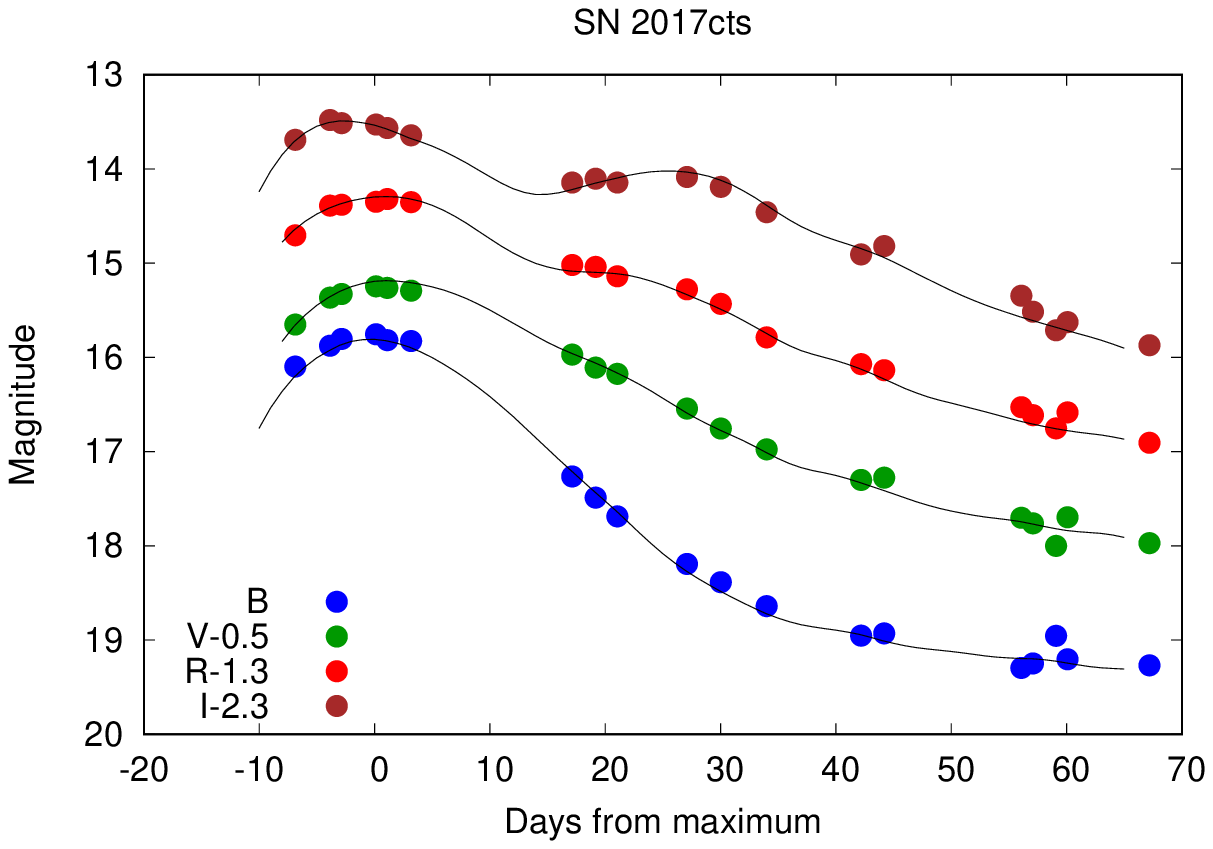}
\includegraphics[width=5cm]{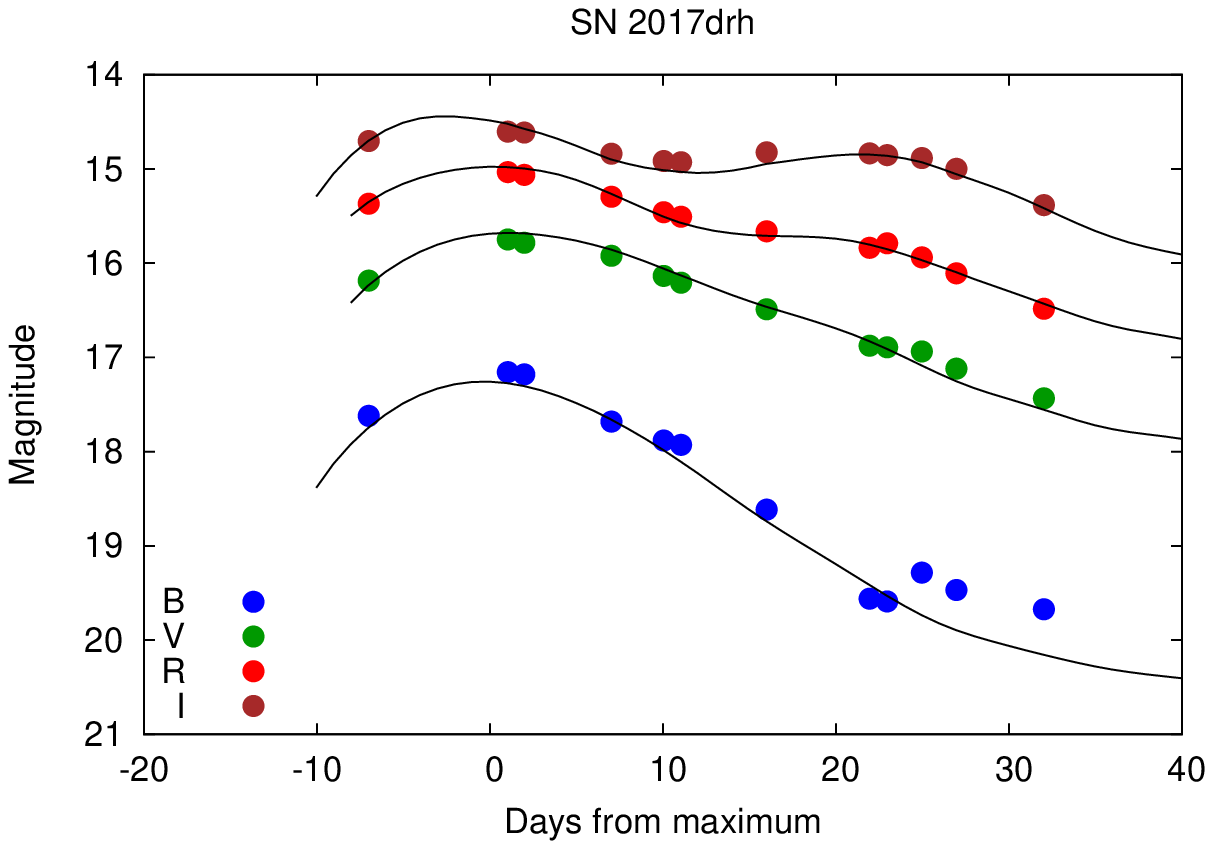}
\includegraphics[width=5cm]{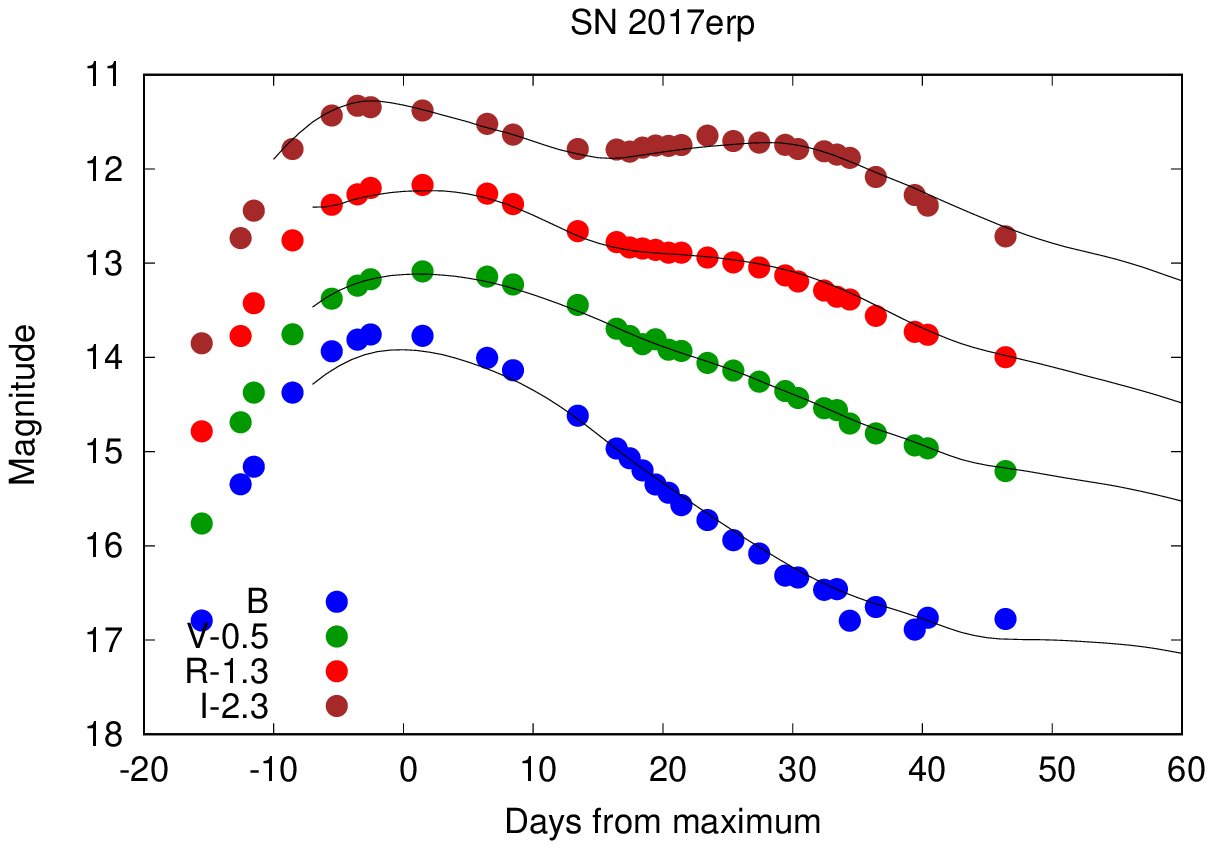}
\includegraphics[width=5cm]{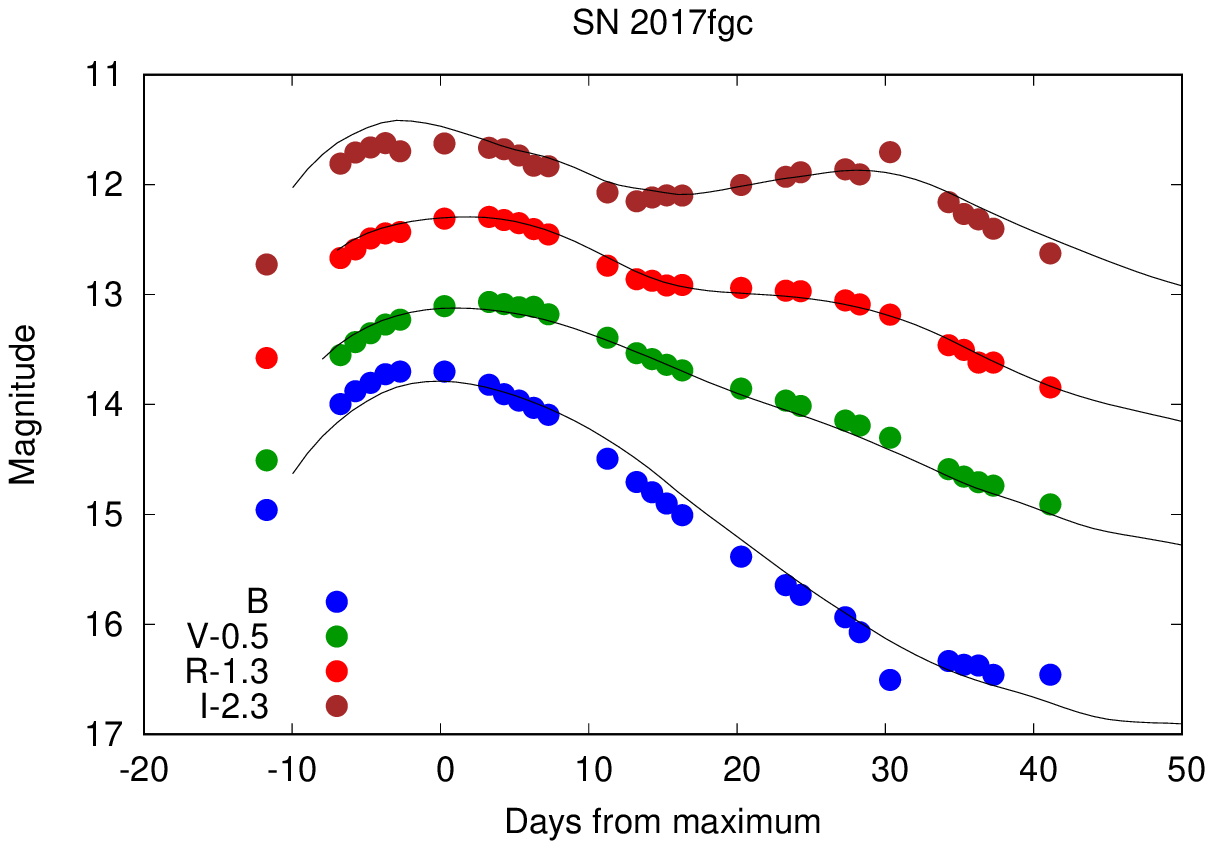}
\includegraphics[width=5cm]{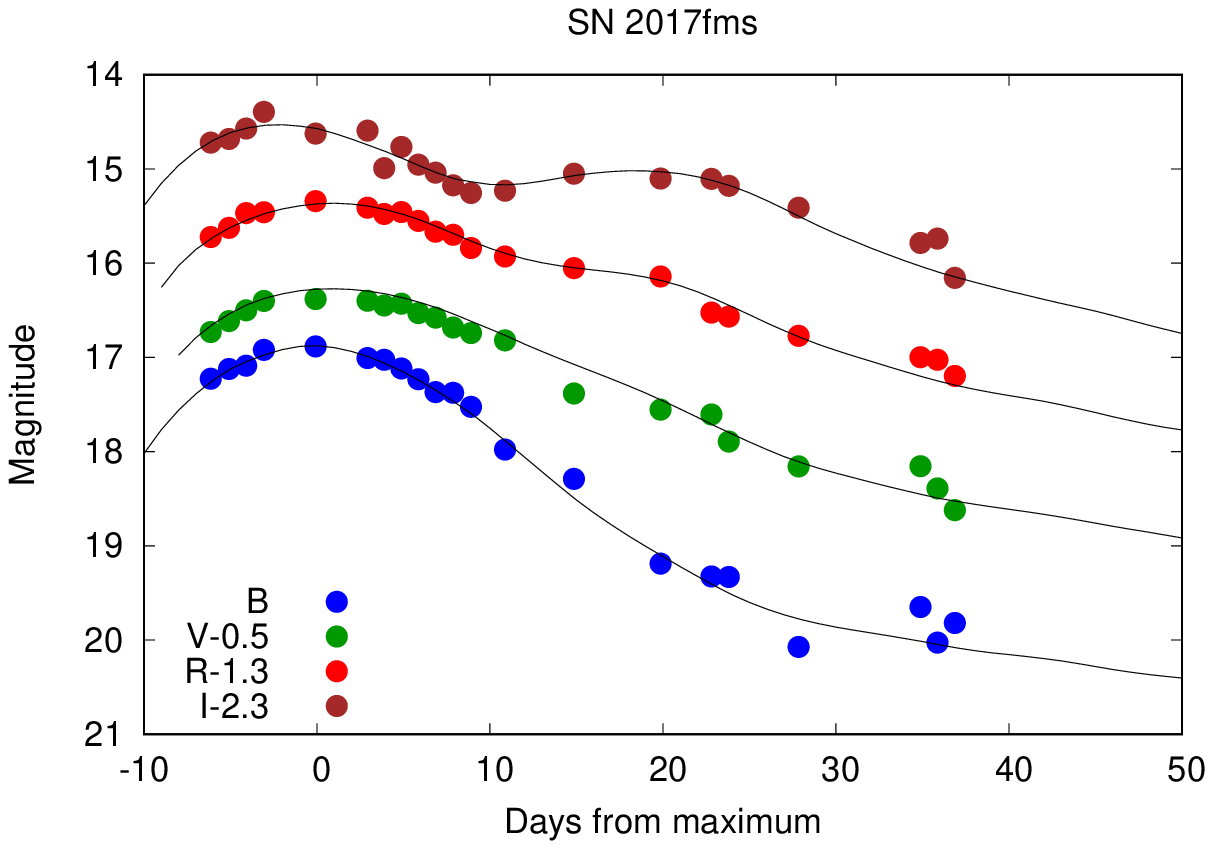}
\includegraphics[width=5cm]{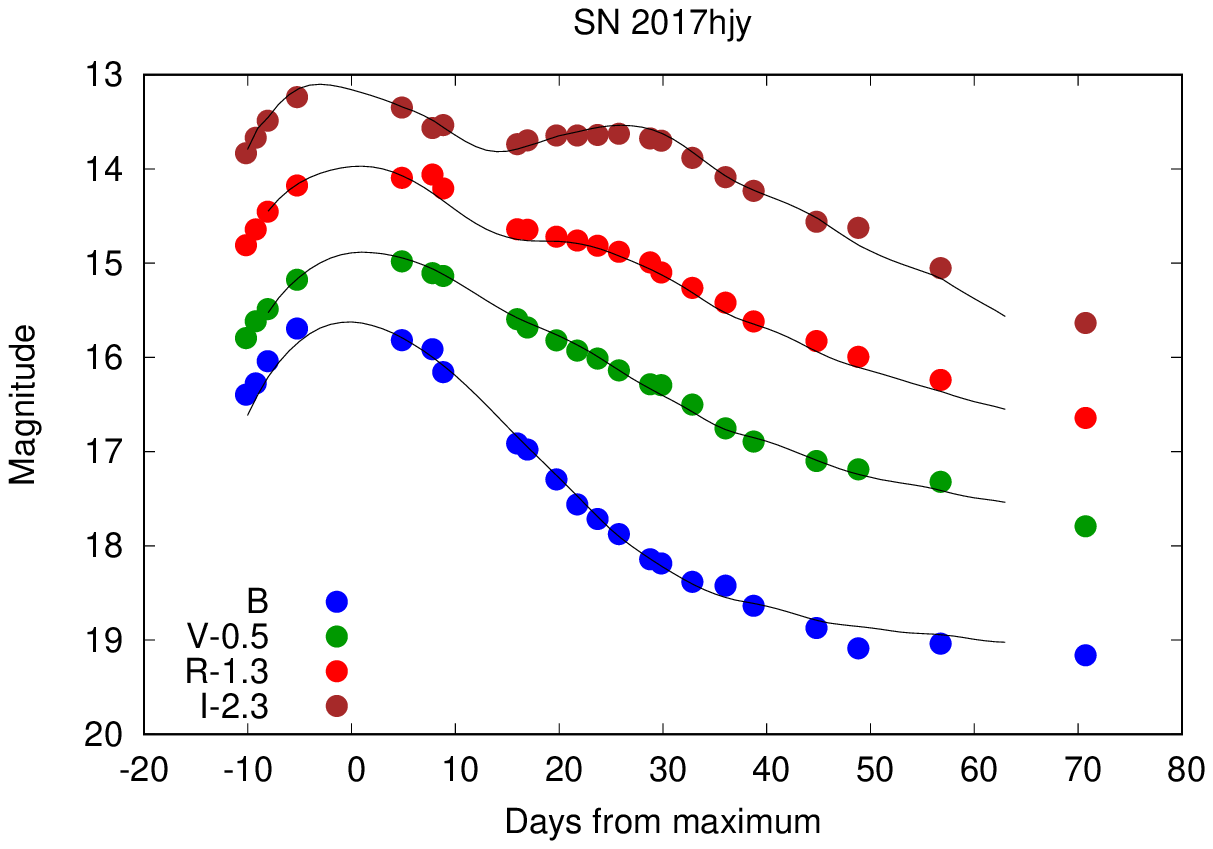}
\includegraphics[width=5cm]{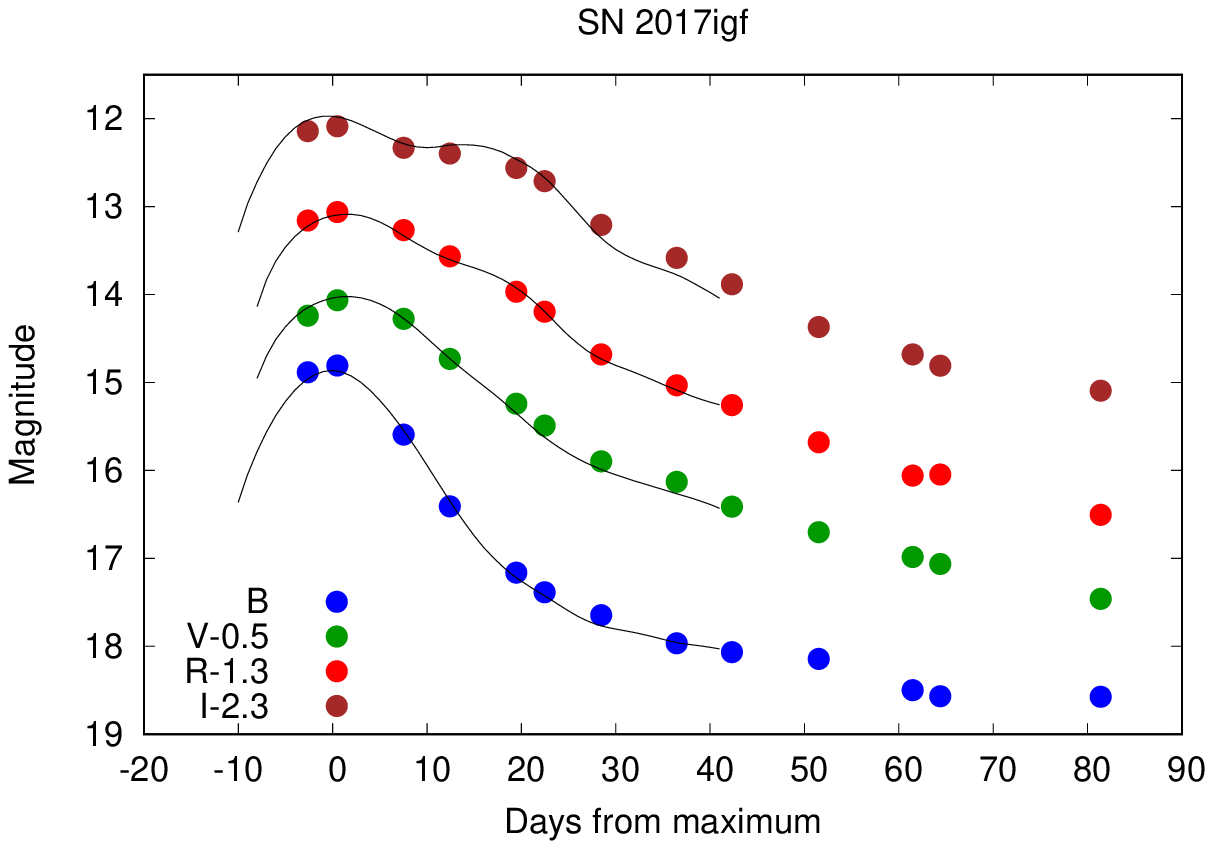}
\includegraphics[width=5cm]{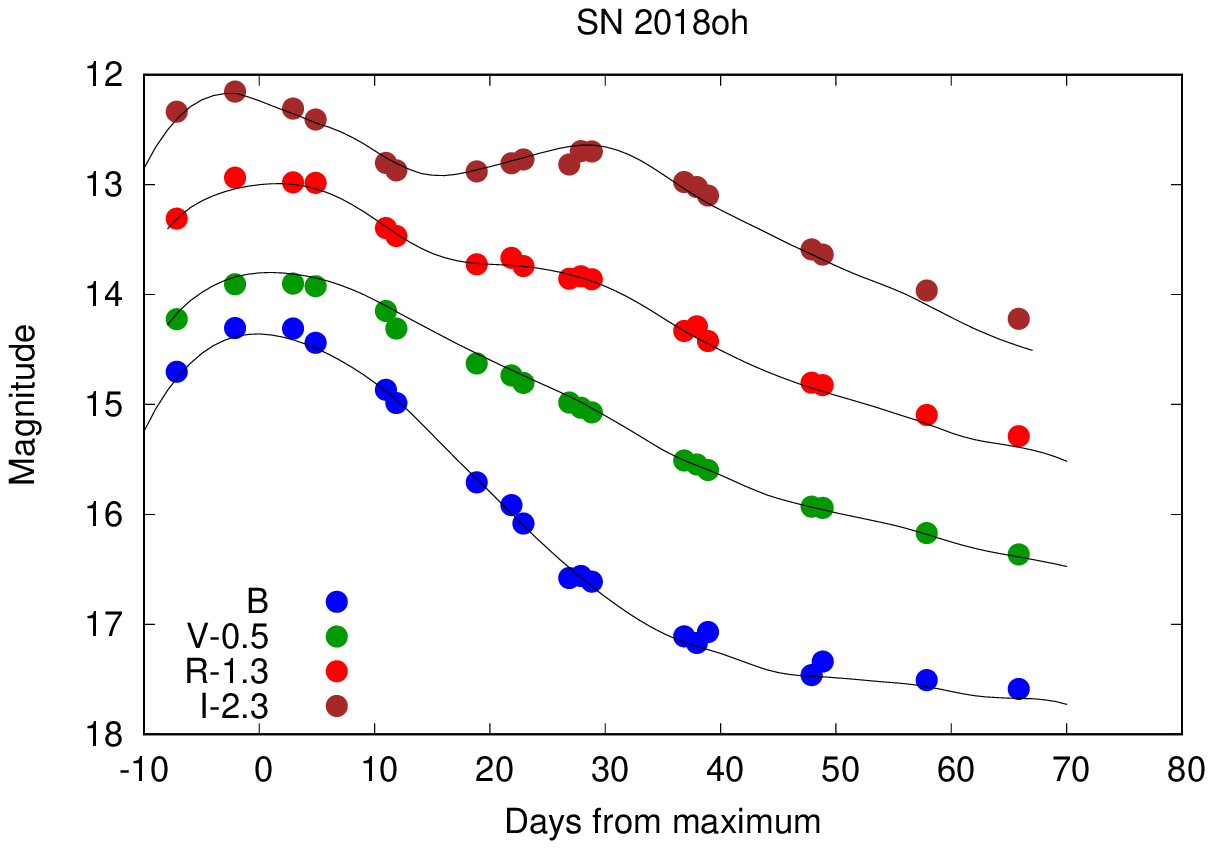}
\caption{BVRI LC-fitting with {\tt SNooPy2}. The curves corresponding to different filters are shifted vertically for better visibility. }
\end{center}
\label{18sne_snpy}
\end{figure*}

\begin{figure*}[h!]
\begin{center}
\includegraphics[width=5cm]{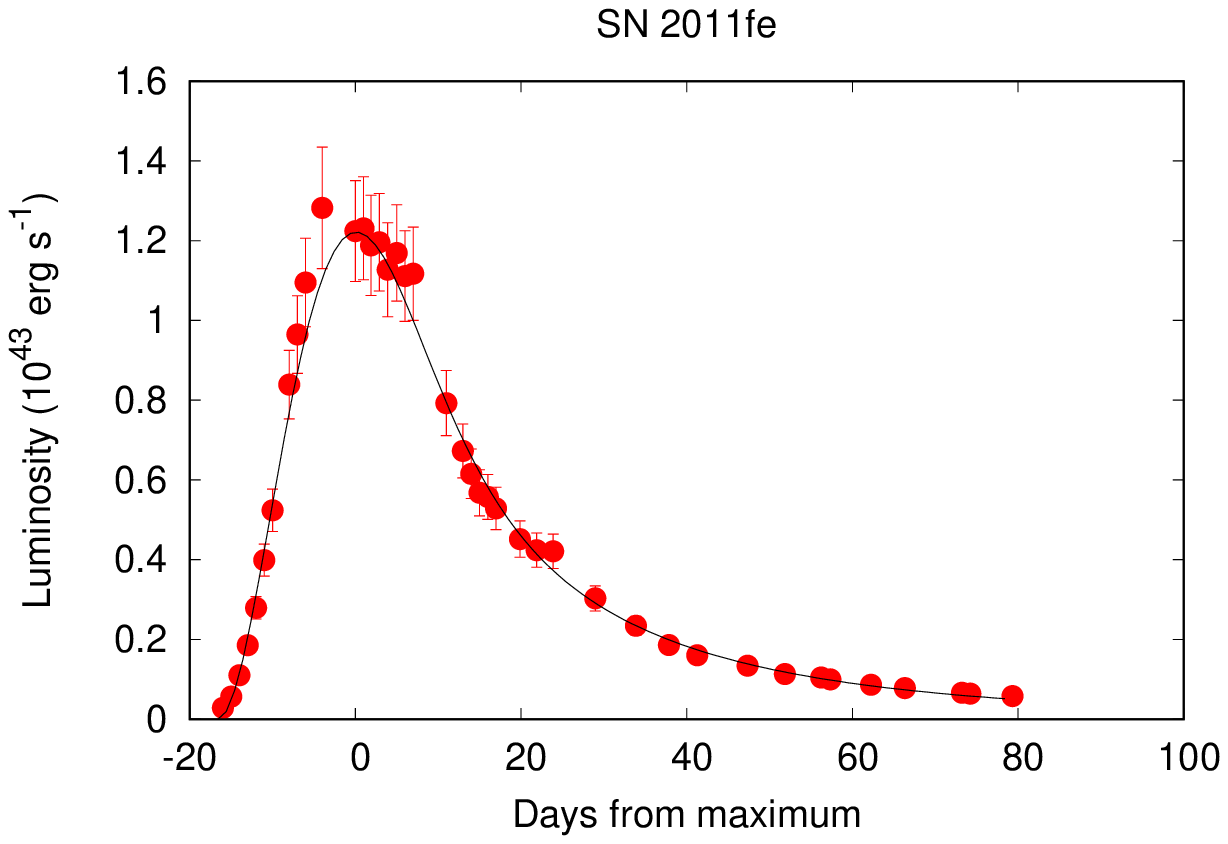}
\includegraphics[width=5cm]{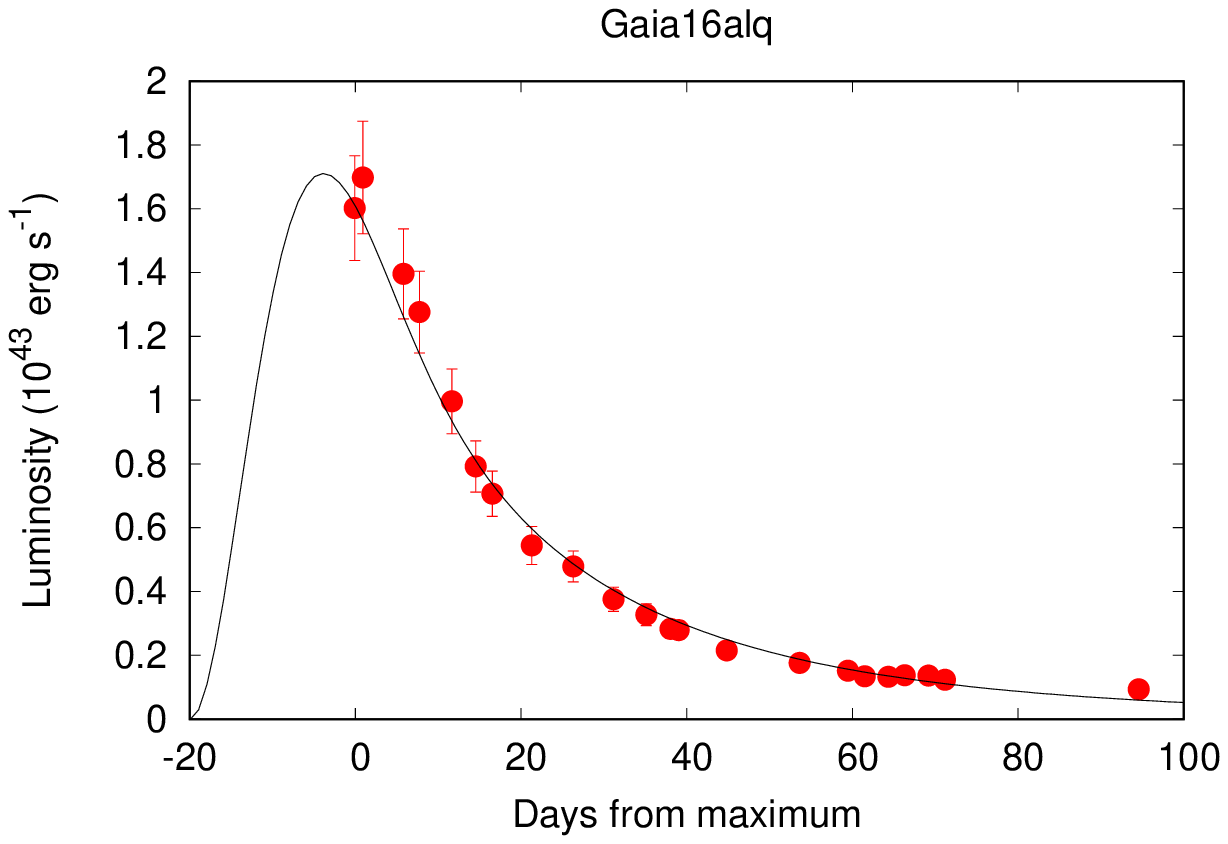}
\includegraphics[width=5cm]{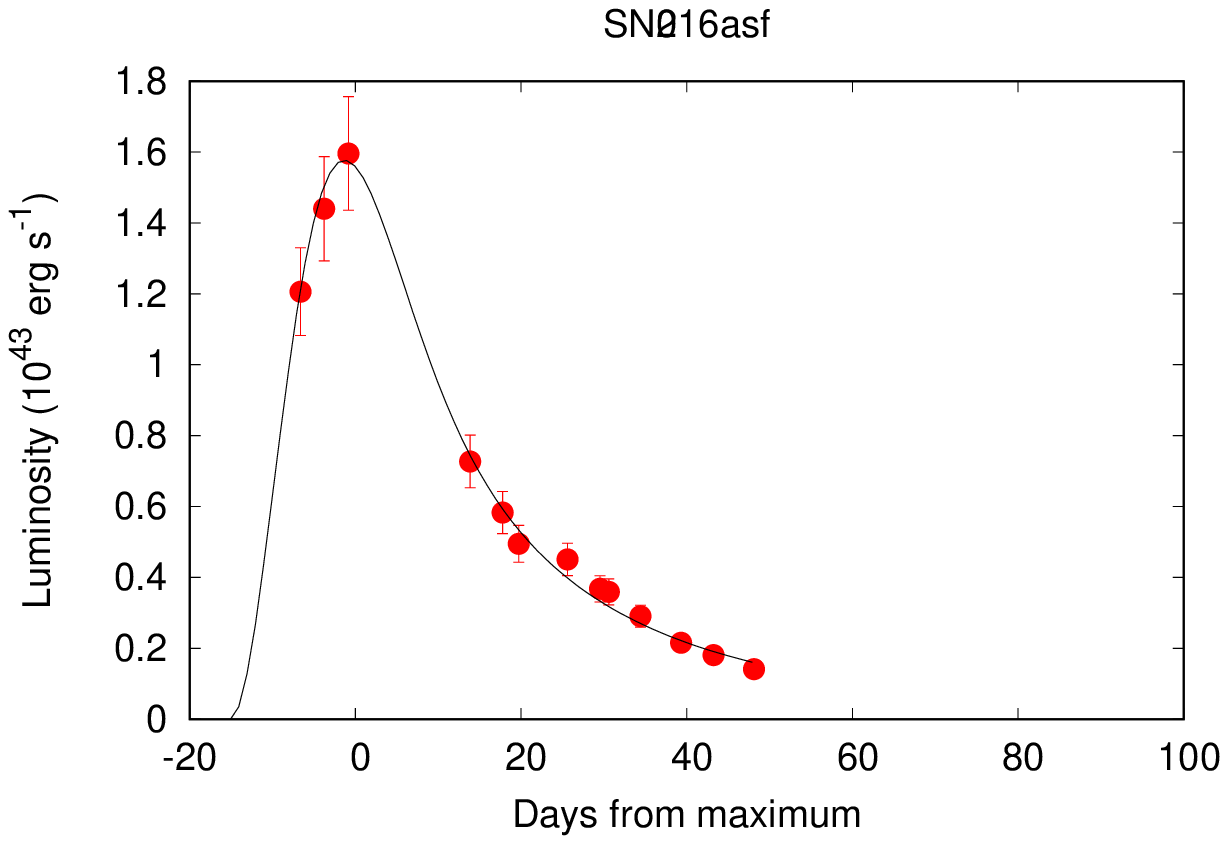}
\includegraphics[width=5cm]{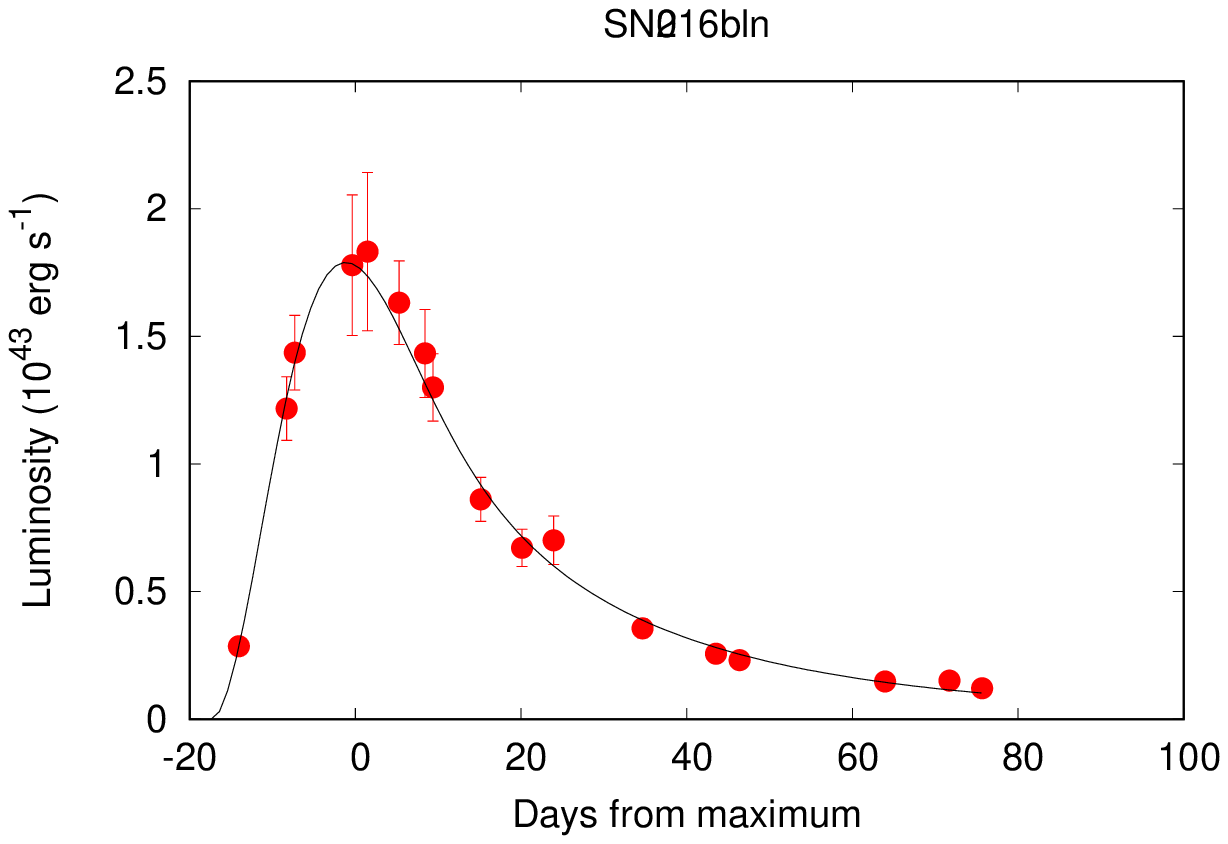}
\includegraphics[width=5cm]{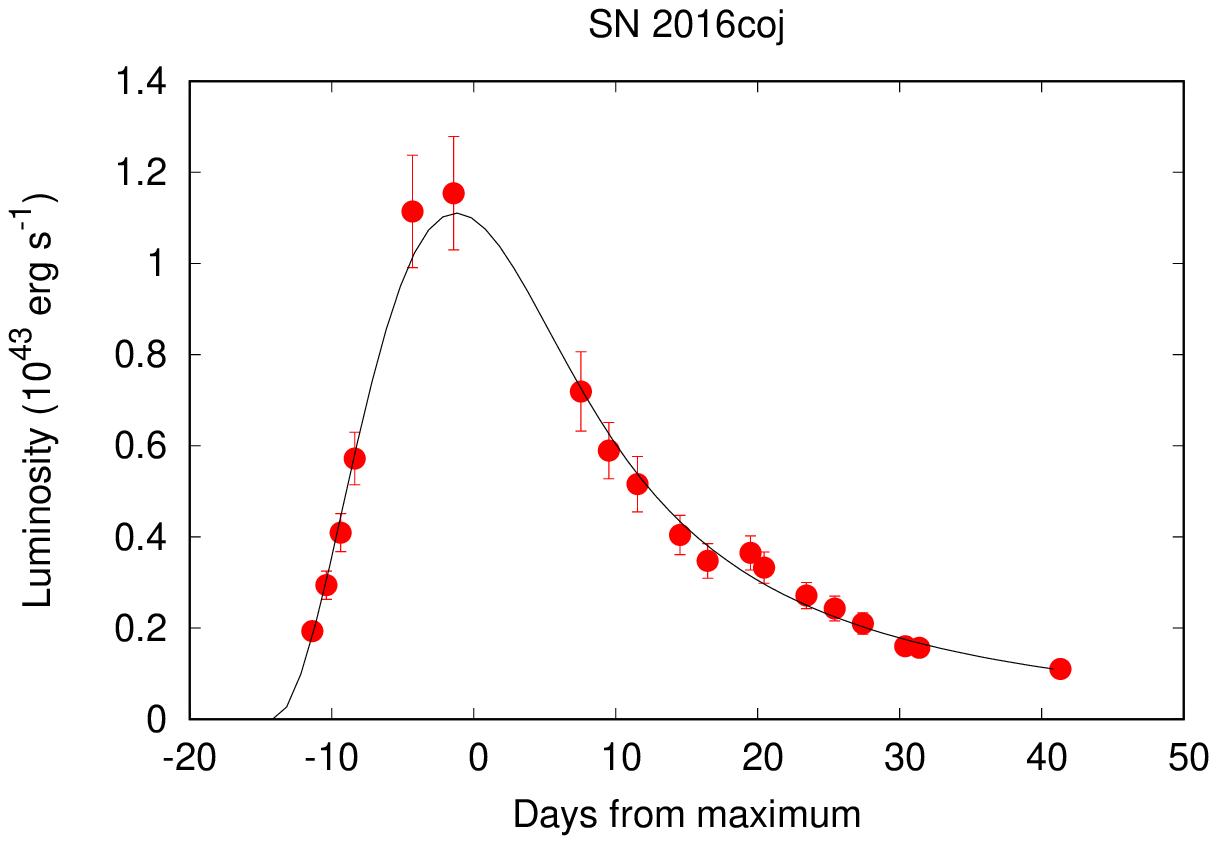}
\includegraphics[width=5cm]{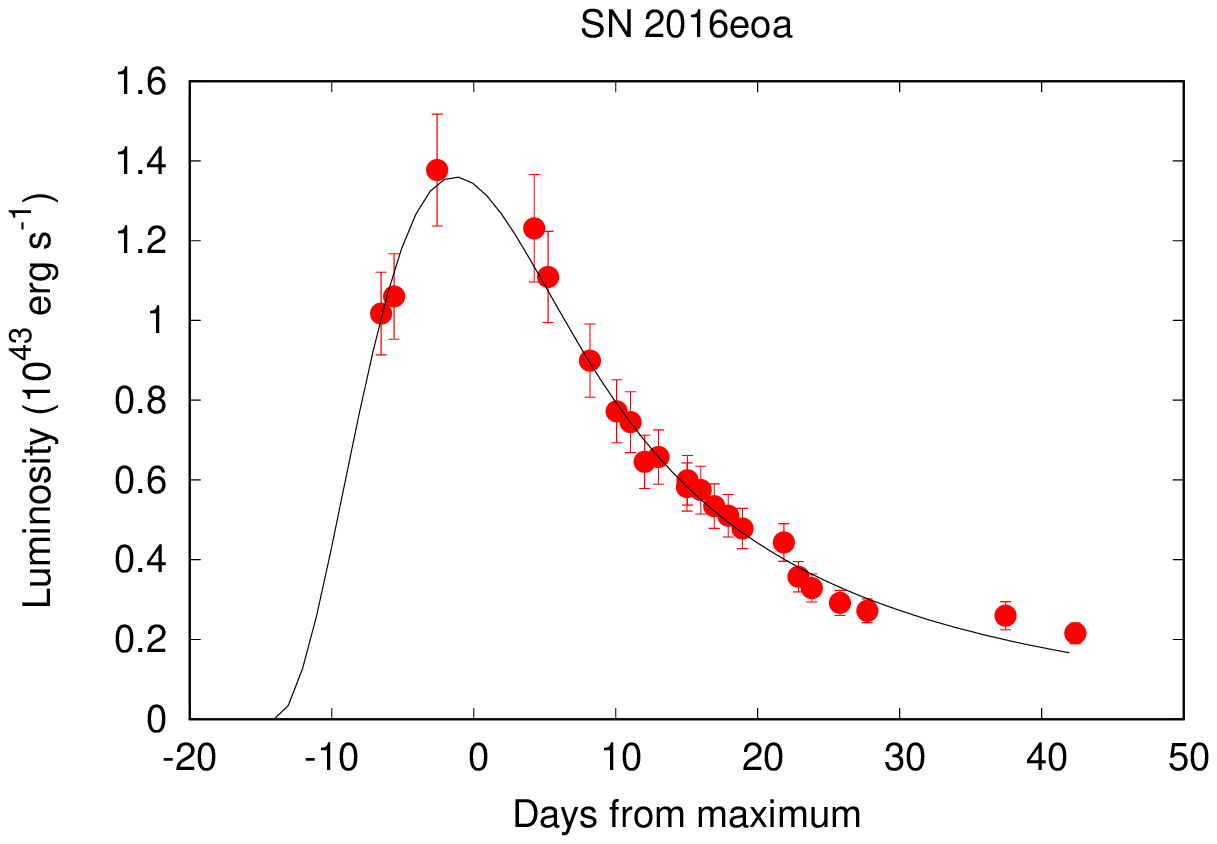}
\includegraphics[width=5cm]{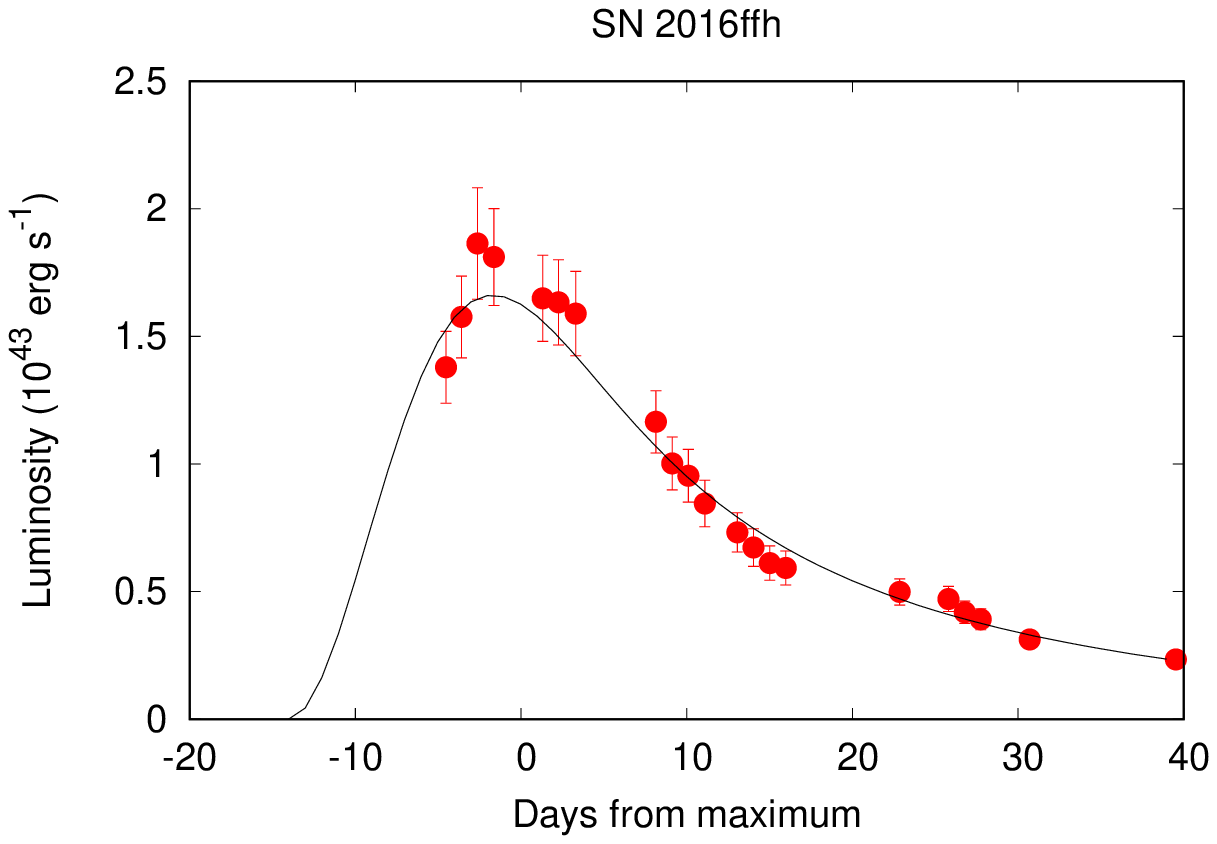}
\includegraphics[width=5cm]{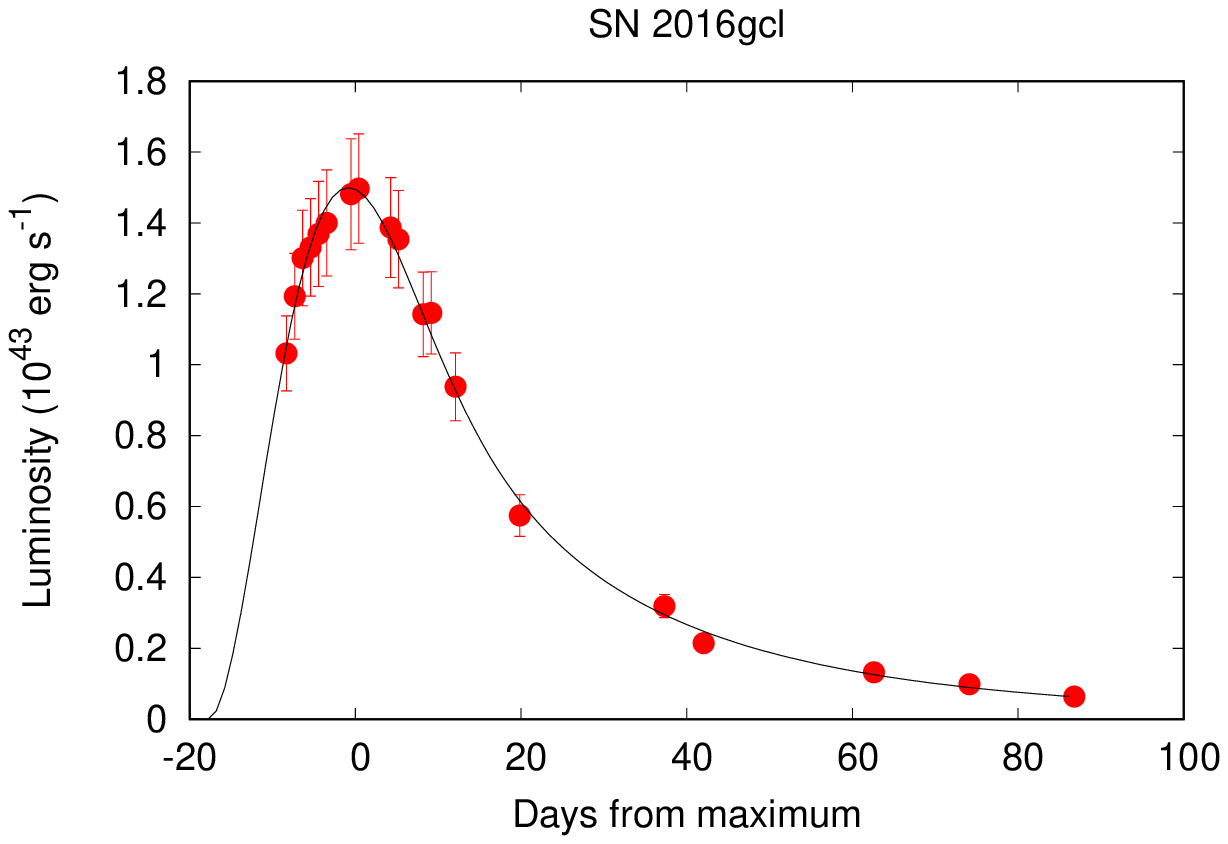}
\includegraphics[width=5cm]{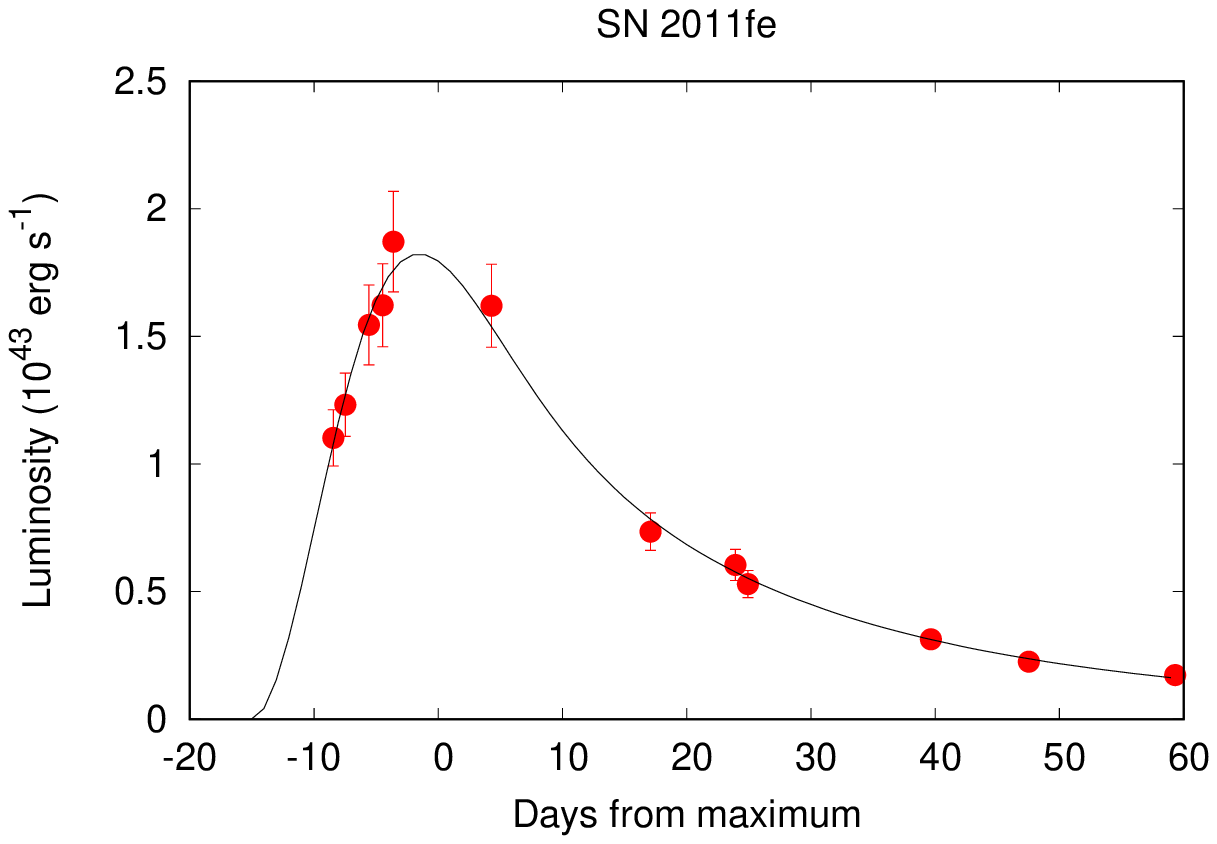}
\includegraphics[width=5cm]{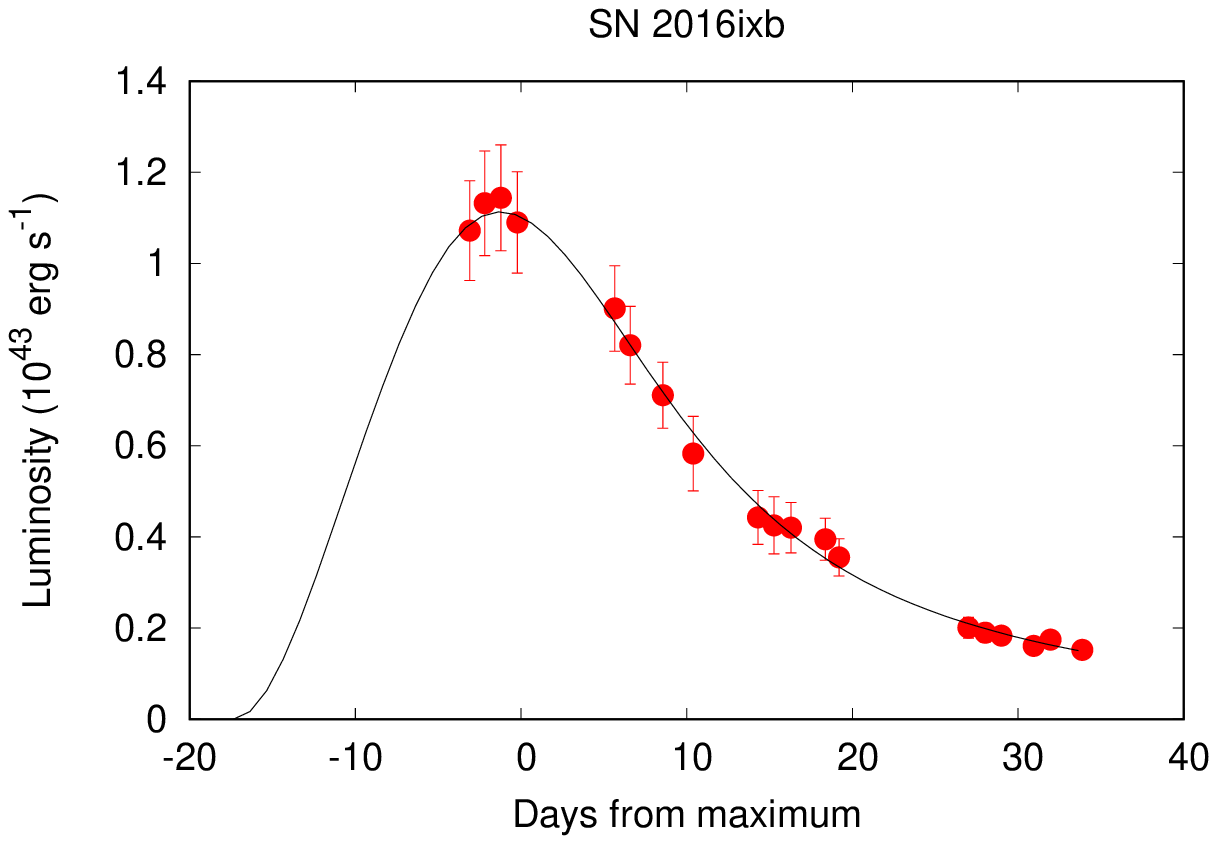}
\includegraphics[width=5cm]{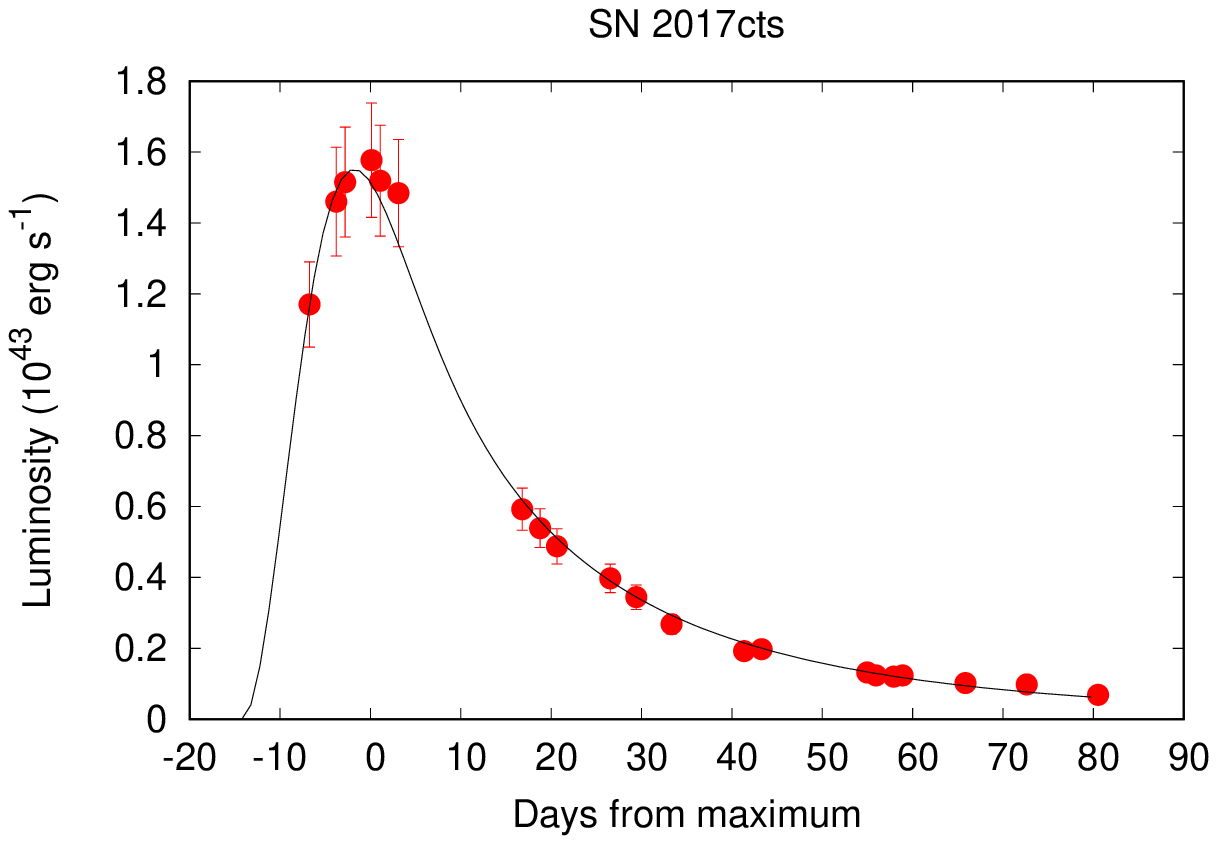}
\includegraphics[width=5cm]{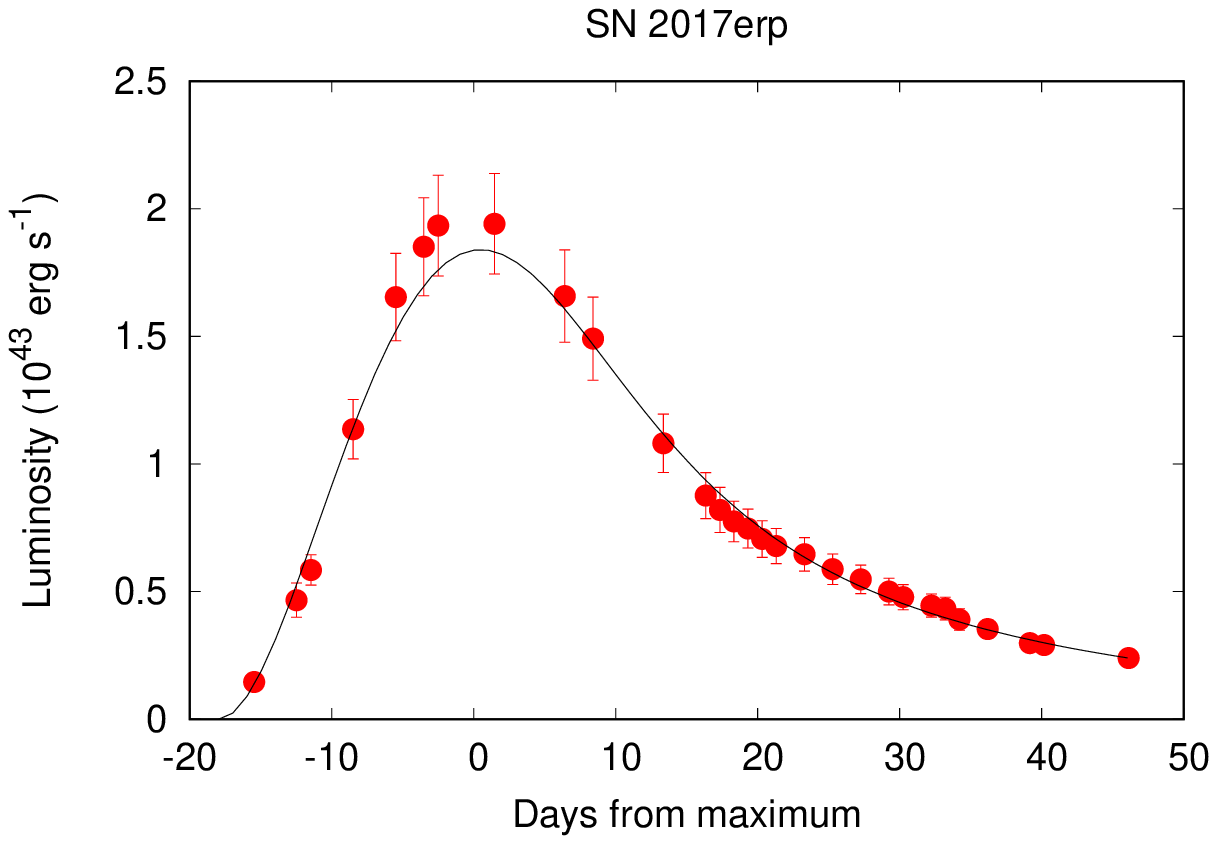}
\includegraphics[width=5cm]{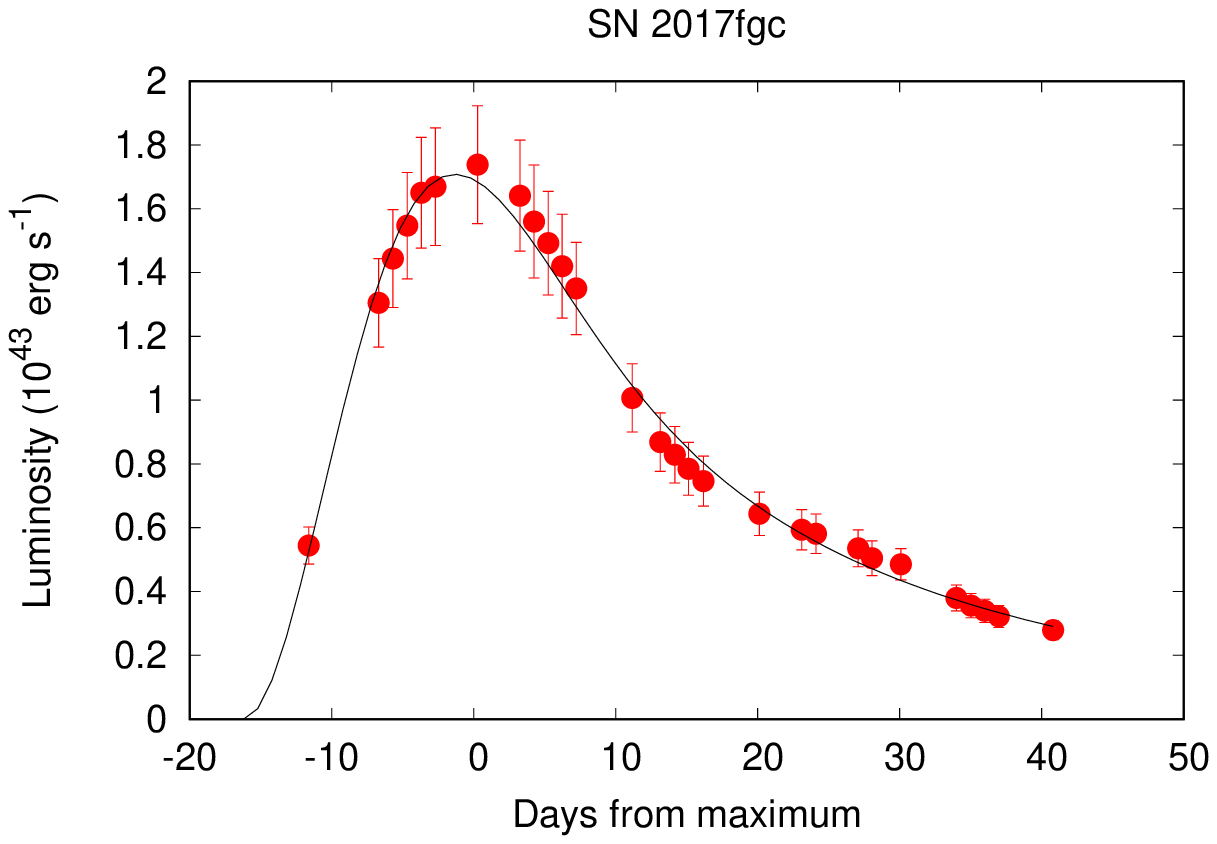}
\includegraphics[width=5cm]{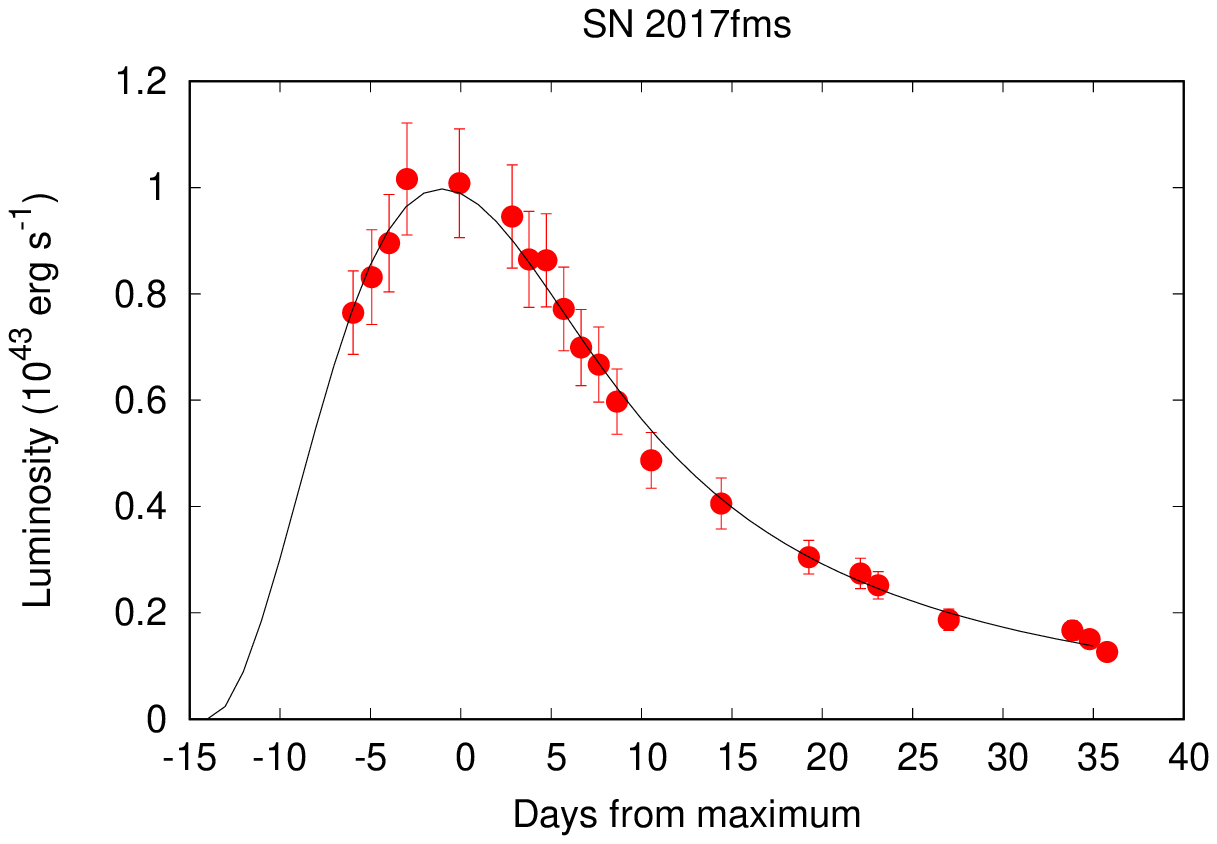}
\includegraphics[width=5cm]{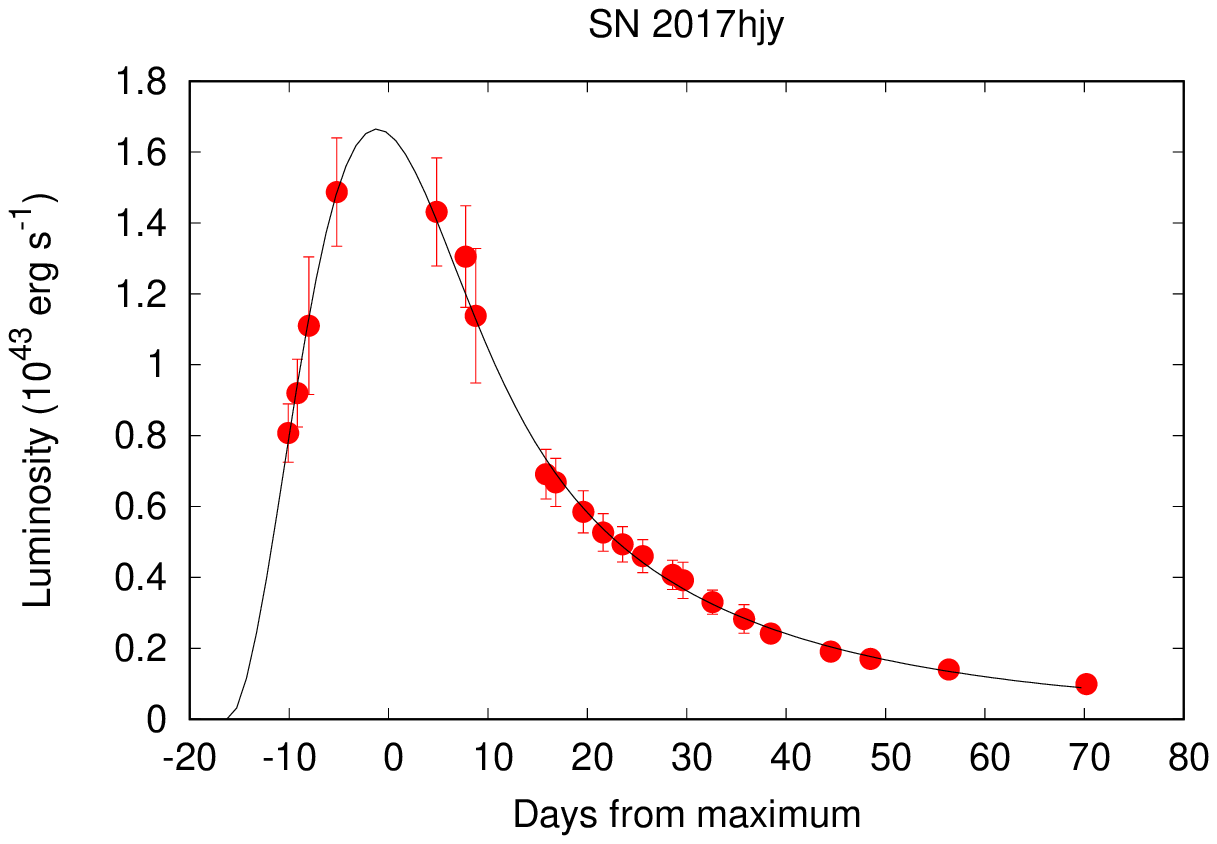}
\includegraphics[width=5cm]{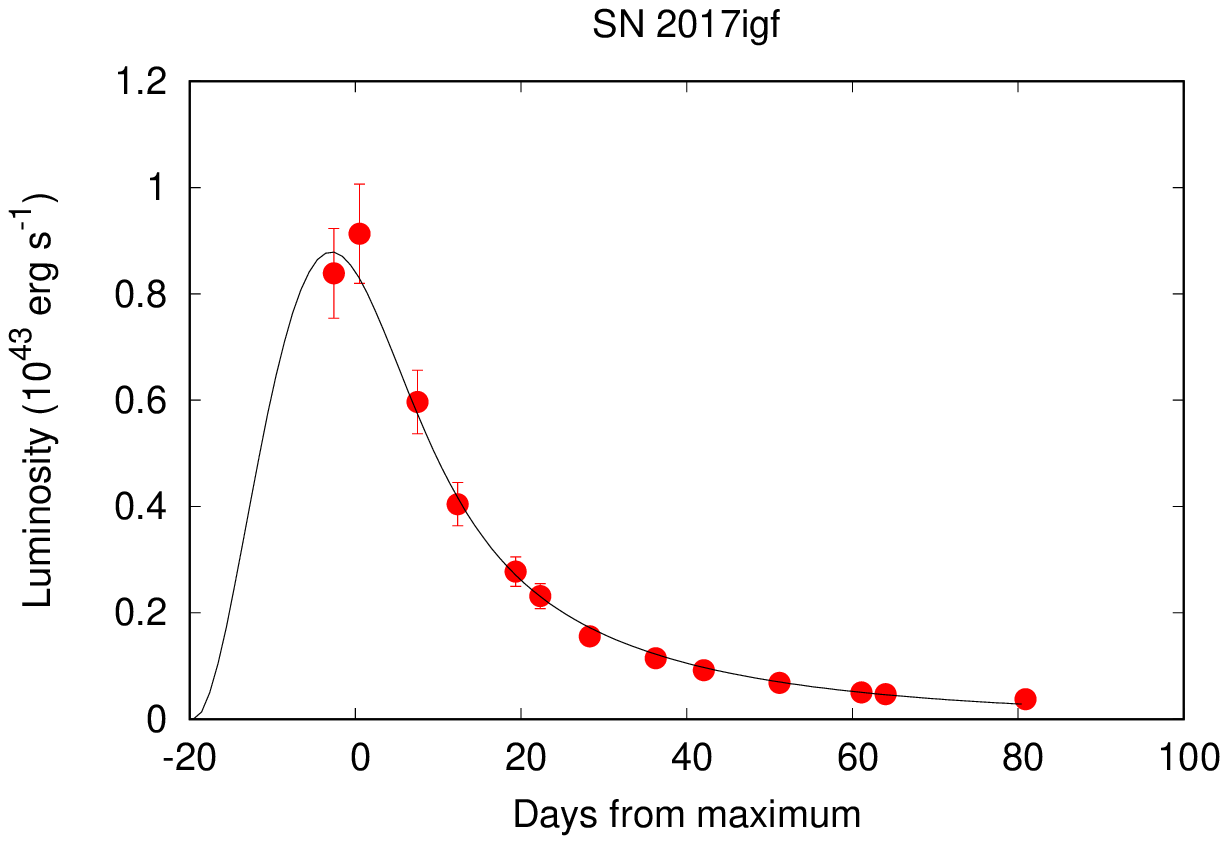}
\includegraphics[width=5cm]{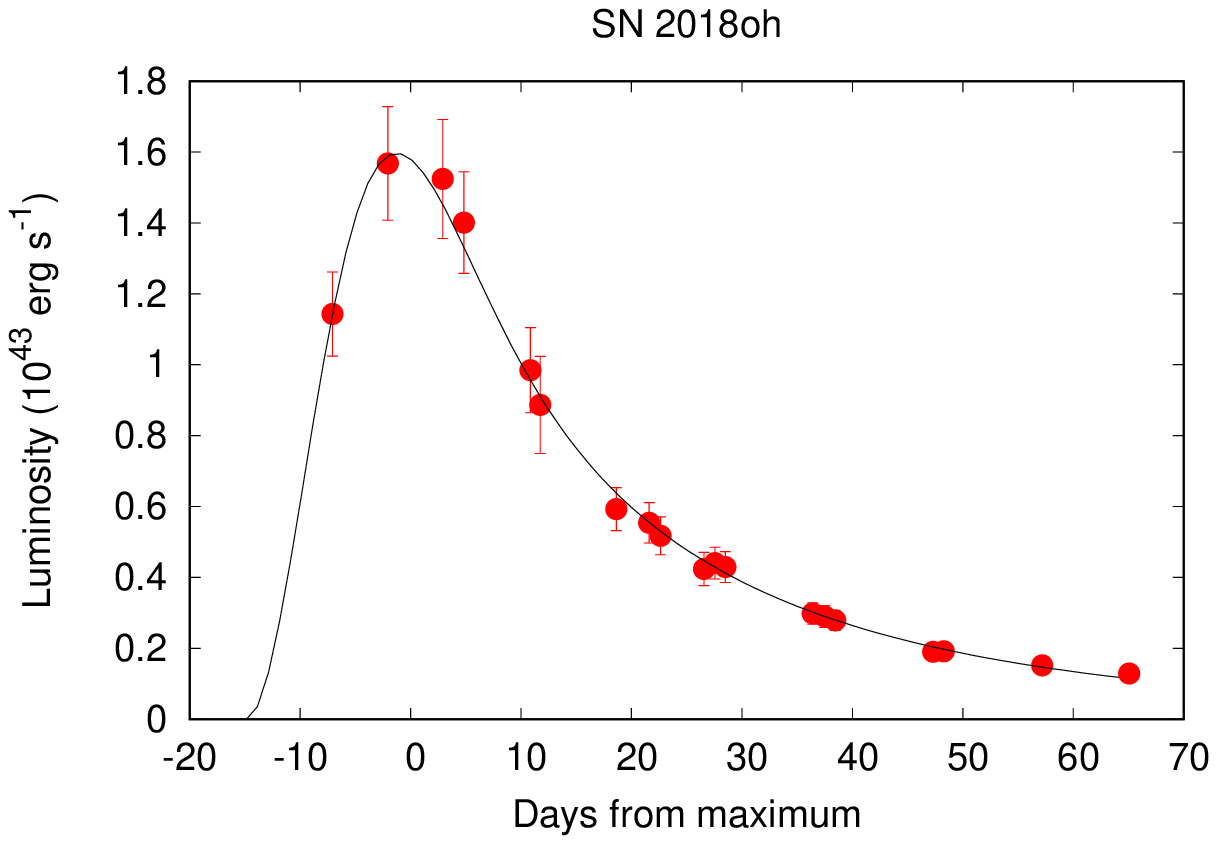}
\caption{The best-fit bolometric LC-s computed with {\tt Minim}. }
\end{center}
\label{18sne_bol}
\end{figure*}

\begin{figure*}[h!]
\begin{center}
\includegraphics[width=5cm]{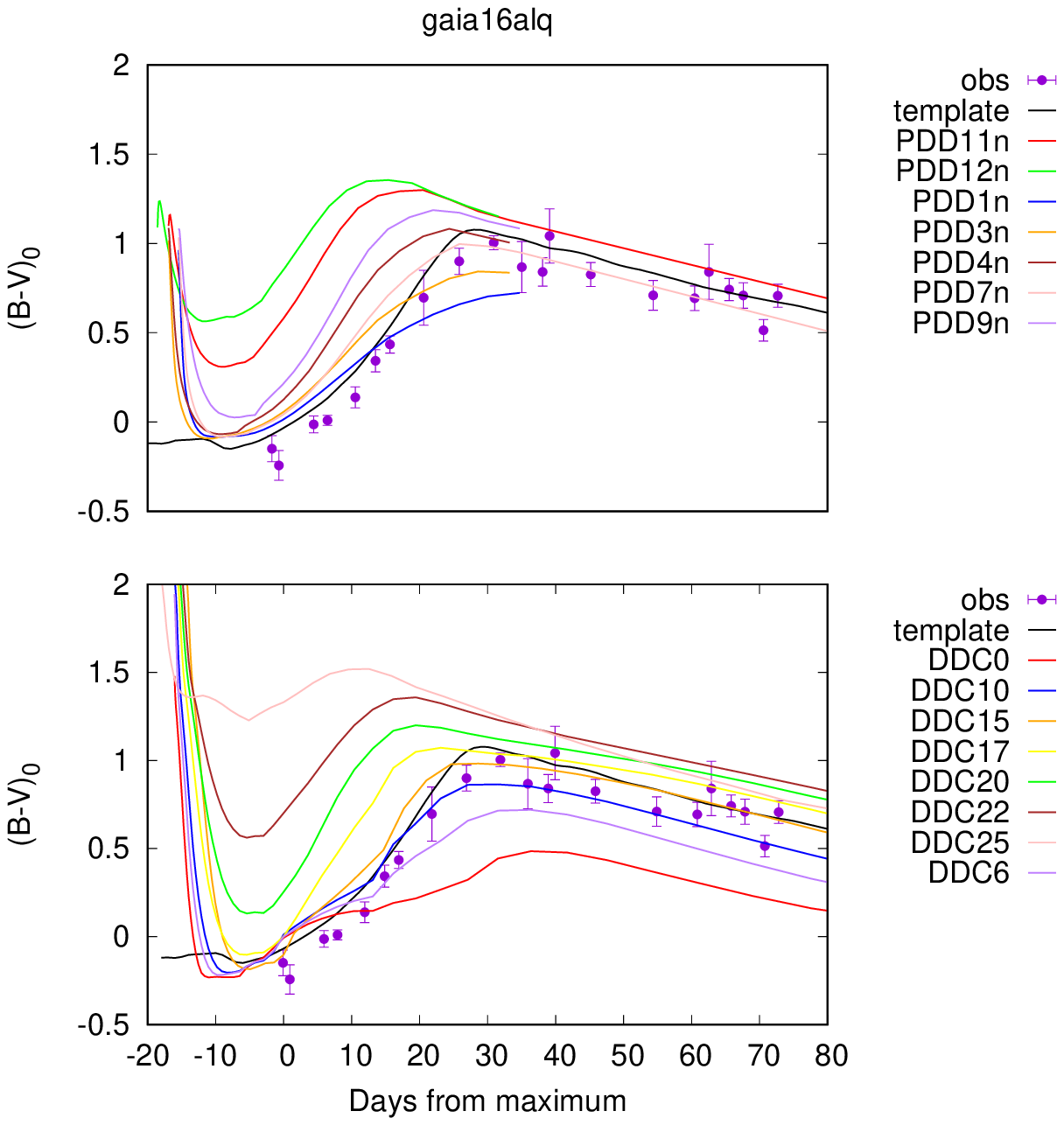}
\includegraphics[width=5cm] {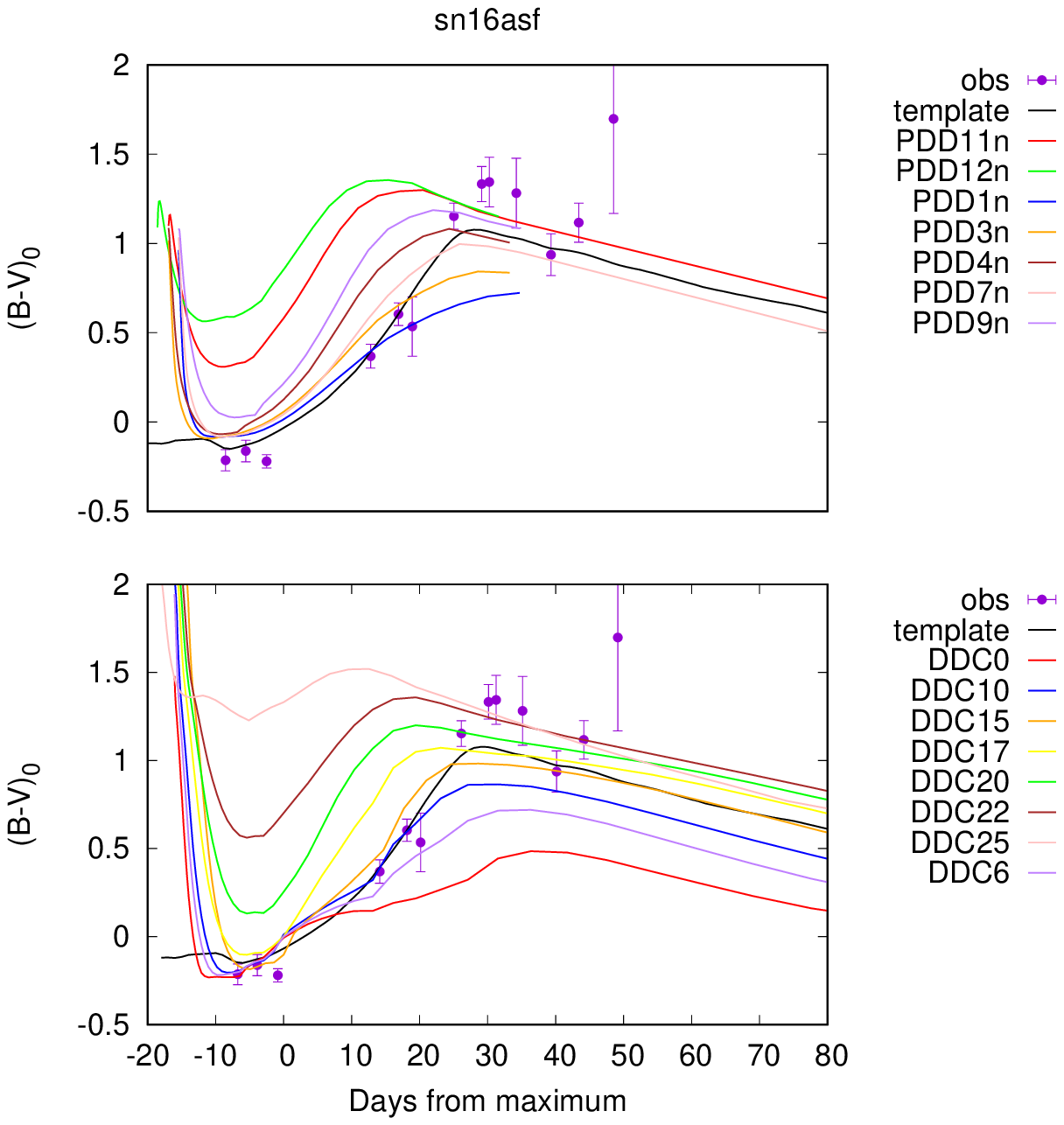}
\includegraphics[width=5cm] {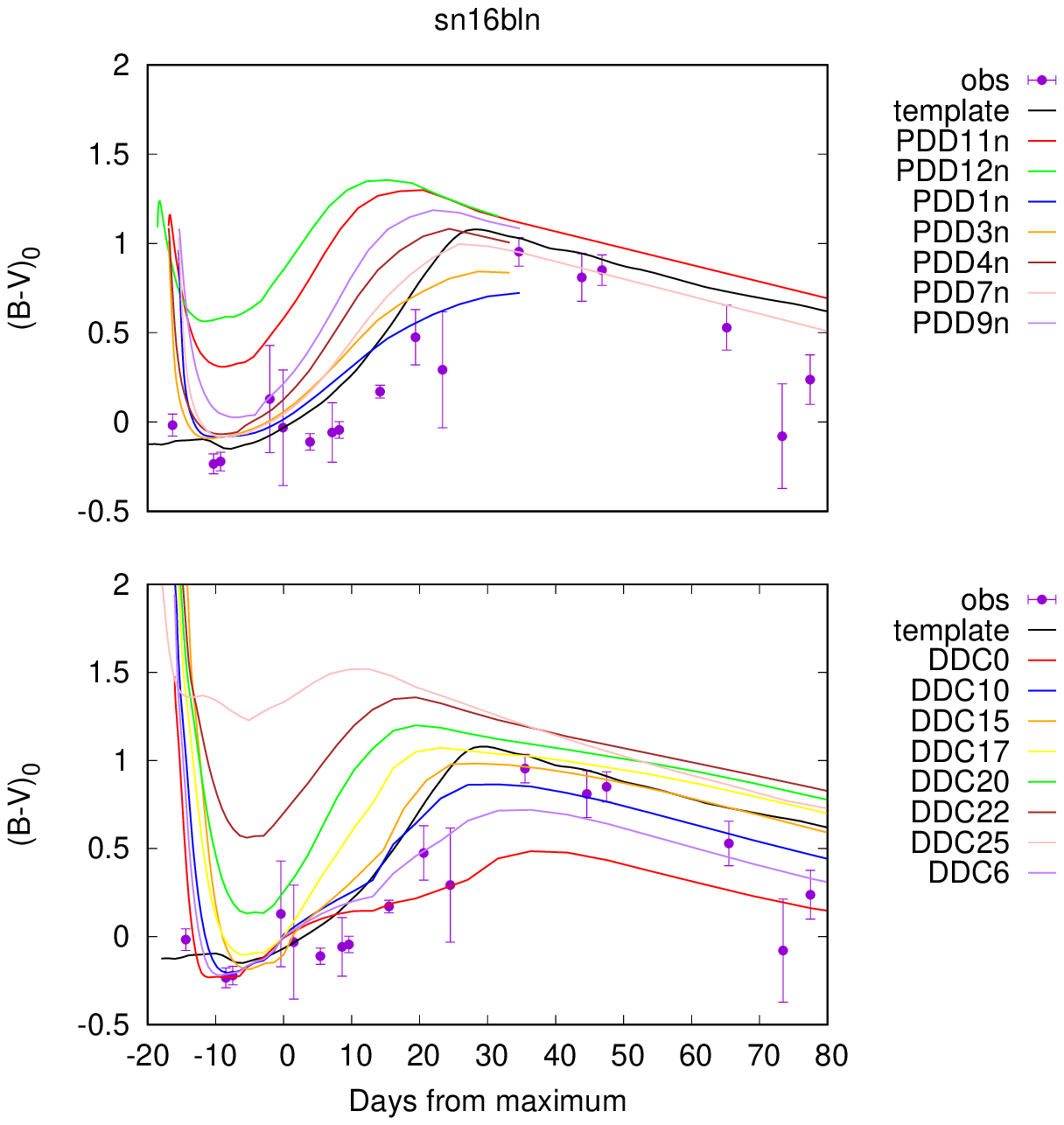}
\includegraphics[width=5cm] {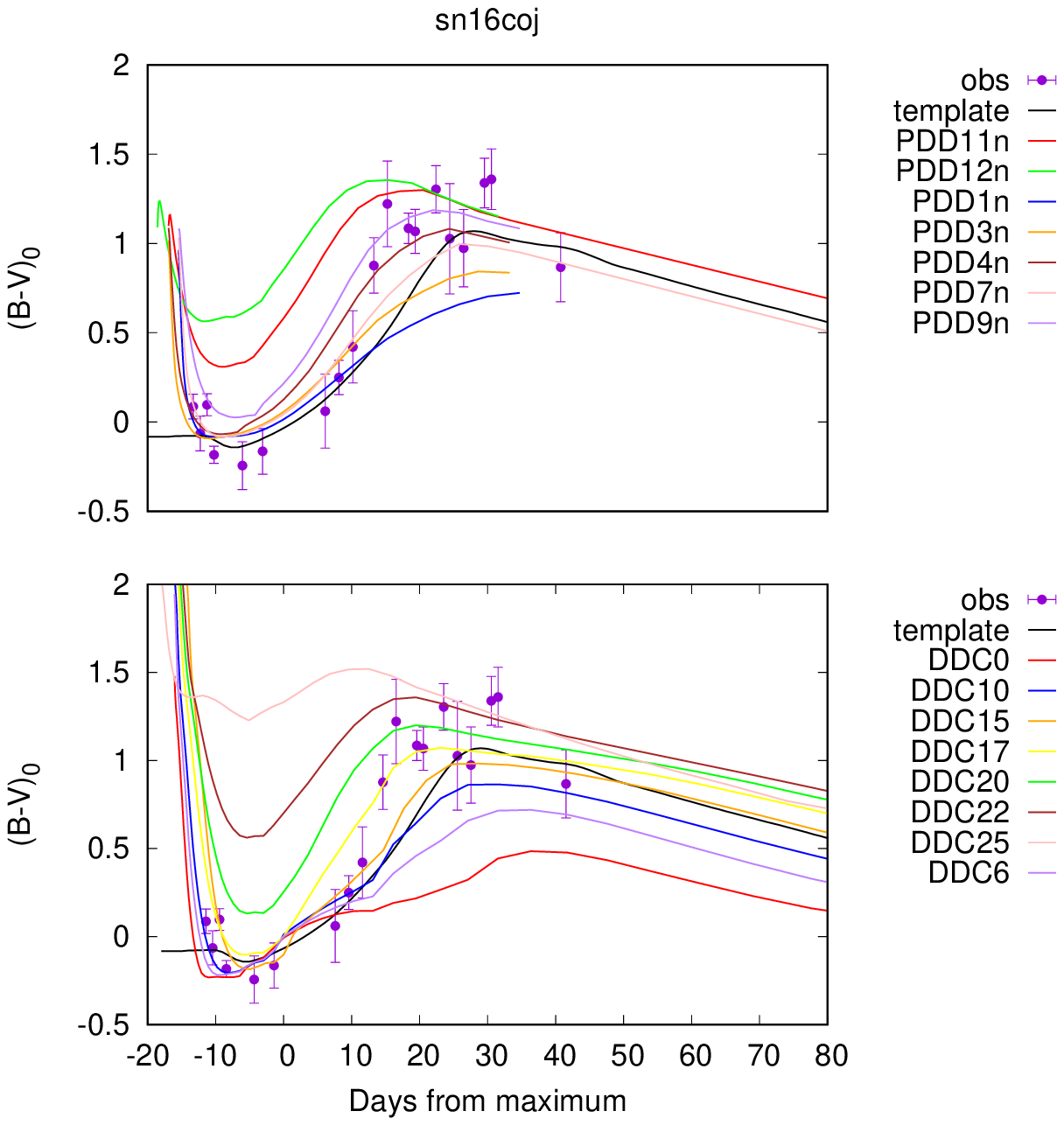}
\includegraphics[width=5cm] {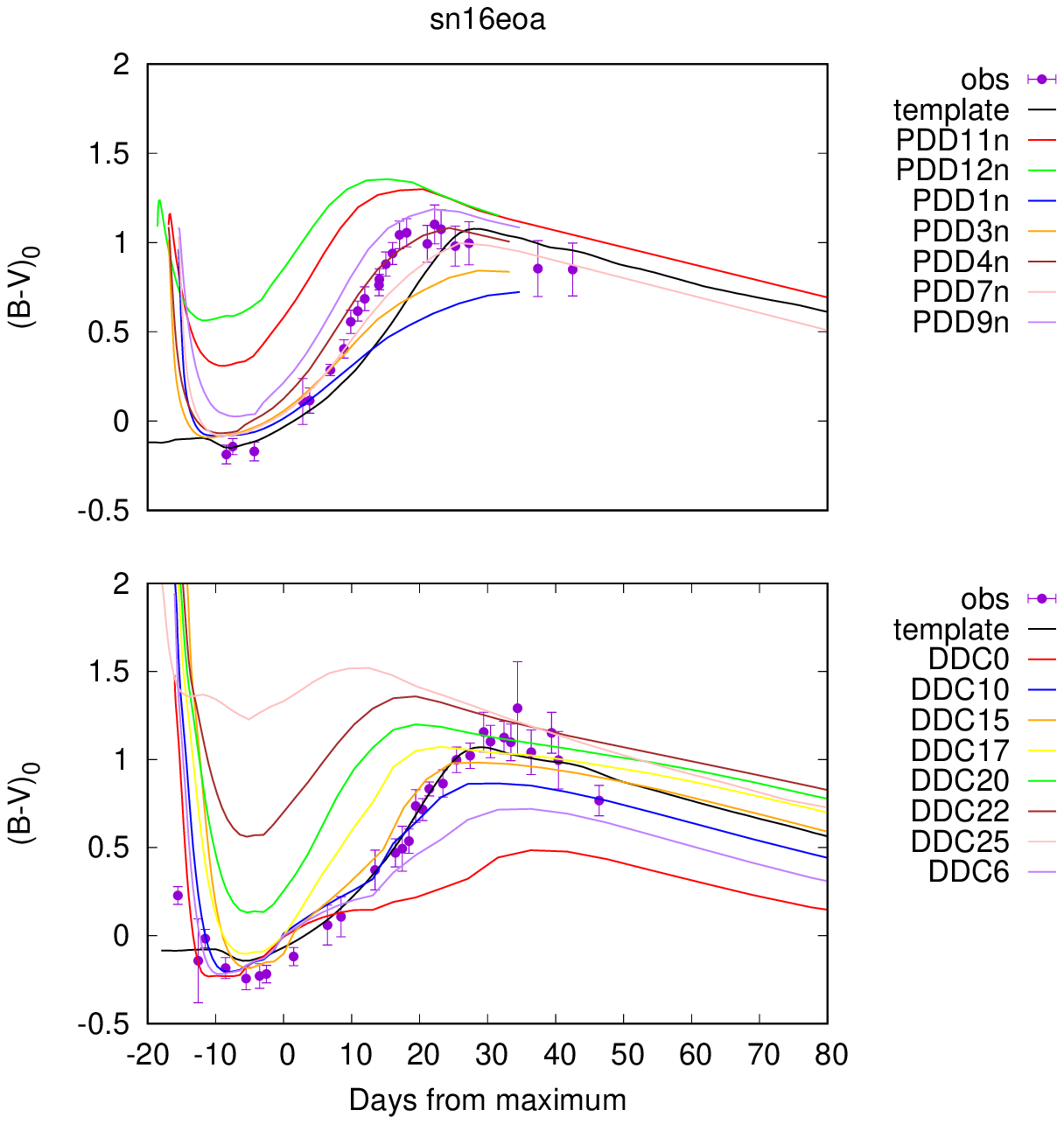}
\includegraphics[width=5cm] {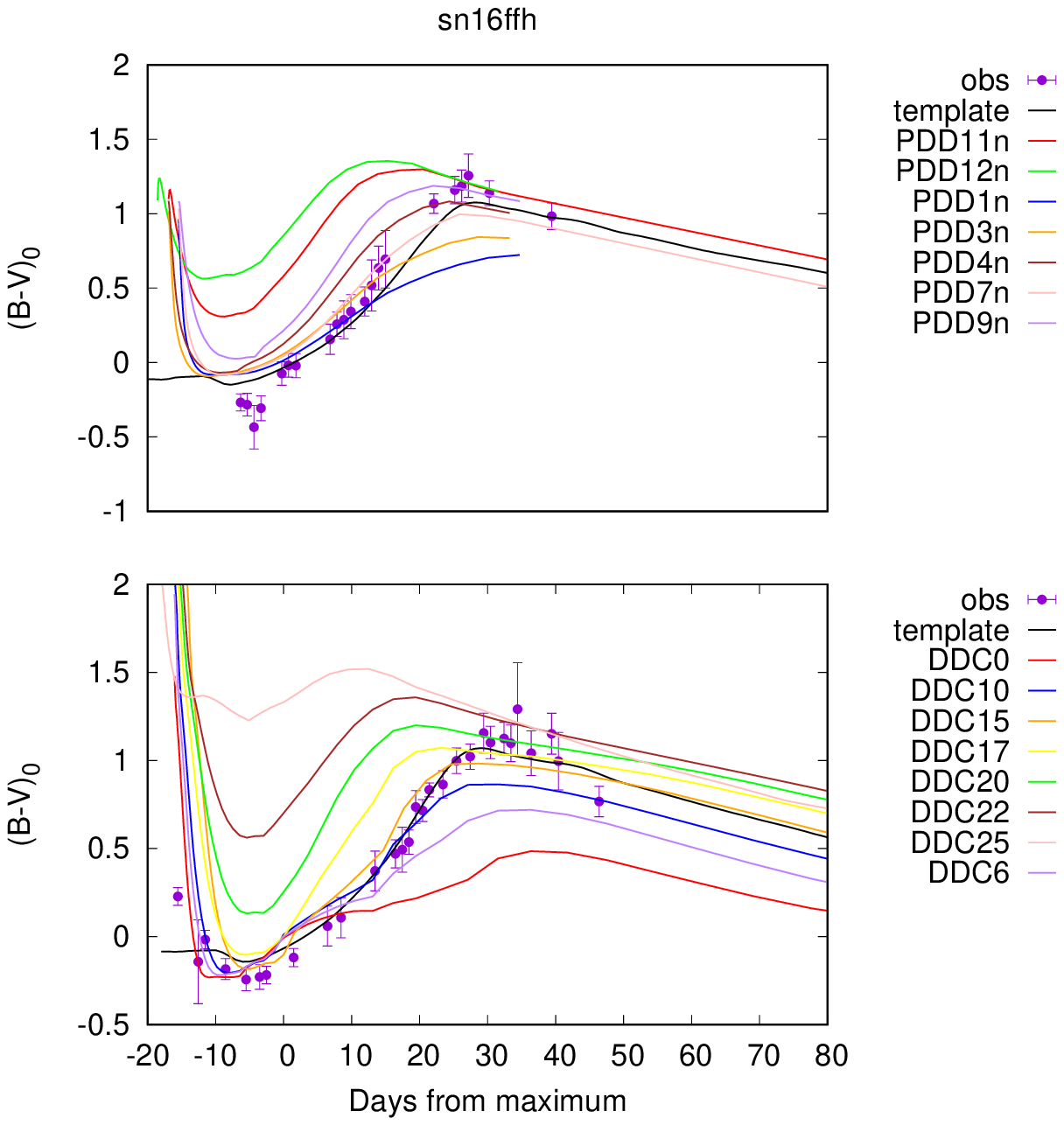}
\includegraphics[width=5cm] {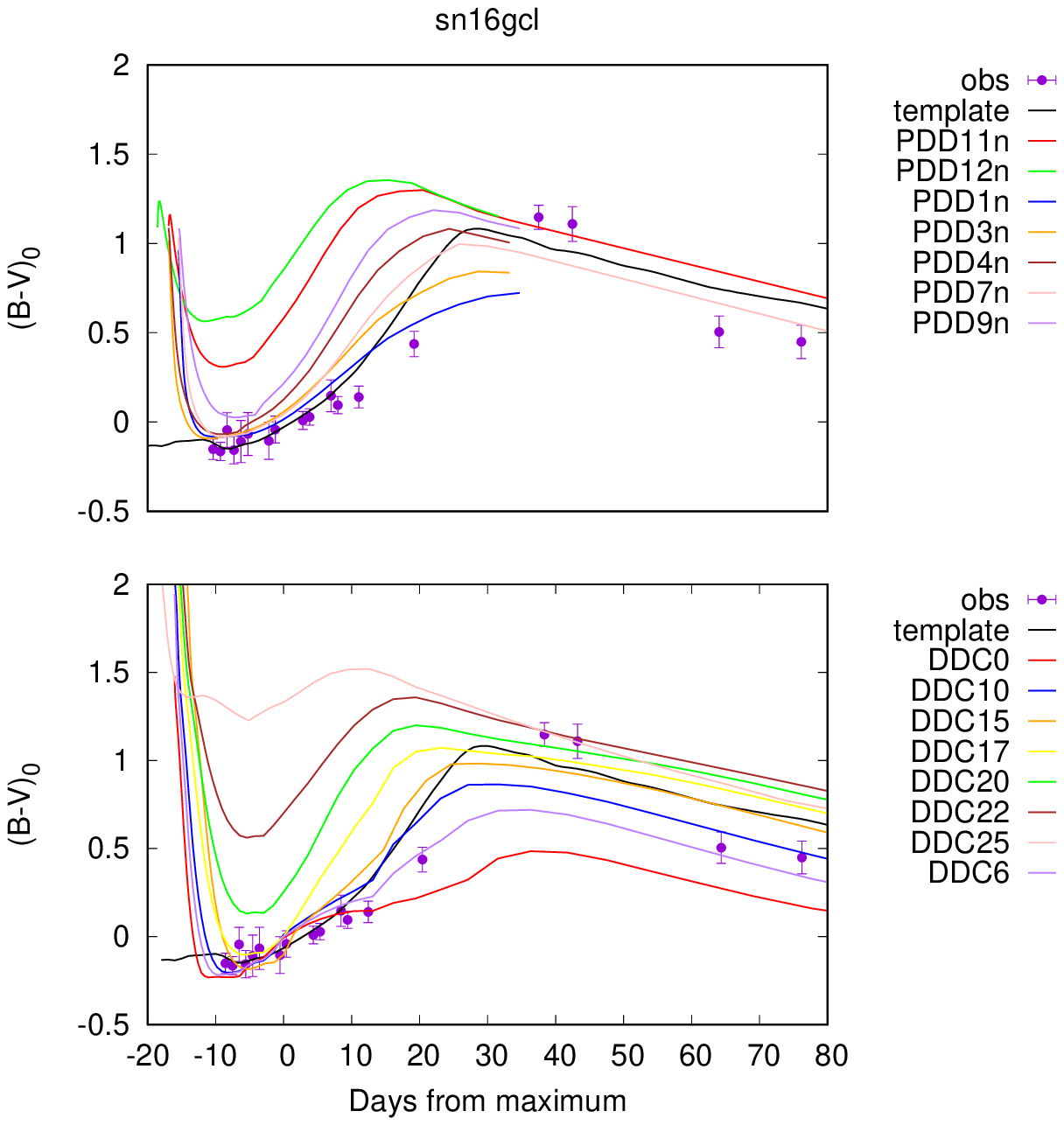}
\includegraphics[width=5cm] {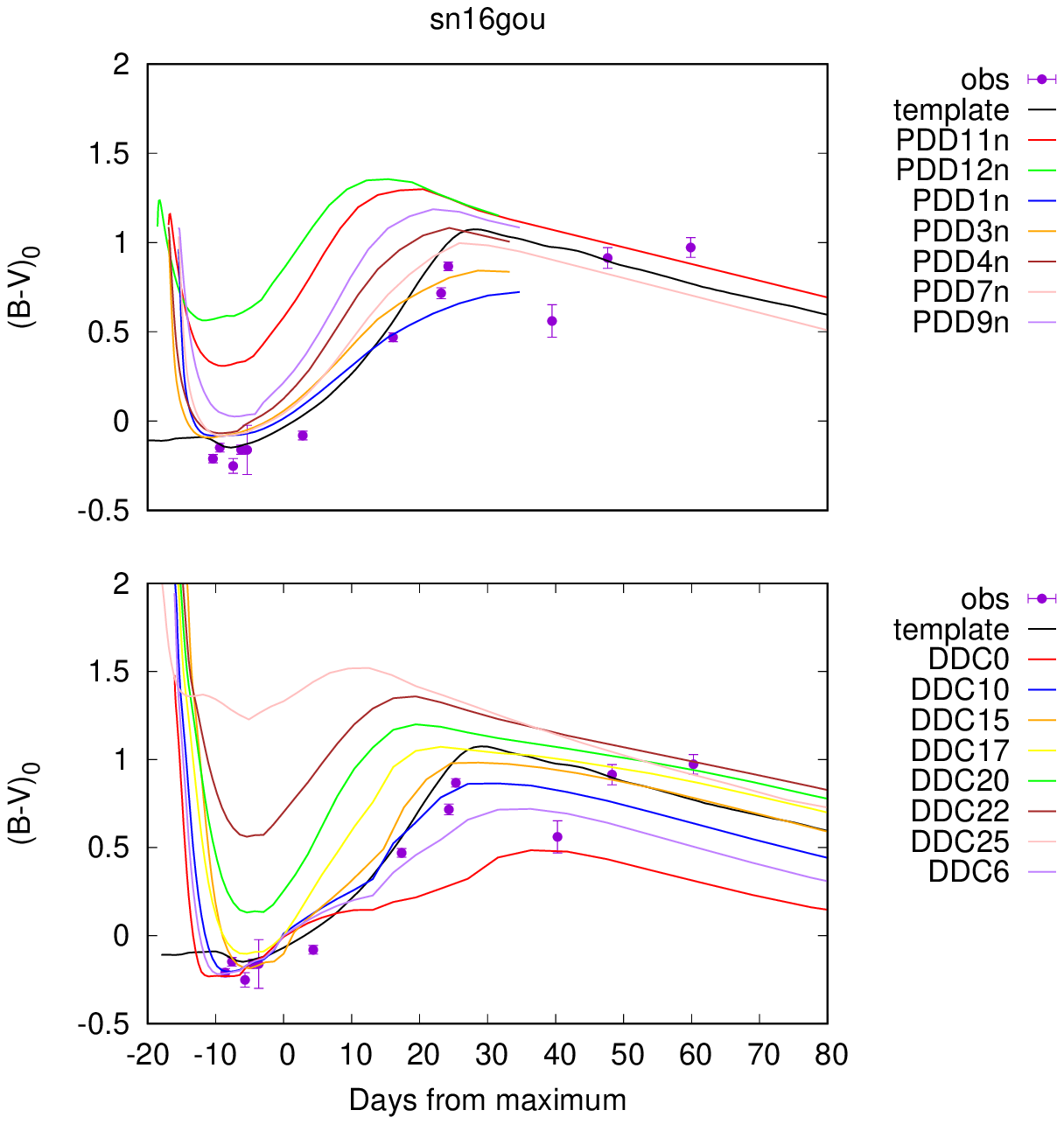}
\includegraphics[width=5cm] {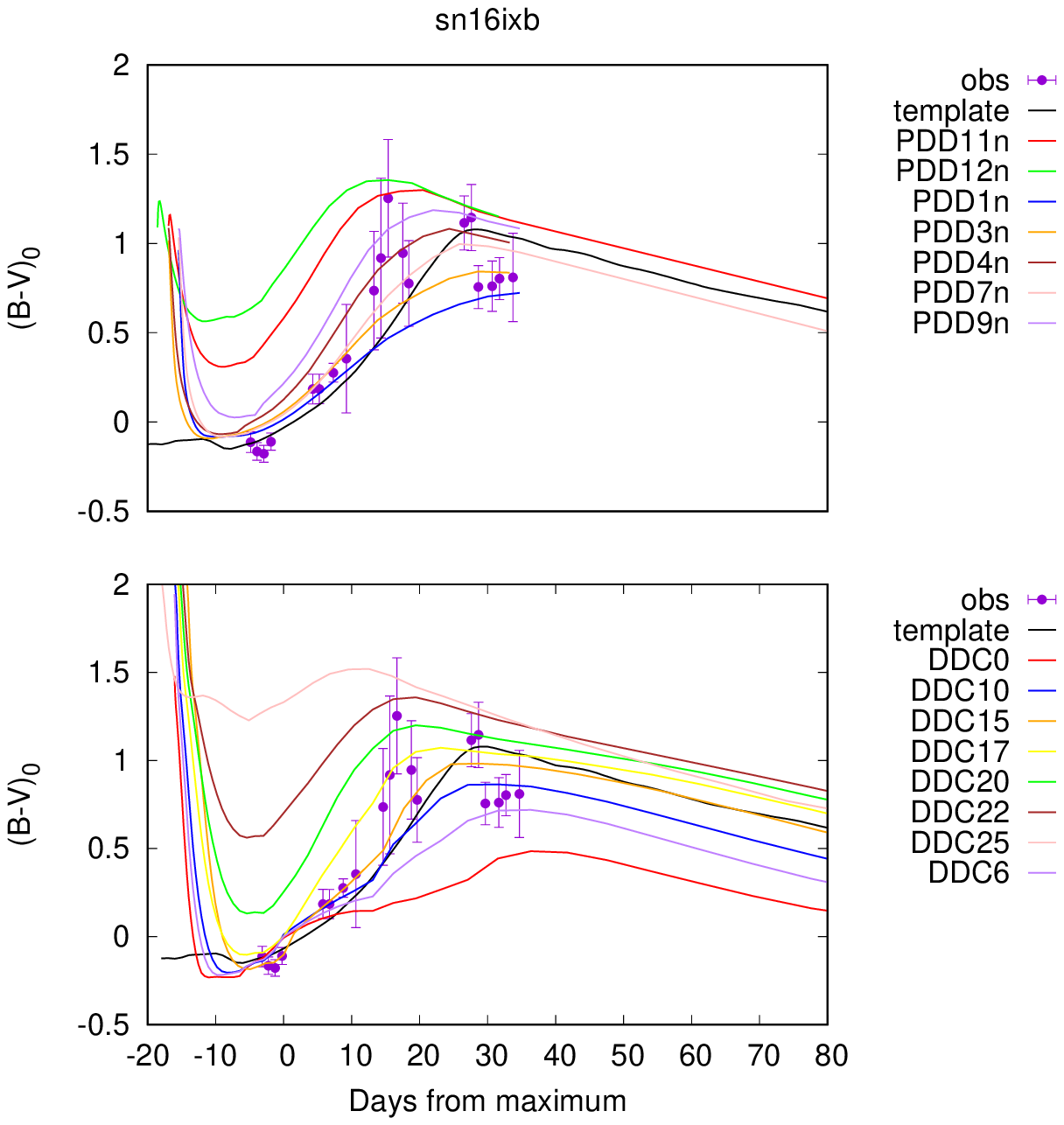}
\end{center}
\caption{Comparison of the observed, de-reddened $(B-V)_0$ colors (filled symbols) with synthetic colors from DDE and PDDE models by \citet{dessart} (colored curves). The black curve corresponds to the synthetic colors inferred from the empirical Hsiao-template \citep{hsiao07}.}
\label{pdd1}
\end{figure*}

\begin{figure*}[h!]
\begin{center}
\includegraphics[width=5cm] {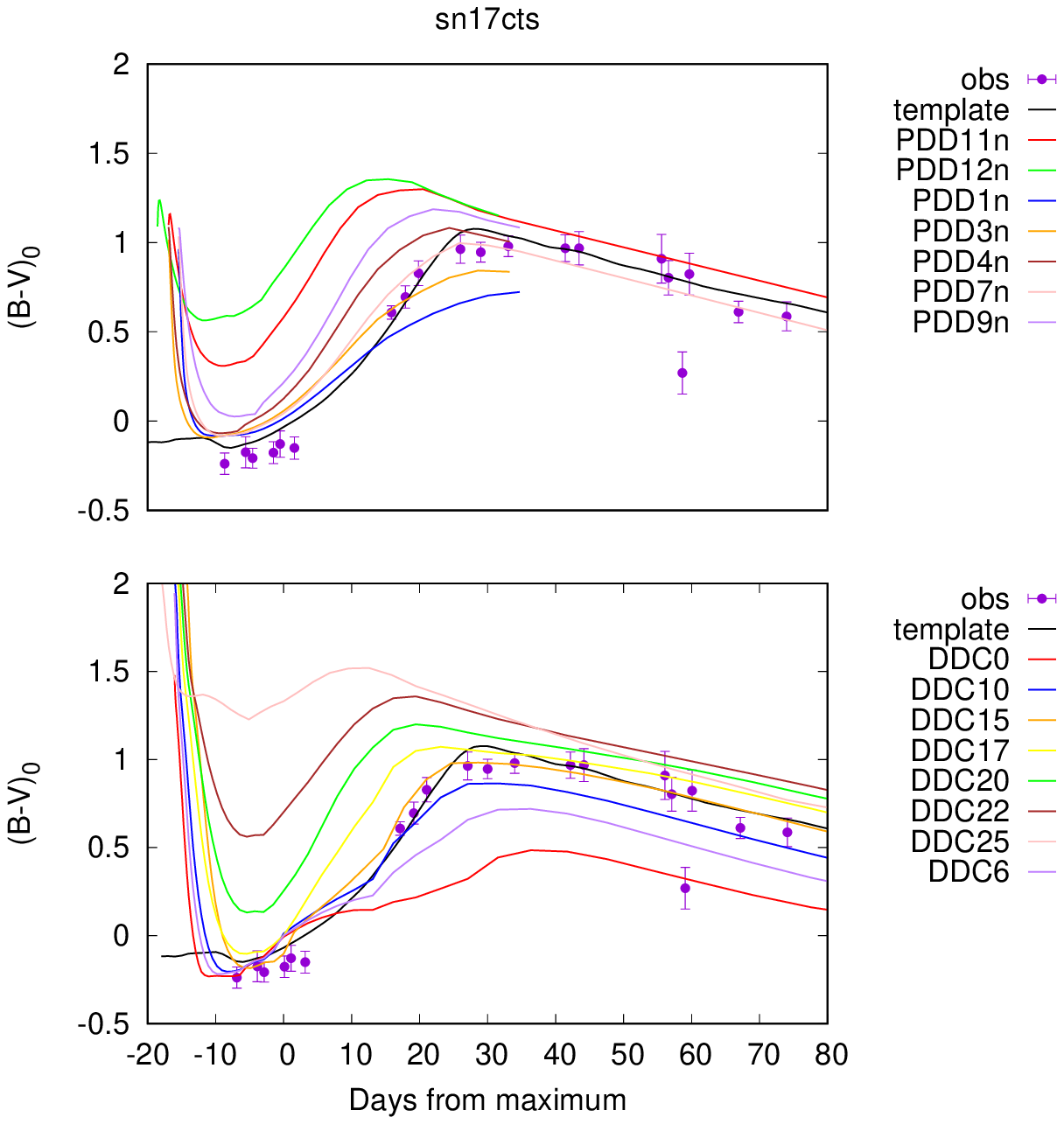}
\includegraphics[width=5cm] {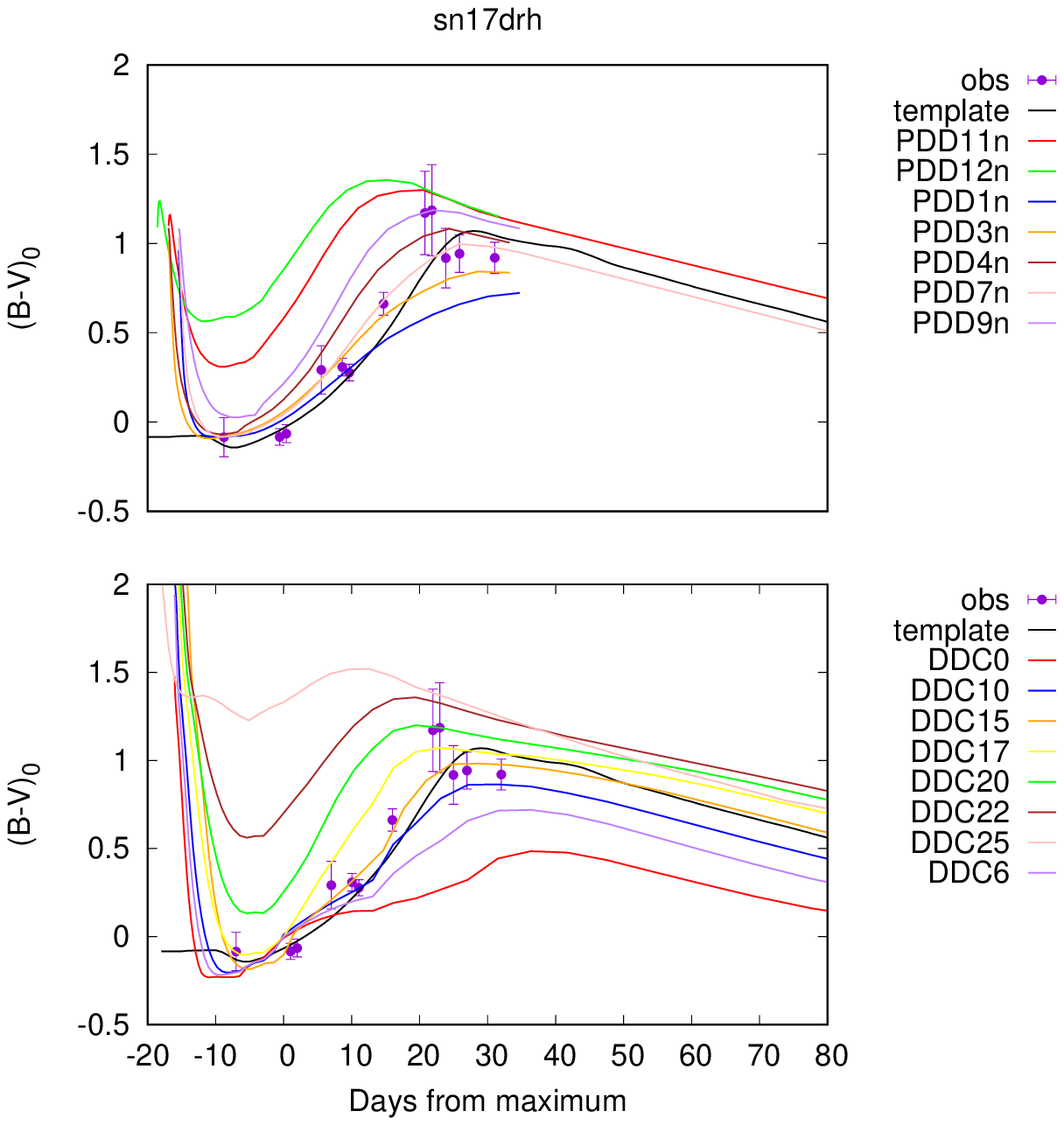}
\includegraphics[width=5cm] {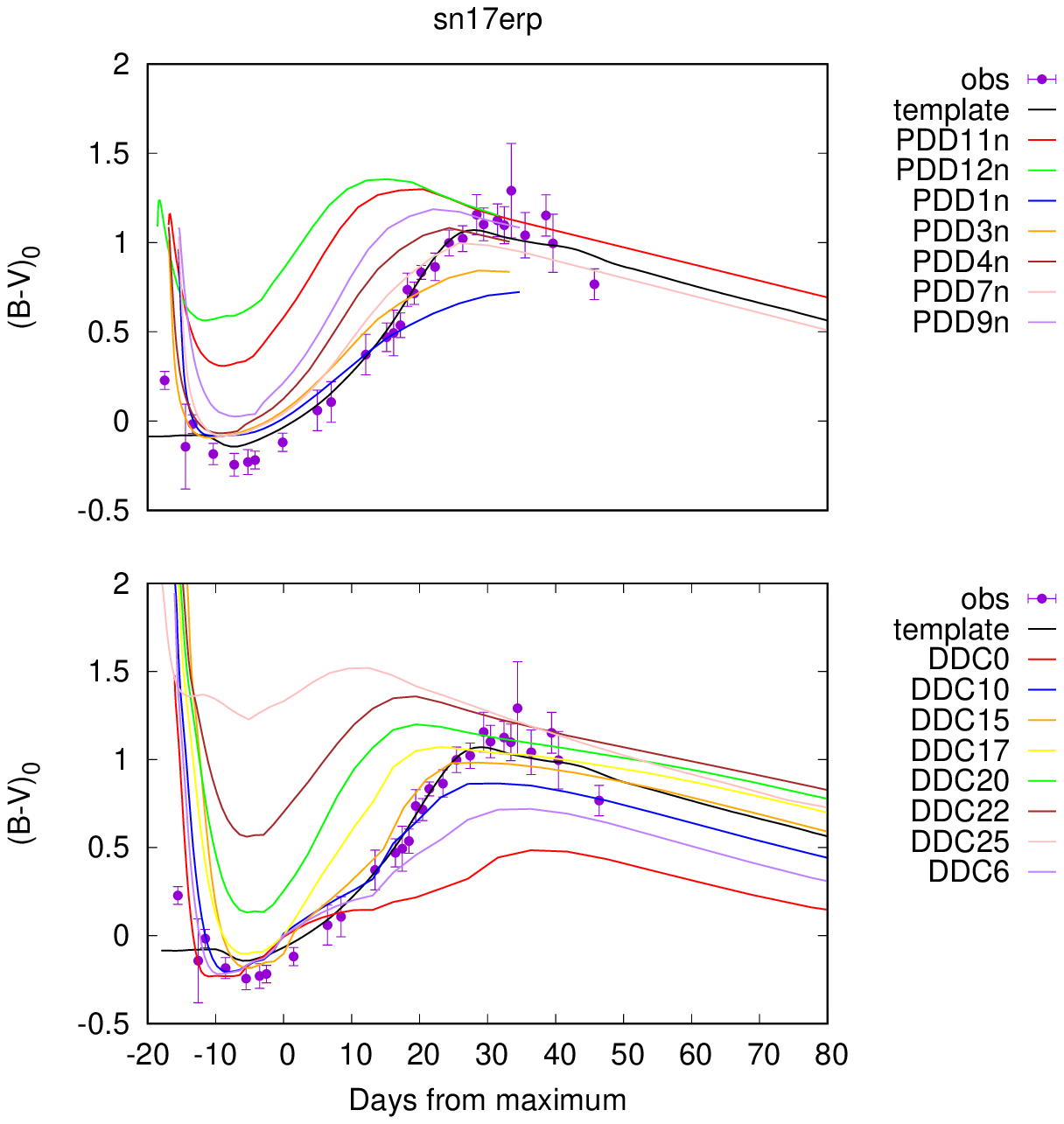}
\includegraphics[width=5cm] {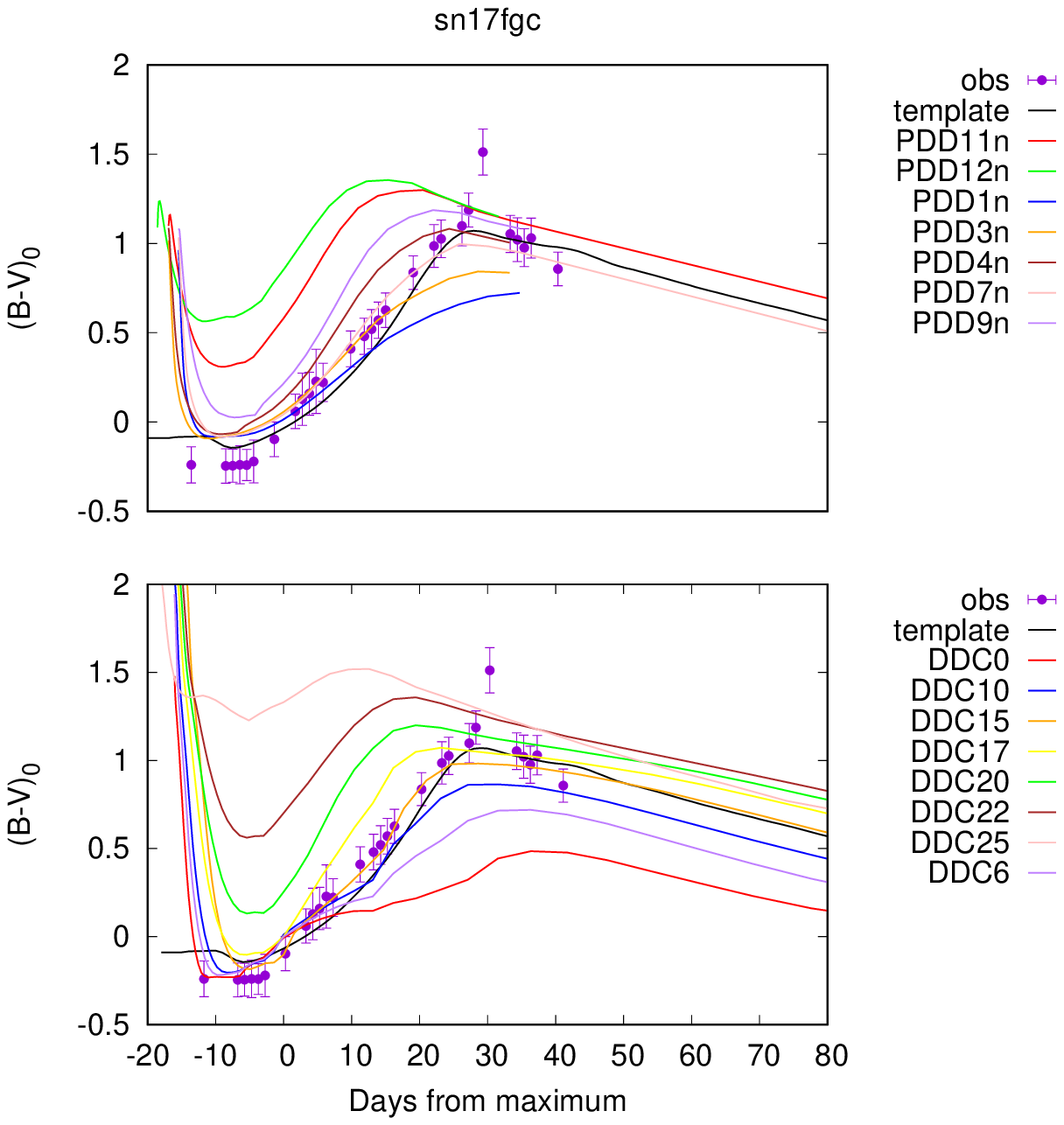}
\includegraphics[width=5cm] {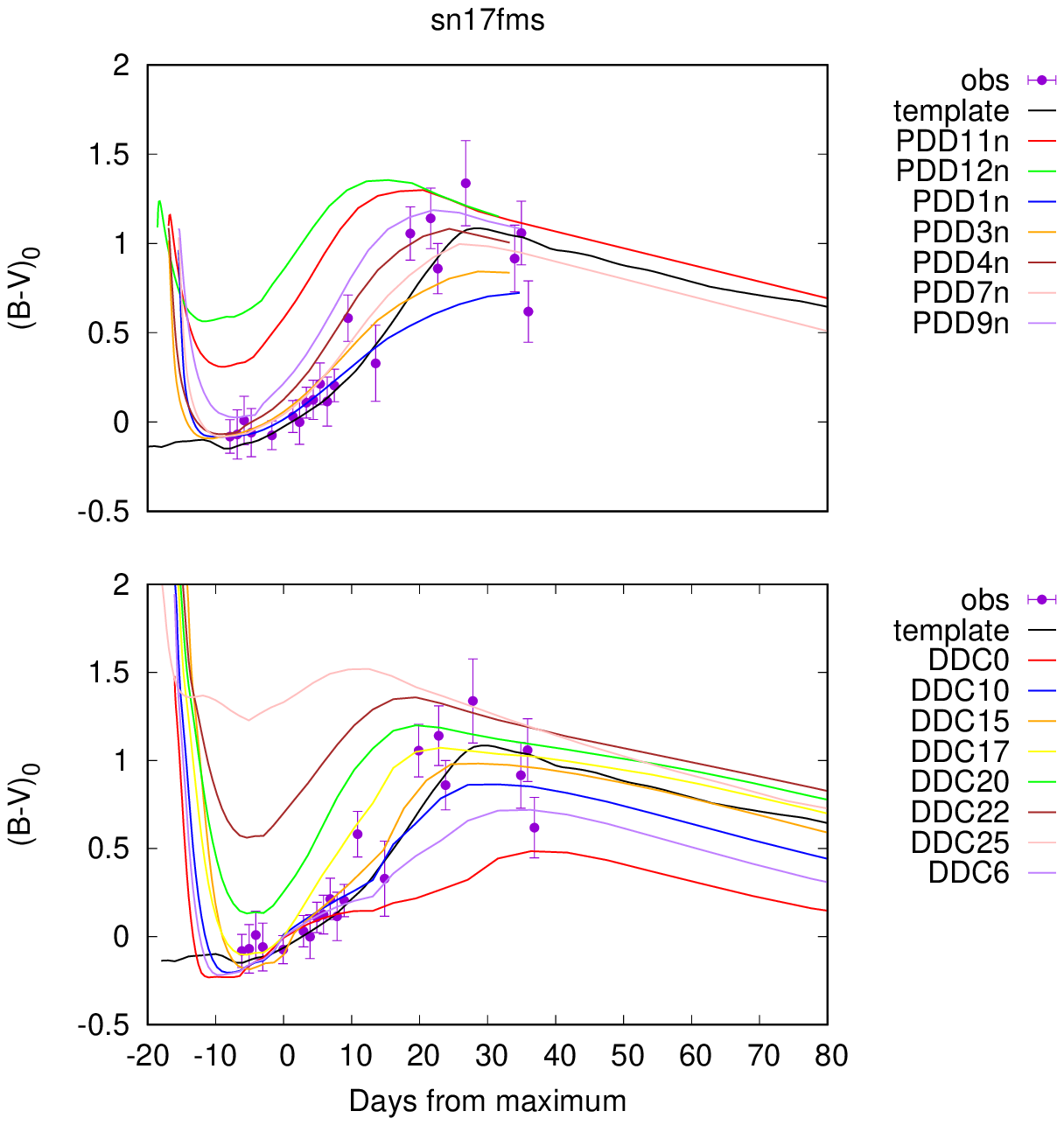}
\includegraphics[width=5cm] {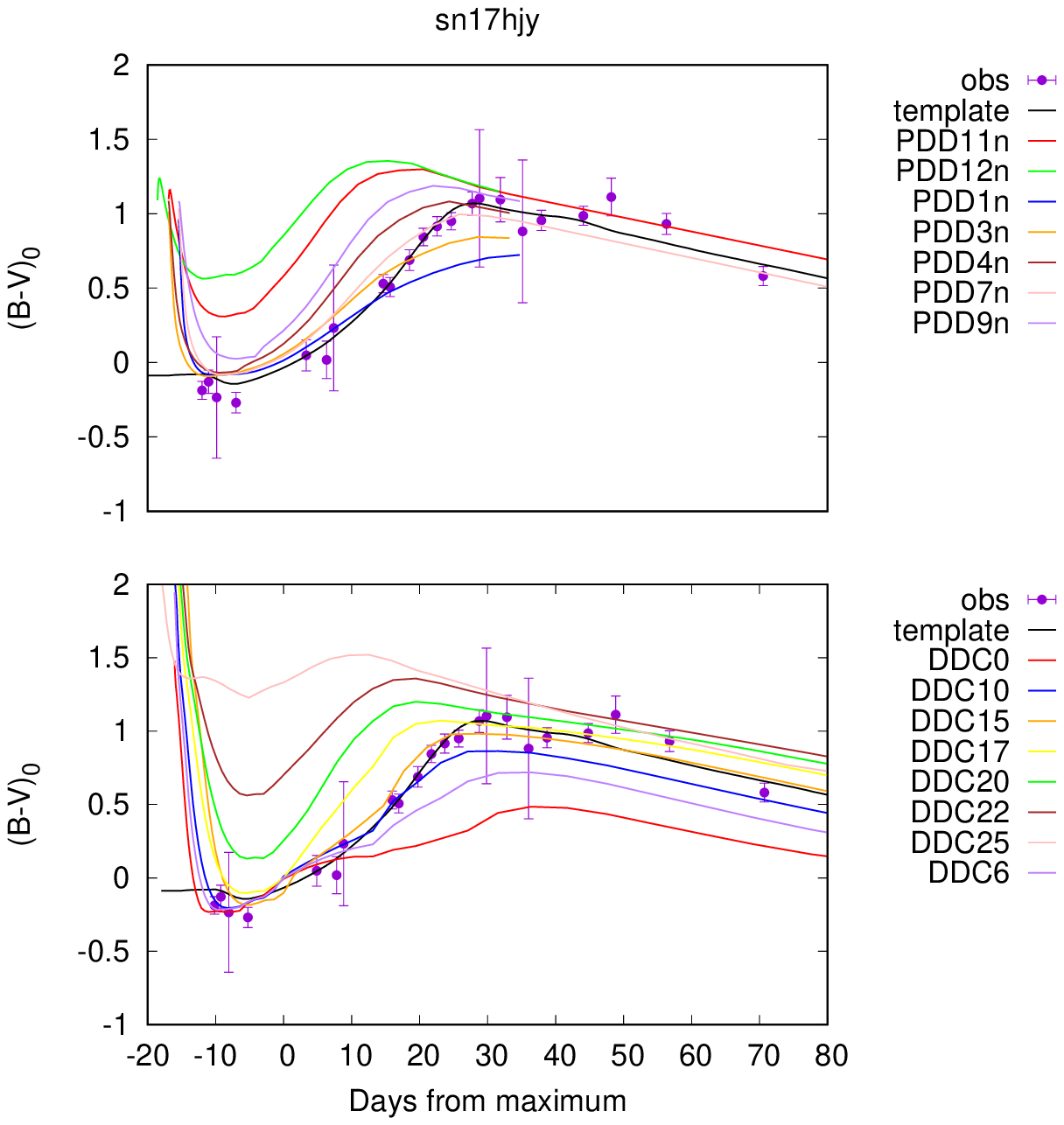}
\includegraphics[width=5cm] {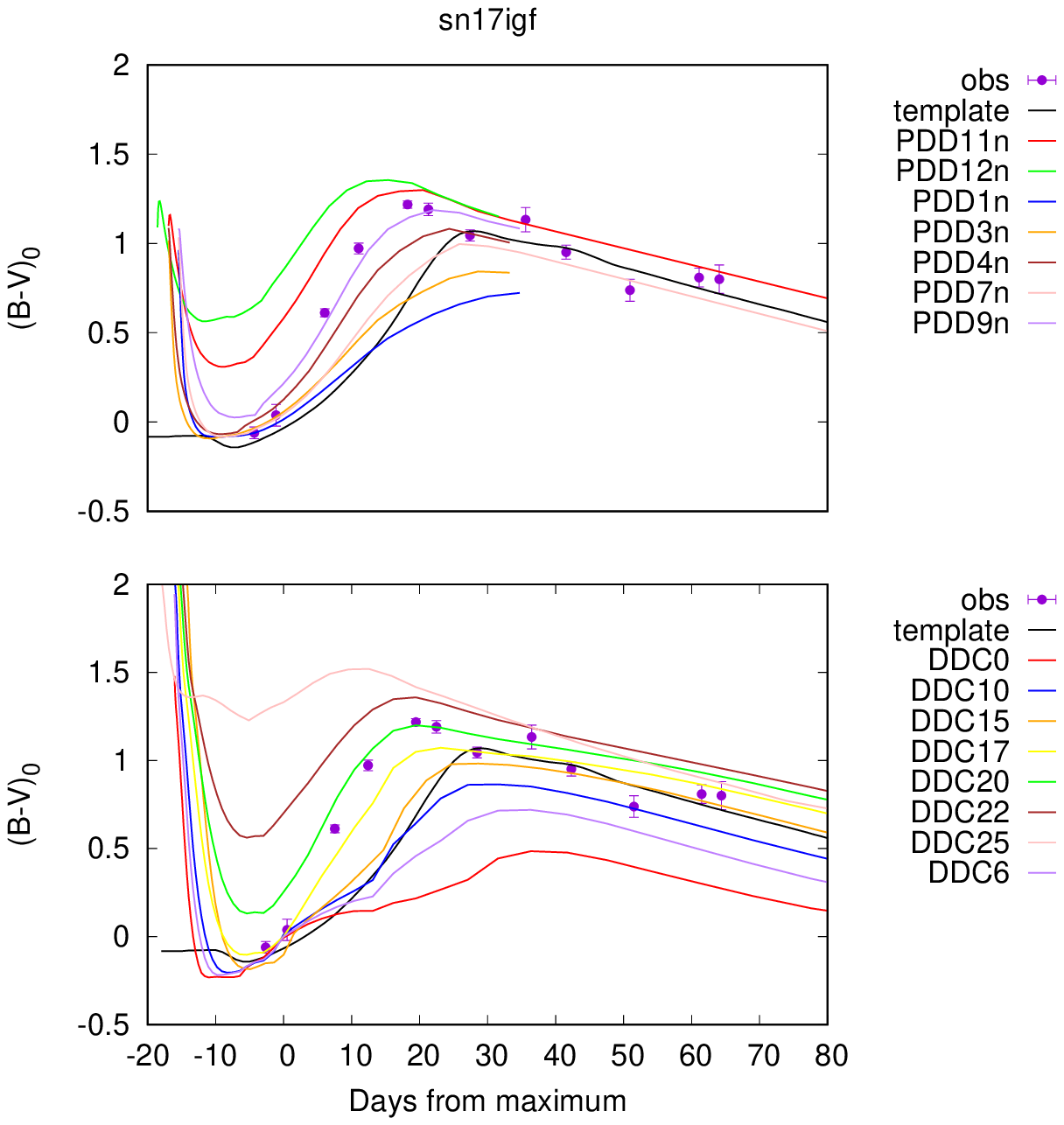}
\includegraphics[width=5cm] {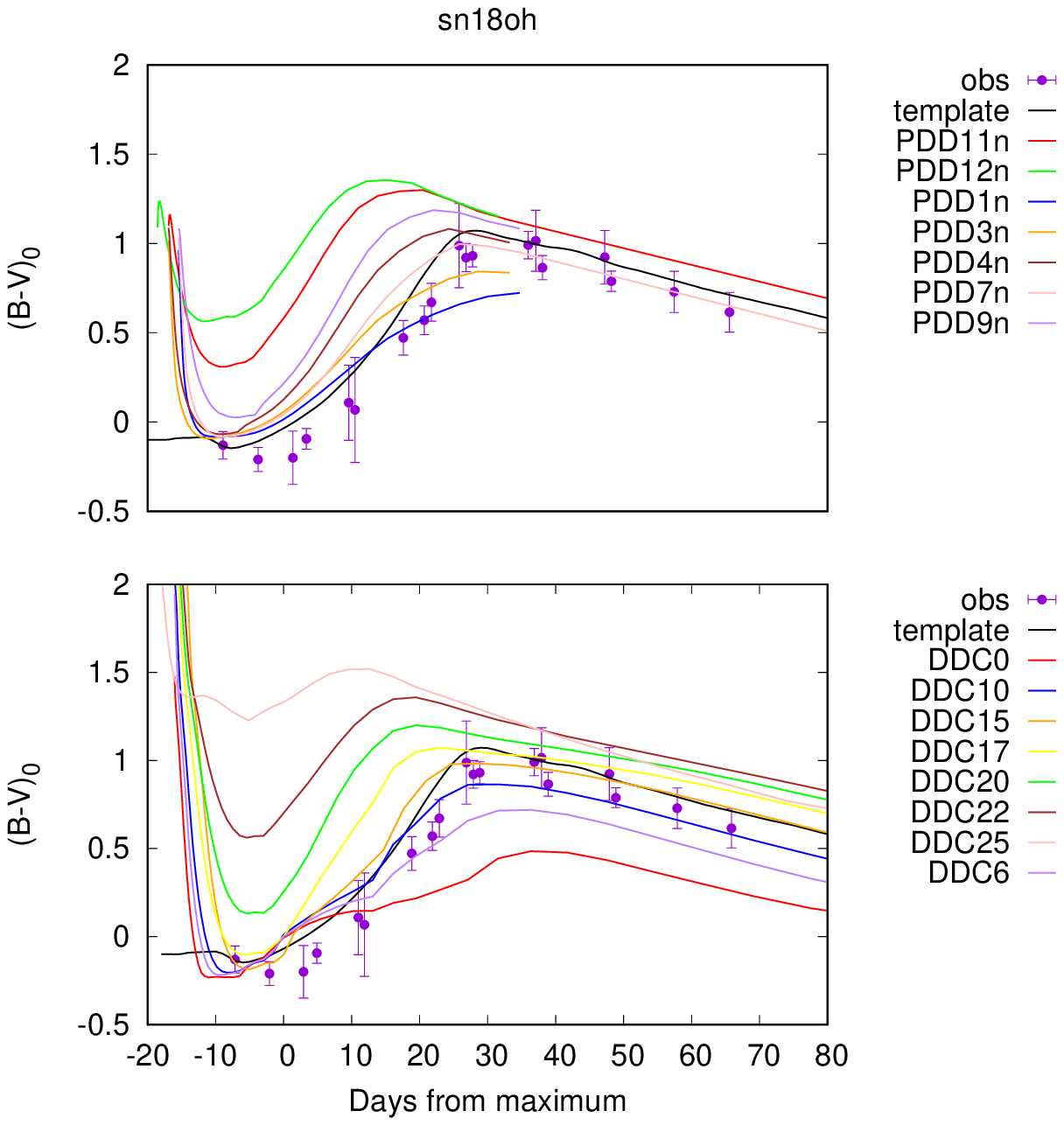}
\end{center}
\caption{The same as Fig.\ref{pdd1} but for additional SNe.}
\label{pdd2}
\end{figure*}


\begin{thebibliography}{}

\bibitem[Amanullah et al.(2015)]{ama15} Amanullah, R., Johansson, J., Goobar, A., et al.\ 2015, \mnras, 453, 3300

\bibitem[Arnett(1982)]{arnett82} Arnett, W.~D.\ 1982, \apj, 253, 785

\bibitem[Astier et al.(2006)]{astier06} Astier, P., Guy, J., Regnault, N., et al.\ 2006, \aap, 447, 31 

\bibitem[Bengaly et al.(2015)]{bengaly15} Bengaly, C.~A.~P., Jr., Bernui, A., \& Alcaniz, J.~S.\ 2015, \apj, 808, 39 


\bibitem[Benitez-Herrera et al.(2013)]{bh13} Benitez-Herrera, S., Ishida, E.~E.~O., Maturi, M., et al.\ 2013, \mnras, 436, 854 


\bibitem[Bertin \& Arnouts (1996)]{sex} Bertin, E. \& Arnouts, S., 1996, \aaps 317, 393

\bibitem[Bessell(1990)]{bessell90} Bessell, M.~S.\ 1990, \pasp, 102, 1181 

\bibitem[Bessell et al.(1998)]{bessell98} Bessell, M.~S., Castelli, F., \& Plez, B.\ 1998, \aap, 333, 231

\bibitem[Betoule et al.(2014)]{beto14} Betoule, M., Kessler, R., Guy, J., et al.\ 2014, \aap, 568, A22 

\bibitem[Brimacombe et al.(2017)]{17cts} Brimacombe, J., Stone, G., Post, R., et al.\ 2017, Transient Name Server Discovery Report 2017-391, 1.

\bibitem[Brown(2016)]{16gcl} Brown, J.\ 2016, Transient Name Server Discovery Report 2016-644, 1.

\bibitem[Brown et al.(2014)]{brown14} Brown, P.~J., Breeveld, A.~A., Holland, S., Kuin, P., \& Pritchard, T.\ 2014, \apss, 354, 89 

\bibitem[Brown et al.(2018)]{erp} Brown, P.~J., Hosseinzadeh, G., Jha, S.~W., et al.\ 2018, arXiv:1808.04729 

\bibitem[Burns et al.(2011)]{burns11} Burns, C.~R., Stritzinger, M., Phillips, M.~M., et al.\ 2011, \aj, 141, 19 

\bibitem[Burns et al.(2014)]{burns14} Burns, C.~R., Stritzinger, M., Phillips, M.~M., et al.\ 2014, \apj, 789, 32 

\bibitem[Burns et al.(2018)]{burns18} Burns, C.~R., Parent, E., Phillips, M.~M., et al.\ 2018, \apj, 869, 56

\bibitem[Cenko et al.(2016)]{bln} Cenko, S.~B., Cao, Y., Kasliwal, M., et al.\ 2016, The Astronomer's Telegram, 8909,

\bibitem[Chatzopoulos et al.(2012)]{manos12} Chatzopoulos, E., Wheeler, J.~C., \& Vinko, J.\ 2012, \apj, 746, 121 

\bibitem[Chatzopoulos et al.(2013)]{minim} Chatzopoulos, E., Wheeler, J.~C., Vinko, J., Horvath, Z.~L., \& Nagy, A.\ 2013, \apj, 773, 76 

\bibitem[Clocchiatti \& Wheeler(1997)]{cw97} Clocchiatti, A., \& Wheeler, J.~C.\ 1997, \apj, 491, 375 

\bibitem[Conley et al.(2008)]{conley08} Conley, A., Sullivan, M., Hsiao, E.~Y., et al.\ 2008, \apj, 681, 482 

\bibitem[Conley et al.(2011)]{conley11} Conley, A., Guy, J., Sullivan, M., et al.\ 2011, \apjs, 192, 1 

\bibitem[Contreras et al.(2018)]{contreras18} Contreras, C., Phillips, M.~M., Burns, C.~R., et al.\ 2018, \apj, 859, 24 

\bibitem[Cruz et al.(2016)]{16asf} Cruz, I., Holoien, T.~W.-S., Stanek, K.~Z., et al.\ 2016, The Astronomer's Telegram, 8784, 1.

\bibitem[Dessart et al.(2014)]{dessart} Dessart, L., Blondin, S., Hillier, D.~J., \& Khokhlov, A.\ 2014, \mnras, 441, 532 

\bibitem[Dhawan et al.(2016)]{dhawan16} Dhawan, S., Leibundgut, B., Spyromilio, J., et al.\ 2016, \aap, 588, A84

\bibitem[Dhawan et al.(2017)]{dhawan17} Dhawan, S., Leibundgut, B., Spyromilio, J., et al.\ 2017, \aap, 602, A118

\bibitem[Dimitriadis et al.(2019)]{dim19} Dimitriadis, G., Foley, R.~J., Rest, A., et al.\ 2019, \apjl, 870, L1 

\bibitem[Dhawan et al.(2018a)]{dhaw18} Dhawan, S., Jha, S.~W., \& Leibundgut, B.\ 2018, \aap, 609, A72 

\bibitem[Dhawan et al.(2018b)]{dhawan18} Dhawan, S., Bulla, M., Goobar, A., et al.\ 2018, \mnras, 480, 1445

\bibitem[Fink et al.(2010)]{fink10} Fink, M., R{\"o}pke, F.~K., Hillebrandt, W., et al.\ 2010, \aap, 514, A53 

\bibitem[Folatelli et al.(2010)]{fola11} Folatelli, G., Phillips, M.~M., Burns, C.~R., et al.\ 2010, \aj, 139, 120

\bibitem[Foley et al.(2012)]{foley12} Foley, R.~J., Challis, P.~J., Filippenko, A.~V., et al.\ 2012, \apj, 744, 38 

\bibitem[Gagliano et al.(2016)]{16eoa} Gagliano, R., Post, R., Weinberg, E., et al.\ 2016, Transient Name Server Discovery Report 2016-516, 1.

\bibitem[Gagliano et al.(2017)]{17fms} Gagliano, R., Post, R., Weinberg, E., et al.\ 2017, Transient Name Server Discovery Report 2017-774, 1.

\bibitem[Goldstein, \& Kasen(2018)]{gold18} Goldstein, D.~A., \& Kasen, D.\ 2018, \apjl, 852, L33

\bibitem[Guillochon et al.(2017)]{james17} Guillochon, J., Parrent, J., Kelley, L.~Z., \& Margutti, R.\ 2017, \apj, 835, 64 

\bibitem[Guy et al.(2007)]{guy07} Guy, J., Astier, P., Baumont, S., et al.\ 2007, \aap, 466, 11 

\bibitem[Guy et al.(2010)]{guy10} Guy, J., Sullivan, M., Conley, A., et al.\ 2010, \aap, 523, A7 

\bibitem[Hosseinzadeh et al.(2017)]{hosse17} Hosseinzadeh, G., Sand, D.~J., Valenti, S., et al.\ 2017, \apjl, 845, L11 


\bibitem[Hsiao et al.(2007)]{hsiao07} Hsiao, E.~Y., Conley, A., Howell, D.~A., et al.\ 2007, \apj, 663, 1187 

\bibitem[Iben \& Tutukov(1984)]{iben84} Iben, I., Jr., \& Tutukov, A.~V.\ 1984, \apjs, 54, 335 

\bibitem[Itagaki(2017)]{17erp} Itagaki, K.\ 2017, Transient Name Server Discovery Report 2017-647, 1.

\bibitem[Jeffery(1999)]{jeff99} Jeffery, D.~J.\ 1999, arXiv e-prints, astro-ph/9907015

\bibitem[Jha et al.(1999)]{jha99} Jha, S., Garnavich, P.~M., Kirshner, R.~P., et al.\ 1999, \apjs, 125, 73 

\bibitem[Jha et al.(2007)]{jha07} Jha, S., Riess, A.~G., \& Kirshner, R.~P.\ 2007, \apj, 659, 122 

\bibitem[Jones et al.(2015)]{jones15} Jones, D.~O., Riess, A.~G., \& Scolnic, D.~M.\ 2015, \apj, 812, 31 


\bibitem[Kasen(2010)]{kasen10} Kasen, D.\ 2010, \apj, 708, 1025 

\bibitem[Kessler et al.(2009)]{kessler09} Kessler, R., Becker, A.~C., Cinabro, D., et al.\ 2009, \apjs, 185, 32 
\bibitem[Khatami \& Kasen(2018)]{kk18} Khatami, D.~K., \& Kasen, D.~N.\ 2018, arXiv:1812.06522 

\bibitem[Khokhlov(1991)]{khok91} Khokhlov, A.~M.\ 1991, \aap, 245, 114 

\bibitem[Kromer et al.(2010)]{kromer10} Kromer, M., Sim, S.~A., Fink, M., et al.\ 2010, \apj, 719, 1067 

\bibitem[Levanon, \& Soker(2019)]{levanon19} Levanon, N., \& Soker, N.\ 2019, \mnras, 486, 5528

\bibitem[Li et al.(2016)]{li16} Li, Z., Gonzalez, J.~E., Yu, H., Zhu, Z.-H., \& Alcaniz, J.~S.\ 2016, \prd, 93, 043014 

\bibitem[Li et al.(2019)]{li18} Li, W., Wang, X., Vink{\'o}, J., et al.\ 2019, \apj, 870, 12

\bibitem[Maoz et al.(2014)]{maoz14} Maoz, D., Mannucci, F., \& Nelemans, G.\ 2014, \araa, 52, 107 

\bibitem[Marion et al.(2016)]{marion16} Marion, G.~H., Brown, P.~J., Vink{\'o}, J., et al.\ 2016, \apj, 820, 92 

\bibitem[Matheson et al.(2012)]{mathe12} Matheson, T., Joyce, R.~R., Allen, L.~E., et al.\ 2012, \apj, 754, 19 

\bibitem[Mazzali et al.(2015)]{mazzali15} Mazzali, P.~A., Sullivan, M., Filippenko, A.~V., et al.\ 2015, \mnras, 450, 2631

\bibitem[Miller et al.(2018)]{miller18} Miller, A.~A., Cao, Y., Piro, A.~L., et al.\ 2018, \apj, 852, 100 

\bibitem[Miller et al.(2016)]{16bln} Miller, A.~A., Laher, R., Masci, F., et al.\ 2016, The Astronomer's Telegram, 8907, 1.

\bibitem[Mink(1997)]{wcst} Mink, D.~J.\ 1997, Astronomical Data Analysis Software and Systems VI, 249

\bibitem[Noebauer et al.(2017)]{noe17} Noebauer, U.~M., Kromer, M., Taubenberger, S., et al.\ 2017, \mnras, 472, 2787 

\bibitem[Nomoto et al.(1984)]{nomoto84} Nomoto, K., Thielemann, F.-K., \& Yokoi, K.\ 1984, \apj, 286, 644 

\bibitem[Pakmor et al.(2012)]{pakmor12} Pakmor, R., Kromer, M., Taubenberger, S., et al.\ 2012, \apjl, 747, L10 

\bibitem[Papadogiannakis et al.(2019)]{papa19} Papadogiannakis, S., Dhawan, S., Morosin, R., et al.\ 2019, \mnras, 485, 2343

\bibitem[Pereira et al.(2013)]{pereira13} Pereira, R., Thomas, R.~C., Aldering, G., et al.\ 2013, \aap, 554, A27

\bibitem[Perlmutter et al.(1999)]{pearl99} Perlmutter, S., Aldering, G., Goldhaber, G., et al.\ 1999, \apj, 517, 565 

\bibitem[Phillips(1993)]{phillips93} Phillips, M.~M.\ 1993, \apjl, 413, L105 

\bibitem[Piascik, \& Steele(2016)]{16alq} Piascik, A.~S., \& Steele, I.~A.\ 2016, The Astronomer's Telegram, 8991, 1.

\bibitem[Prieto et al.(2006)]{prieto06} Prieto, J.~L., Rest, A., \& Suntzeff, N.~B.\ 2006, \apj, 647, 501 

\bibitem[Pskovskii(1977)]{psk} Pskovskii, I.~P.\ 1977, \azh, 54, 1188 

\bibitem[Reindl et al.(2005)]{reindl05} Reindl, B., Tammann, G.~A., Sandage, A., et al.\ 2005, \apj, 624, 532

\bibitem[Rest et al.(2014)]{rest14} Rest, A., Scolnic, D., Foley, R.~J., et al.\ 2014, \apj, 795, 44 


\bibitem[Riess et al.(1998)]{riess98} Riess, A.~G., Filippenko, A.~V., Challis, P., et al.\ 1998, \aj, 116, 1009

\bibitem[Riess et al.(2012)]{riess12} Riess, A.~G., Macri, L., Casertano, S., et al.\ 2012, \apj, 752, 76 

\bibitem[Riess et al.(2016)]{riess16} Riess, A.~G., Macri, L.~M., Hoffmann, S.~L., et al.\ 2016, \apj, 826, 56 

\bibitem[Riess et al.(2007)]{riess07} Riess, A.~G., Strolger, L.-G., Casertano, S., et al.\ 2007, \apj, 659, 98 

\bibitem[Sand et al.(2017)]{17fgc} Sand, D.~J., Valenti, S., Tartaglia, L., et al.\ 2017, The Astronomer's Telegram, 10569, 1.

\bibitem[Scalzo et al.(2014)]{scalzo14a} Scalzo, R., Aldering, G., Antilogus, P., et al.\ 2014, \mnras, 440, 1498 


\bibitem[Scalzo et al.(2019)]{scalzo18} Scalzo, R.~A., Parent, E., Burns, C., et al.\ 2019, \mnras, 483, 628 


\bibitem[Scolnic et al.(2014)]{scolnic14} Scolnic, D., Rest, A., Riess, A., et al.\ 2014, \apj, 795, 45 


\bibitem[Seitenzahl et al.(2013)]{seite13} Seitenzahl, I.~R., Ciaraldi-Schoolmann, F., R{\"o}pke, F.~K., et al.\ 2013, \mnras, 429, 1156 

\bibitem[Shappee et al.(2019)]{shap19} Shappee, B.~J., Holoien, T.~W.-S., Drout, M.~R., et al.\ 2019, \apj, 870, 13 

\bibitem[Sim et al.(2010)]{sim10} Sim, S.~A., R{\"o}pke, F.~K., Hillebrandt, W., et al.\ 2010, \apjl, 714, L52 

\bibitem[Sim et al.(2012)]{sim12} Sim, S.~A., Fink, M., Kromer, M., et al.\ 2012, \mnras, 420, 3003 

\bibitem[Stanek(2016)]{16ixb} Stanek, K.~Z.\ 2016, Transient Name Server Discovery Report 2016-1049, 1

\bibitem[Stanek(2017)]{17igf} Stanek, K.~Z.\ 2017, Transient Name Server Discovery Report 2017-1273, 1.

\bibitem[Stanek(2018)]{18oh} Stanek, K.~Z.\ 2018, Transient Name Server Discovery Report 2018-150, 1.

\bibitem[Stritzinger et al.(2006)]{strici06} Stritzinger, M., Leibundgut, B., Walch, S., et al.\ 2006, \aap, 450, 241

\bibitem[Stritzinger et al.(2018)]{strici18} Stritzinger, M.~D., Shappee, B.~J., Piro, A.~L., et al.\ 2018, arXiv:1807.07576

\bibitem[Tody(1986)]{iraf1} Tody, D.\ 1986, \procspie, 733

\bibitem[Tody(1993)]{iraf2} Tody, D.\ 1993, Astronomical Data Analysis Software and Systems II, 173

\bibitem[Tonry et al.(2016)]{16ffh} Tonry, J., Denneau, L., Stalder, B., et al.\ 2016, Transient Name Server Discovery Report 2016-583, 1.

\bibitem[Tonry et al.(2016)]{16gou} Tonry, J., Denneau, L., Stalder, B., et al.\ 2016, Transient Name Server Discovery Report 2016-718, 1.

\bibitem[Tonry et al.(2017)]{17hjy} Tonry, J., Stalder, B., Denneau, L., et al.\ 2017, Transient Name Server Discovery Report 2017-1123, 1.

\bibitem[Valenti et al.(2008)]{valenti08} Valenti, S., Benetti, S., Cappellaro, E., et al.\ 2008, \mnras, 383, 1485 

\bibitem[Valenti et al.(2017)]{17drh} Valenti, S., Sand, D.~J., \& Tartaglia, L.\ 2017, Transient Name Server Discovery Report 2017-513, 1.

\bibitem[van Rossum et al.(2016)]{vanrossum16} van Rossum, D.~R., Kashyap, R., Fisher, R., et al.\ 2016, \apj, 827, 128 

\bibitem[Vink{\'o} et al.(2012)]{vinko12} Vink{\'o}, J., S{\'a}rneczky, K., Tak{\'a}ts, K., et al.\ 2012, \aap, 546, A12 


\bibitem[Vink{\'o} et al.(2018)]{vinko18} Vink{\'o}, J., Ordasi, A., Szalai, T., et al.\ 2018, \pasp, 130, 064101 


\bibitem[Wang et al.(2013)]{wang13} Wang, X., Wang, L., Filippenko, A.~V., et al.\ 2013, Science, 340, 170

\bibitem[Wheeler et al.(2015)]{wheeler15} Wheeler, J.~C., Johnson, V., \& Clocchiatti, A.\ 2015, \mnras, 450, 1295 

\bibitem[Whelan \& Iben(1973)]{whelan73} Whelan, J., \& Iben, I., Jr.\ 1973, \apj, 186, 1007 

\bibitem[Wilk et al.(2018)]{wilk18} Wilk, K.~D., Hillier, D.~J., \& Dessart, L.\ 2018, \mnras, 474, 3187

\bibitem[Wood-Vasey et al.(2007)]{wv07} Wood-Vasey, W.~M., Miknaitis, G., Stubbs, C.~W., et al.\ 2007, \apj, 666, 694 

\bibitem[Woosley \& Weaver(1994)]{woosley94} Woosley, S.~E., \& Weaver, T.~A.\ 1994, \apj, 423, 371 

\bibitem[Zhang et al.(2016)]{zhang16} Zhang, K., Wang, X., Zhang, J., et al.\ 2016, \apj, 820, 67

\bibitem[Zhang et al.(2017)]{zhang17} Zhang, B.~R., Childress, M.~J., Davis, T.~M., et al.\ 2017, \mnras, 471, 2254 

\bibitem[Zheng et al.(2017)]{16coj} Zheng, W., Filippenko, A.~V., Mauerhan, J., et al.\ 2017, \apj, 841, 64.

\end{thebibliography}
\end{document}